\documentclass[titlepage,12pt]{article}
\usepackage{amsmath,amssymb,amscd,latexsym}
\usepackage[all]{xy}
\def\hybrid{
        \topmargin -20pt    
        \oddsidemargin 0pt
        \headheight 0pt
        \headsep 0pt
        \textwidth 6.25in       % A4 paper
        \textheight 9.5in       % A4 paper
        \marginparwidth .875in
        \parskip 5pt plus 1pt   
        \jot = 1.5ex}
\hybrid
\numberwithin{equation}{section}
\numberwithin{table}{section}
\numberwithin{figure}{section}

\DeclareMathOperator{\tr}{tr}
\DeclareMathOperator{\coker}{coker}
\DeclareMathOperator{\Ext}{Ext}
\DeclareMathOperator{\Imag}{Im}
\DeclareMathOperator{\Real}{Re}
\newcommand{\id}{\mathbf{1}}
\newcommand{\dd}{\textrm{d}}
\newcommand{\vol}{\textrm{vol}}
\newcommand{\rel}[1]{\underline{#1}}
\newcommand{\ins}[1]{\mathbf{i}_{#1}}

\newcommand{\ket}[1]{\lvert #1 \rangle}
\newcommand{\tbi}[3]{#1^{#2}_{\hphantom{#2}#3}}
\newcommand{\bti}[3]{#1_{#2}^{\hphantom{#2}#3}}
\newcommand{\tbundle}[1]{\textrm{T}#1}
\newcommand{\nbundle}[1]{\textrm{N}#1}
\newcommand{\sheaf}[1]{\mathcal{#1}}
\newcommand{\inv}[1]{{#1}^\mathbf{-1}}
\newcommand{\trans}[1]{{#1}^\textbf{T}}

\newcommand{\Riem}[4]{R_{#1\hphantom{#2}#3#4}^{\hphantom{#1}#2}}

\newcommand{\db}{\zeta}            % D7-brane fluctuations
\newcommand{\dbf}{f}               % D7-brane fluxes
\newcommand{\dbbf}{\mathcal{B}}    % D7-brane fluxes plus B-field
\newcommand{\cov}{\nabla}          % Gauge covariant derivative
\newcommand{\dil}{S}               % Modified dilaton

\begin{document}

%%%%%%%%%%%%%%%%%%%%%%%%%%%%%%%%%%%%%%%%%%%%%%%%%%%%%%%%%%%%%%

\begin{titlepage}
\begin{center}

\hfill hep-th/0409098\\
\vskip 3.5cm

{\large \bf The effective action of D7-branes\\[8pt]
in $\mathcal{N}=1$ Calabi-Yau orientifolds}\footnote{%
Work supported by: DFG -- The German Science Foundation,
European RTN Program HPRN-CT-2000-00148 and the
DAAD -- the German Academic Exchange Service.}\\

\vskip 2cm

{\bf Hans Jockers and Jan Louis}  \\
\vskip 1cm

{\em II. Institut f{\"u}r Theoretische Physik\\
Universit{\"a}t Hamburg\\
Luruper Chaussee 149\\
D-22761 Hamburg, Germany}\\
\vskip 10pt

{\tt  hans.jockers@desy.de, jan.louis@desy.de} \\

\end{center}

\vskip 2cm

\begin{center} {\bf ABSTRACT } \end{center}
%\vspace{-2mm}
Using a Kaluza-Klein reduction of the Dirac-Born-Infeld and Chern-Simons action we compute the four dimensional $\mathcal{N}=1$ effective action for the massless modes of a D7-brane which is wrapped on a four-cycle of a compact Calabi-Yau orientifold. We do not consider a specific orientifold but instead determine the K\"ahler potential, the gauge kinetic functions and the scalar potential in terms of geometrical data of a generic orientifold and its wrapped four-cycle. In particular we derive the couplings of the D-brane excitations to the bulk moduli of the orientifold as they are important for the study of soft supersymmetry breaking terms. We relate the resulting K\"ahler geometry to the $\mathcal{N}=1$ special geometry of Lerche, Mayr and Warner. Finally we comment on the structure of the D-term which is induced by a Green-Schwarz term in the Chern-Simons action. 

\noindent
\vfill

%\noindent \today
\noindent September 2004

\end{titlepage}

%%%%%%%%%%%%%%%%%%%%%%%%%%%%%%%%%%%%%%%%%%%%%%%%%%%%%%%%%%%%%%%%%%

\section{Introduction}

%%%%%%%%%%%%%%%%%%%%%%%%%%%%%%%%%%%%%%%%%%%%%%%%%%%%%%%%%%%%%%%%%%

Since the discovery of D-branes as non-perturbative states in string theory \cite{Polchinski:1995mt} they have been studied from many different perspectives. In particular D-branes have been used as a new ingredient in string model building in order to improve the link of string theory with particle physics and/or cosmology \cite{reviewPP,reviewcosmo}.  In these models one considers a stack of space-time filling D-branes which give rise to the Standard Model or rather some supersymmetric generalization thereof.  The D-branes are inserted into a space-time background which is a warped product of a flat four dimensional Minkowski space times a compact Calabi-Yau orientifold.  Including orientifold planes in the compactification is a consistency requirement and necessary in order to  satisfy the tadpole cancellation condition \cite{JP,PradisiGimon}.

The massless spectrum of such backgrounds consists of the excitations in the Calabi-Yau bulk and those on the D-branes.  The bulk contributes a gravitational multiplet, a set of Abelian vector multiplets and gauge neutral moduli multiplets.  The D-branes give rise to a non-Abelian gauge theory with charged matter multiplets.  Additional Wilson line moduli can appear when the wrapped part of the D-brane world-volume contains non-trivial one-cycles.  

The amount of supersymmetry, the mechanism of its spontaneous breaking and the communication of the breaking to the Standard Model are important phenomenological ingredients. Currently the most promising set-up chooses an $\mathcal{N}=1$ bulk theory with the D-branes respecting the same supersymmetry.\footnote{D-brane models with more supersymmetry have been analyzed, for example, in \cite{Frey:2002hf,Ferrara,BHK}.} This $\mathcal{N}=1$ is then spontaneously broken  by including background fluxes in the orientifold bulk \cite{Bachas:1995ik,Polchinski:1995sm,Michelson:1996pn,Gukov:1999ya,Dasgupta:1999ss,Taylor:1999ii,Mayr:2000hh,Curio:2000sc,Giddings:2001yu,BB,Kachru:2002he,BBHL,DWG,Giryavets:2003vd}. The supersymmetry  breaking is communicated to the Standard Model sector by the couplings of the matter fields on the branes to the bulk moduli and a set of soft supersymmetry breaking terms is generated \cite{Grana:2002,Kors:2003wf,Camara:2003ku,Grana:2003ek,Lawrence:2004zk,Lust:2004fi,Camara:2004jj}.

In addition the fluxes also stabilize some but generically not all of the moduli fields \cite{Polchinski:1995sm,Taylor:1999ii,Mayr:2000hh,Curio:2000sc,Giddings:2001yu,BB,Kachru:2002he,Blumenhagen:2002mf,Blumenhagen:2003vr,Cascales:2003pt,Cascales:2003wn,D'Auria:2004qv}. This potential instability can be cured by considering hidden D-branes which are geometrically separated from the Standard Model branes and thus all interactions are $\alpha'$-suppressed. For example, gaugino condensation on hidden D7-branes and/or Euclidean D3-brane instantons have been suggested as additional non-perturbative contributions to the potential \cite{KKLMT:2003,Denef:2004dm,Gorlich:2004qm}.

Recently branes have also received some attention in inflationary models of string cosmology \cite{reviewcosmo}.  It has been suggested in ref.~\cite{KKLMT:2003} that metastable deSitter vacua in type~IIB Calabi-Yau compactifications can be manufactured by adding anti-D3-branes. Another possibility, pointed out in ref.~\cite{Burgess:2003ic}, replaces the  anti-D3-branes by  D7-branes with internal background fluxes on the world-volume of the D7-brane. These fluxes are supposed to generate the positive energy necessary for deSitter vacua.  

In order to reliably determine the soft supersymmetry breaking terms and the vacuum energy it is necessary to compute the $\mathcal{N}=1$ low energy effective action of D-branes and in particular their couplings to the bulk excitations. The purpose of this paper is to perform such a computation for D7-branes in type IIB compactifications on generic Calabi-Yau orientifolds. For simplicity we will concentrate on a single D7-brane or in other words on an Abelian $U(1)$ gauge theory leaving the non-Abelian generalization to a future publication. In our analysis we do not specify a specific orientifold and derive the K\"ahler potential, the gauge kinetic function and the scalar potential in terms of geometrical data of the orientifold and its wrapped four-cycle by using a Kaluza-Klein reduction of the Dirac-Born-Infeld and Chern-Simons action. For space-time filling D3-branes a similar analysis has been performed in refs.~\cite{Camara:2003ku,Grana:2003ek}. We work at leading order in $\alpha'$ and do not include perturbative string corrections, which, however, are important for phenomenological applications \cite{BBHL,Balasubramanian:2004uy}.

In this paper we confine our attention to the derivation of the effective action without turning on three-form fluxes in the bulk and as a consequence we find no corresponding superpotential in the effective theory. We do, however, consider turning on two-form background flux on the wrapped D7-brane which contributes to the D-term potential. The situation where three-form fluxes are included has recently been studied in refs.~\cite{Lust:2004fi,Camara:2004jj}. In these papers, however, a K\"ahler potential of a specific model is used. In refs.~\cite{Lust:2004cx,Lust:2004fi} properties of the K\"ahler potential are computed using string scattering amplitudes but not a Kaluza-Klein reduction as we do here. Thus our paper can be viewed as closely related but complementary to refs.~\cite{Lust:2004fi,Camara:2004jj,Lust:2004cx}.

Specifically the organization of this paper is as follows. To set the stage for our analysis we review in section~\ref{sec:bulk} the massless $\mathcal{N}=1$ spectrum and its four dimensional low energy effective supergravity action resulting from IIB string theory with O3/O7 orientifold planes \cite{BBHL,Grimm:2004uq}. However, we slightly reformulated the effective theory in that we use the democratic IIB supergravity action of ref.~\cite{Bergshoeff:2001pv}. This form of the action will prove to be more convenient when D7-branes are coupled  to the bulk theory. 

In section~\ref{sec:brane} we analyze the D7-brane theory. Section~\ref{sec:spec} reviews the massless spectrum of the D7-brane while section~\ref{sec:DBIflux} discusses the D7-brane world-volume fluxes which are turned on. Section~\ref{sec:susycal} determines the geometrical condition (the calibration condition) on the four-cycle in order to have $\mathcal{N}=1$ supersymmetry. In section \ref{sec:D7} we perform the Kaluza-Klein reduction of the Dirac-Born-Infeld and Chern-Simons action in the Calabi-Yau orientifold bulk in the approximation where the complex structure deformations of the orientifold and the matter fields are considered to be small. 

In section~\ref{sec:total} we determine the standard form of the $\mathcal{N}=1$ effective low energy action. To do so we first need to impose the tadpole cancellation condition (section~\ref{sec:tadpole}) and then eliminate the auxiliary degrees of freedom in the democratic version of the action used so far (section~\ref{sec:effact}). Section~\ref{sec:chiral} then determines the K\"ahler coordinates, the K\"ahler potential, the gauge kinetic term and the D-term potential. We find that the dilaton has to be redefined and as a consequence couples non-trivially to the matter fields. Furthermore, both the matter fields and the Wilson line moduli couple to the K\"ahler moduli of the orientifold bulk. Analogously to orientifold compactifications with D3-branes \cite{BBHL,Grana:2003ek,Grimm:2004uq} the K\"ahler potential is implicitly defined. We discuss specific models where an explicit form of the K\"ahler potential can be obtained. The gauge coupling of the D7-branes is not the dilaton but the geometrical modulus which controls the size of the wrapped four-cycle. Finally a D-term arises (even in the absence of any fluxes) due to the presence of a Green-Schwarz term or equivalently due to a gauged translational isometry.

In section~\ref{sec:geom} we discuss the geometry of the K\"ahler potential and in particular consider the common moduli space of the D7-brane fluctuations and the complex structure deformations of the bulk orientifold theory. The resulting geometrical structures can be related to the $\mathcal{N}=1$ special geometry of refs.~\cite{Mayr:2001} where aspects of the combined moduli space of D5-brane moduli and bulk complex structure moduli on non-compact Calabi-Yau manifolds have been uncovered. For D7-brane orientifolds a similar analysis allows us to further generalize the K\"ahler potential to include the interplay of D7-brane matter fields and the bulk complex structure deformations. In section~\ref{sec:relform} and section~\ref{sec:var} we introduce the mathematical framework to address the geometry of the underlying moduli space, which we use in section~\ref{sec:kcs} to deduce the K\"ahler potential of the moduli space for these fields. Finally the results are implemented into the supergravity K\"ahler potential of section~\ref{sec:chiral}.

Section~\ref{sec:conc} contains our conclusions and some of the technical details are relegated to three appendices. In appendix~\ref{sec:normal} we review the normal coordinate expansion of tensor fields which is needed in the reduction of the Dirac-Born-Infeld and Chern-Simons action. In appendix~\ref{sec:ints} we present some details of the computation of the effective four dimensional D7-brane action while appendix~\ref{sec:3form} supplements the analysis of the gauge kinetic coupling functions.

%%%%%%%%%%%%%%%%%%%%%%%%%%%%%%%%%%%%%%%%%%%%%%%%%%%%%%%%%%%%%%%%%%

\section{Type IIB compactified on Calabi-Yau orientifolds} \label{sec:bulk}

%%%%%%%%%%%%%%%%%%%%%%%%%%%%%%%%%%%%%%%%%%%%%%%%%%%%%%%%%%%%%%%%%%

This analysis begins with a review of compactifying type IIB string theory on a Calabi-Yau orientifold $Y$ in the presence of O3/O7 orientifold planes as has been considered in refs.~\cite{Acharya:2002ag,Brunner:2003zm,Giddings:2001yu,BBHL,DWG,Grimm:2004uq}. Usually in order to get a low energy effective description of theories originating from type IIB string theory, a natural starting point is the chiral ten dimensional $\mathcal{N}=2$ type IIB supergravity \cite{Marcus:1982yu,Dall'Agata:1998va}. In our case, however, it is more convenient to start with the democratic type IIB supergravity action \cite{Bergshoeff:2001pv}, because it facilitates the derivation of the whole action containing both the bulk and brane fields.\footnote{The advantage of using the democratic action has also been observed in ref.~\cite{Gukov:2002iq}.} This type IIB supergravity action is comprised of all RR~form fields of type IIB string theory at the cost of introducing additional self-duality constraints. 

%%%%%%%%%%%%%%%%%%%%%%%%%%%%%%%%%%%%%%%%%%%%%%%%%%%%%%%%%%%%%%%%%%

\subsection{Type IIB orientifold spectrum with O3/O7 planes}

%%%%%%%%%%%%%%%%%%%%%%%%%%%%%%%%%%%%%%%%%%%%%%%%%%%%%%%%%%%%%%%%%%

The massless bosonic spectrum of type IIB string theory in ten dimensions consists of fields from the NS-NS~sector and the RR~sector. The former sector gives rise to the ten dimensional metric $g_{10}$, the anti-symmetric two tensor $B$ and the dilaton $\phi$. The RR~sector contributes all even dimensional anti-symmetric tensors, i.e. the form fields $C^{(0)}$, $C^{(2)}$, $C^{(4)}$, $C^{(6)}$, and $C^{(8)}$. The field strengths of these form fields, as they appear in the low energy effective action, are given by \cite{Bergshoeff:2001pv}
\begin{equation} \label{eq:fs}
   G^{(p)}=\begin{cases} 
              \dd C^{(0)} & p=1 \\ 
              \dd C^{(p-1)}-\dd B\wedge C^{(p-2)} & \text{else} \ .
           \end{cases} 
\end{equation}
Note, however, that the number of physical degrees of freedom is reduced by imposing the duality conditions
\begin{align} \label{eq:dual}
   G^{(1)}=*_{10}G^{(9)} \ , && G^{(3)}=(-1)*_{10}G^{(7)} \ , && G^{(5)}=*_{10}G^{(5)} \ .
\end{align}

As the ten dimensional space-time background we take the product ansatz $\mathbb{R}^{3,1}\times Y/\mathcal{O}$ where the internal Calabi-Yau orientifold $Y/\mathcal{O}$ is a compact Calabi-Yau manifold~$Y$ moded out by the orientifold projection $\mathcal{O}$.\footnote{For ease of notation we denote in the following both the Calabi-Yau orientifold and the Calabi-Yau manifold by $Y$.} The corresponding metric takes the form\footnote{For Calabi-Yau compactifications with localized sources such as orientifold planes and/or D-branes one really has to make a warped ansatz for the metric so as to capture the back-reaction of these localized sources to geometry \cite{Giddings:2001yu,DWG,deAlwisBuchel}. This, however, is beyond the scope of this paper. We perform our analysis in a regime, where the internal space is large enough that this back-reaction can be treated as a negligible perturbation to our product ansatz \eqref{eq:met}.} 
\begin{equation} \label{eq:met}
   \dd s_{10}^2 = \hat\eta_{\mu\nu}\:\dd x^\mu\dd x^\nu+
                  2\:\hat g_{i\bar\jmath}(y)\:\dd y^i \dd \bar y^{\bar\jmath} \ ,
\end{equation}
where $\hat\eta_{\mu\nu}$ is the flat metric of the four dimensional Minkowski space and $\hat g_{i\bar\jmath}(y)$ is the metric of the internal Calabi-Yau manifold~$Y$. The orientifold projection $\mathcal{O}$ acting on type IIB string states is given by \cite{Acharya:2002ag,Brunner:2003zm}
\begin{equation} \label{eq:proj}
   \mathcal{O}=(-1)^{F_L}\Omega_p \sigma^* \ ,
\end{equation}
where $F_L$ is the fermion number for the left-movers and $\Omega_p$ is the world-sheet parity operator. Finally $\sigma$ is an isometric involution on the Calabi-Yau manifold $Y$ acting via pullback on the type IIB fields. In order to obtain a Calabi-Yau orientifold model with O3/O7 planes the involution $\sigma$ must be holomorphic with the additional property \cite{Acharya:2002ag,Brunner:2003zm}
\begin{equation} \label{eq:invOmega}
   \sigma^*\Omega=-\Omega \ ,
\end{equation}
where $\Omega$ denotes the unique $(3,0)$-form of the Calabi-Yau manifold~$Y$.

The massless Kaluza-Klein spectrum of type~IIB string theory compactified on a six dimensional Calabi-Yau orientifold~$Y$ can essentially be obtained in two steps: First of all, before taking the orientifold projection, the ten dimensional fields are expanded in harmonics of the Calabi-Yau threefold~$Y$ and one obtains the massless $\mathcal{N}=2$ Kaluza-Klein spectrum in four dimensions. The information about this spectrum is encoded in the Hodge diamond of the Calabi-Yau threefold $Y$. The next step is to truncate the $\mathcal{N}=2$ spectrum further to $\mathcal{N}=1$ by just keeping the states invariant under the orientifold projection~$\mathcal{O}$ \cite{Acharya:2002ag,Brunner:2003zm,Grimm:2004uq,Andrianopoli:2001}. The resulting spectrum is the massless $\mathcal{N}=1$ Kaluza-Klein spectrum in four dimensions for type~IIB string theory compactified on the Calabi-Yau orientifold~$Y$. 

As $\sigma$ is holomorphic the harmonic forms of $H^{(p,q)}_{\bar\partial}(Y)$ split naturally into positive and negative eigenforms under the pullback $\sigma^*$ \cite{Brunner:2003zm,Grimm:2004uq}. In the following we split the cohomology groups $H^{(p,q)}_{\bar\partial}(Y)$ into its eigenspaces $H^{(p,q)}_{\bar\partial,\pm}(Y)$ under $\sigma^*$. Our chosen basis for all the cohomology spaces $H^{(p,q)}_{\bar\partial,\pm}(Y)$ are summarized in Table~\ref{tab:coh}
\begin{table}
\begin{center}
\begin{tabular}{|c|c|c||c|c|c|}
   \hline
      \bf space  &  \bf basis  &  \bf dimension  &
      \bf space  &  \bf basis  &  \bf dimension  
      \rule[-1.5ex]{0pt}{4.5ex} \\
   \hline
   \hline
      $H^{(1,1)}_{\bar\partial,+}(Y)$  &  $\omega_\alpha$
      &  $\alpha=1,\ldots,h^{1,1}_+$ 
      &  $H^{(1,1)}_{\bar\partial,-}(Y)$  &  $\omega_a$
      &  $a=1,\ldots,h^{1,1}_-$
      \rule[-1.5ex]{0pt}{4.5ex} \\
   \hline
      $H^{(2,2)}_{\bar\partial,+}(Y)$  &  $\tilde\omega^\alpha$
      &  $\alpha=1,\ldots,h^{2,2}_+$ 
      &  $H^{(2,2)}_{\bar\partial,-}(Y)$  &  $\tilde\omega^a$
      &  $a=1,\ldots,h^{1,1}_-$
      \rule[-1.5ex]{0pt}{4.5ex} \\
   \hline
      $H^{3}_+(Y)$  &  $\alpha_{\hat\alpha},\beta^{\hat\alpha}$ 
      &  $\hat\alpha=1,\ldots,h^{2,1}_+$
      &  $H^{3}_-(Y)$  &  $\alpha_{\hat a},\beta^{\hat a}$ 
      &  $\hat a=0,\ldots,h^{2,1}_-$
      \rule[-1.5ex]{0pt}{4.5ex} \\
    \hline
      $H^{(2,1)}_{\bar\partial,+}(Y)$  &  $\chi_{\tilde\alpha}$  
      &  $\tilde\alpha=1,\ldots,h^{2,1}_+$
      &  $H^{(2,1)}_{\bar\partial,-}(Y)$  &  $\chi_{\tilde a}$
      &  $\tilde a=1,\ldots,h^{2,1}_-$
      \rule[-1.5ex]{0pt}{4.5ex} \\
    \hline  
      $H^{(1,2)}_{\bar\partial,+}(Y)$  &  $\bar\chi_{\tilde\alpha}$  
      &  $\tilde\alpha=1,\ldots,h^{2,1}_+$
      &  $H^{(1,2)}_{\bar\partial,-}(Y)$  &  $\bar\chi_{\tilde a}$
      &  $\tilde a=1,\ldots,h^{2,1}_-$
      \rule[-1.5ex]{0pt}{4.5ex} \\
    \hline  
\end{tabular} 
\caption{Cohomology basis} \label{tab:coh} 
\end{center}
\end{table}
and it obeys
\begin{equation}
\begin{aligned}
   \int_Y\omega_\alpha\wedge\tilde\omega^\beta
       =\delta_\alpha^\beta \ , \qquad && \qquad
   \int_Y\omega_a\wedge\tilde\omega^b
       =\delta_a^b \ , \\
   \int_Y\alpha_{\hat\alpha}\wedge\beta^{\hat\beta}
       =\delta_{\hat\alpha}^{\hat\beta} \ , \qquad && \qquad
   \int_Y\alpha_{\hat a}\wedge\beta^{\hat b}
       =\delta_{\hat a}^{\hat b} \ ,  
\end{aligned}
\end{equation}
with all other pairings vanishing. 

Now we have introduced all the technical tools to expand the type~IIB fields in harmonics invariant under the orientifold projection $\mathcal{O}$. The dilaton $\phi$, the metric $g$ (and therefore the K\"ahler form $J$) and the RR~zero-form $C^{(0)}$, four-form $C^{(4)}$ and eight-form $C^{(8)}$ are even under the world-sheet parity operator $\Omega_p$. The remaining fields, i.e. $B$, $C^{(2)}$ and $C^{(6)}$, have odd parity. Thus for the NS-NS~fields we arrive at the expansion
\begin{align} \label{eq:NS}
   J&=v^\alpha(x)\:\omega_\alpha \ , & B&=b^a(x)\:\omega_a \ , & \phi&=\phi(x) \ ,
\end{align}
and for the RR~fields
\begin{equation} \label{eq:C}
\begin{split}
   C^{(8)}&=\tilde l^{(2)}(x)\wedge\frac{\Omega\wedge\bar\Omega }{\int_Y\Omega\wedge\bar\Omega}
            \ , \\
   C^{(6)}&=\tilde c^{(2)}_a(x)\wedge\tilde\omega^a \ , \\
   C^{(4)}&=D_{(2)}^\alpha(x)\wedge\omega_\alpha
            +V^{\hat\alpha}(x)\wedge \alpha_{\hat\alpha}
            +U_{\hat\alpha}(x)\wedge \beta^{\hat\alpha}
            +\rho_\alpha(x)\:\tilde\omega^\alpha \ , \\
   C^{(2)}&=c^a(x)\:\omega_a \ , \\
   C^{(0)}&=l(x) \ .
\end{split}
\end{equation}
In the effective four dimensional theory $v^\alpha$, $b^a$, $\phi$, $\rho_\alpha$, $c^a$ and $l$ are scalar fields, $V^{\hat\alpha}$ and $U_{\hat\alpha}$ are vector fields, and $\tilde l^{(2)}$, $\tilde c_a^{(2)}$ and $D^\alpha_{(2)}$ are two-form tensor fields.

In addition to the above fields we have complex scalars, which arise from the complex structure deformations of the internal space. For the case of O3/O7 orientifold compactifications the complex structure deformations are in one-to-one correspondence with the elements of $H_{\bar\partial,-}^{(2,1)}(Y)$ \cite{Brunner:2003zm,Candelas:1990pi}, and we denote the corresponding four dimensional scalar fields by $z^{\tilde a}$. 

The resulting four dimensional $\mathcal{N}=1$ supergravity spectrum of the O3/O7 orientifold model is summarized in Table~\ref{tab:sp}. Note that in this table only the physical degrees of freedom are listed, i.e.\ the duality conditions \eqref{eq:dual} are taken into account. In four dimensions this
duality relates a massless scalar to a massless two-form or a gauge boson to its magnetic dual. In terms of the fields given in eqs.~\eqref{eq:C} the duality \eqref{eq:dual} corresponds to the dual pairs $\tilde l^{(2)}\sim\tilde l, c^{(2)}_a\sim c^a, D^\alpha_{(2)}\sim \rho_\alpha,V^{\hat\alpha}\sim U_{\hat\alpha}$. 
\begin{table}
\begin{center}
\begin{tabular}{|c|c|c||c|c|c|}
   \hline
      \bf multiplet  &  \bf multiplicity &  \bf bos. fields &
      \bf multiplet  &  \bf multiplicity &  \bf bos. fields \rule[-1.5ex]{0pt}{4.5ex} \\
   \hline
   \hline
      gravity  &  $1$  &  $g_{\mu\nu}$  &
      chiral &  $h_-^{1,1}$  &  $(b^a,c^a)$ \rule[-1.5ex]{0pt}{4.5ex} \\
   \hline
      vector  &  $h_+^{2,1}$  &  $V^{\hat\alpha}_\mu$  & 
      chiral  &  $h_+^{1,1}$  &  $(\rho_\alpha,v^\alpha)$ \rule[-1.5ex]{0pt}{4.5ex} \\
   \hline
      chiral &  $1$  &  $(l,\phi)$  &
      chiral &  $h_-^{2,1}$  &  $z^{\tilde a}$ \rule[-1.5ex]{0pt}{4.5ex} \\
   \hline
\end{tabular} 
\caption{$\mathcal{N}=1$ multiplets} \label{tab:sp} 
\end{center}
\end{table}

%%%%%%%%%%%%%%%%%%%%%%%%%%%%%%%%%%%%%%%%%%%%%%%%%%%%%%%%%%%%%%%%%%

\subsection{Democratic low energy effective bulk action}

%%%%%%%%%%%%%%%%%%%%%%%%%%%%%%%%%%%%%%%%%%%%%%%%%%%%%%%%%%%%%%%%%%

As the Chern-Simons action of the D7-brane, which we discuss in detail in section~\ref{sec:D7}, couples to all RR~forms, we start from the democratic action of type~IIB supergravity \cite{Bergshoeff:2001pv}. This action contains both the usual RR~forms of IIB~supergravity and also their dual RR~forms. Therefore the equations of motion resulting from this action have to be supplemented by the duality constraints \eqref{eq:dual}.\footnote{Note that already for the usual ten dimensional IIB~supergravity the same phenomenon appears, namely for the self-dual four-form $C^{(4)}$ the self-duality condition on its five-form field strength must be imposed by hand \cite{Marcus:1982yu,Dall'Agata:1998va}.} The bosonic part of this action in the string frame reads \cite{Bergshoeff:2001pv}
\begin{equation} \label{eq:IIBdemo}
\begin{split}
   \mathcal{S}_\text{IIB}^\text{sf}
   \ =&\ \frac{1}{2\kappa_{10}^2}\int\dd^{10}x\sqrt{-g_{10}}\: e^{-2\phi} R 
   \ - \ \frac{1}{4\kappa_{10}^2}\int e^{-2\phi}
         \left(8\: \dd\phi\wedge *_{10}\dd\phi-H\wedge *_{10}H\right) \\
   \ +&\ \frac{1}{8\kappa_{10}^2}\int\sum_{p=1,3,5,7,9} G^{(p)}\wedge *_{10}G^{(p)} \ ,
\end{split}
\end{equation}
where $\kappa_{10}$ is the ten dimensional gravitational coupling constant. The field strengths $G^{(p)}$ are defined in \eqref{eq:fs} and $H$ is the field strength $H = \dd B$.

After having inserted the expansions \eqref{eq:NS} and \eqref{eq:C} into the ten dimensional action \eqref{eq:IIBdemo}, we integrate out the internal Calabi-Yau space $Y$ and perform a Weyl rescaling of the metric \eqref{eq:met} 
\begin{equation} \label{eq:Weyl}
   \hat\eta\ =\ \frac{6}{\mathcal{K}}\, e^{\phi/2}\,  \eta \ ,
   \qquad\hat g\ =\ e^{\phi/2}\, g \ .
\end{equation}
Then the effective four dimensional action Weyl-rescaled to the Einstein frame is found to be
\begin{equation} \label{eq:4Dbulk}
\begin{split}
   \mathcal{S}_\text{Bulk}^\text{E} 
   =&\frac{1}{2\kappa_4^2}\int \left[-R\:*_4 1
        +2\mathcal{G}_{\tilde a\tilde b}\dd z^{\tilde a}\wedge*_4\dd\bar z^{\tilde b}
        +2G_{\alpha\beta}\dd v^\alpha \wedge *_4 \dd v^\beta \right. \\
   &+\frac{1}{2}\dd(\ln \mathcal{K})\wedge *_4 \dd(\ln \mathcal{K})
        +\frac{1}{2}\dd\phi\wedge *_4 \dd\phi 
        +\frac{1}{4} e^{2\phi} \dd l\wedge *_4\dd l \\
   &+2e^{-\phi}G_{ab}\dd b^a\wedge\dd b^b
        +e^\phi G_{ab}(\dd c^a-l\dd b^a) \wedge *_4(\dd c^b-l\dd b^b)  \\
   &+\frac{\mathcal{K}^2}{36} G_{\alpha\beta} \dd D_{(2)}^\alpha\wedge 
        *_4 \dd D_{(2)}^\beta + \frac{9}{4\mathcal{K}^2} 
        G^{\alpha\beta}\dd\rho_\alpha\wedge *_4 \dd \rho_\beta  \\
   &+\frac{9}{4\mathcal{K}^2}\left(\mathcal{K}_{ab\gamma}\mathcal{K}_{cd\delta}G^{\gamma\delta}
        +\mathcal{K}_{abe}\mathcal{K}_{cdf}G^{ef}\right) (db^a\wedge c^b)
        \wedge *_4(db^c\wedge c^d)  \\
   &-\frac{9}{2\mathcal{K}^2}\mathcal{K}_{ab\beta}G^{\beta\gamma}\dd\rho_\gamma
        \wedge *_4(\dd b^a\wedge c^b) 
        +\frac{1}{16} e^{-\phi} G^{ab} \dd\tilde c^{(2)}_a \wedge 
        *_4 \dd\tilde c^{(2)}_b \\
   &+\frac{1}{16}e^{-\phi}\mathcal{K}_{ab\gamma}\mathcal{K}_{cd\delta}G^{bd}
        \left(\dd b^a\wedge D^\gamma_{(2)}\right)\wedge *_4
        \left(\dd b^c\wedge D^\delta_{(2)}\right) \\
   &-\frac{1}{8}e^{-\phi}\mathcal{K}_{ab\gamma}G^{bc}
        \left(\dd b^a\wedge D^\gamma_{(2)}\right)\wedge *_4\dd\tilde c^{(2)}_c
        +\frac{1}{4}e^{-2\phi} \dd\tilde l^{(2)}\wedge *_4\dd\tilde l^{(2)} \\
   &+\frac{1}{4}e^{-2\phi} \left(\dd b^a\wedge\tilde c^{(2)}_a\right)
        \wedge *_4\left(\dd b^b \wedge \tilde c^{(2)}_b\right)
        +\frac{1}{2} e^{-2\phi} \left(\dd b^a\wedge\tilde c^{(2)}_a\right)
        \wedge *_4 \dd\tilde l^{(2)} \\
   &+\left.\frac{1}{4} B_{\hat\alpha\hat\beta} \dd V^{\hat\alpha}\wedge *_4\dd V^{\hat\beta}-
        \frac{1}{4} C^{\hat\alpha\hat\beta} \dd U_{\hat\alpha}\wedge *_4\dd U_{\hat\beta} 
        -\frac{1}{2}\bti{A}{\hat\beta}{\hat\alpha} 
        \dd U_{\hat\alpha}\wedge *_4\dd V^{\hat\beta} \right] \ .
\end{split}
\end{equation}
$\kappa_4$ is now the four dimensional gravitational coupling constant and
\begin{align} \label{eq:triple}
   \mathcal{K}_{\alpha\beta\gamma}
     =\int_Y \omega_{\alpha}\wedge\omega_{\beta}\wedge\omega_{\gamma} \ , 
   &&\mathcal{K}_{ab\gamma}
     =\int_Y \omega_a\wedge\omega_b\wedge\omega_\gamma \ ,
\end{align}
are the non-vanishing triple intersection numbers of the Calabi-Yau manifold~$Y$. Note that these intersection numbers are topological invariants of the Calabi-Yau manifold~$Y$ which are symmetric in their indices. The intersection numbers $\mathcal{K}_{\alpha\beta c}$ vanish because the volume form $\dd\vol(Y)$ of the Calabi-Yau is even whereas $\omega_\alpha\wedge\omega_\beta\wedge\omega_c$ is odd with respect to the pullback $\sigma^*$ \cite{Grimm:2004uq,Andrianopoli:2001}. Additionally we define contractions of these intersection numbers with the fields $v^\alpha$ and obtain with \eqref{eq:NS} all non-vanishing combinations 
\begin{equation} \label{eq:K}
\begin{aligned}
   \mathcal{K}&=\int_Y J\wedge J\wedge J
      =\mathcal{K}_{\alpha\beta\gamma}v^\alpha v^\beta v^\gamma \ , 
   & \mathcal{K}_\alpha&=\int_Y \omega_\alpha\wedge J\wedge J
      =\mathcal{K}_{\alpha\beta\gamma}v^\beta v^\gamma \ , \\
   \mathcal{K}_{\alpha\beta}&=\int_Y \omega_\alpha\wedge\omega_\beta\wedge J
      =\mathcal{K}_{\alpha\beta\gamma}v^\gamma \ , 
   & \mathcal{K}_{ab}&=\int_Y \omega_a\wedge\omega_b\wedge J
      =\mathcal{K}_{ab\gamma}v^\gamma \ ,
\end{aligned}
\end{equation}
where $\mathcal{K}$ is proportional to the volume of the internal Calabi-Yau manifold~$Y$, i.e. $6\:\vol (Y)=\mathcal{K}$.

In the action \eqref{eq:4Dbulk} there appear also various metrics. On the space of harmonic two-forms one defines the metrics \cite{Strominger:1985ks,Candelas:1990pi}
\begin{equation} \label{eq:metK}
\begin{split}
   G_{\alpha\beta}&=\frac{3}{2\mathcal{K}}\int_Y\omega_\alpha\wedge *_6\omega_\beta 
      =-\frac{3}{2}\left(\frac{\mathcal{K}_{\alpha\beta}}{\mathcal{K}}-\frac{3}{2}
       \frac{\mathcal{K}_\alpha\mathcal{K}_\beta}{\mathcal{K}^2}\right) \ , \\
   G_{ab}&=\frac{3}{2\mathcal{K}}\int_Y\omega_a\wedge *_6\omega_b
      =-\frac{3}{2}\frac{\mathcal{K}_{ab}}{\mathcal{K}} \ , 
\end{split}
\end{equation}
which is just the usual metric for the space of K\"ahler deformations split into odd and even part with respect to the involution $\sigma$. The inverse metrics of \eqref{eq:metK} are denoted by $G^{\alpha\beta}$ and $G^{ab}$. Similarly, for the complex structure deformations $z^{\tilde a}$ one defines the special K\"ahler metric \cite{Candelas:1990pi}
\begin{align} \label{eq:CSt}
   \mathcal{G}_{\tilde a\tilde b}\
      =\ \frac{\partial^2}{\partial  z^{\tilde a}\partial \bar z^{\tilde b}} \
      K_\text{CS}(z,\bar z) \ , &&
      K_\text{CS}(z,\bar z)\ =\ -\ln \left(-i\int_Y\Omega \wedge \bar\Omega\right) \ ,
\end{align}
which is the metric on the complex structure moduli space of the Calabi-Yau manifold~$Y$ restricted to the complex structure deformations compatible with the holomorphic involution $\sigma$ \cite{Brunner:2003zm}.

Finally we introduce the coefficient matrices of the kinetic terms of the vector fields $V^{\hat\alpha}$ and $U^{\hat\alpha}$. They are given by \cite{SuzukiCeresole}
\begin{equation} \label{eq:mat3}
\begin{aligned}
   \bti{A}{\hat\beta}{\hat\alpha}
       &=-\int_Y \beta^{\hat\alpha}\wedge *_6 \alpha_{\hat\beta} \ , 
   \qquad & \qquad B_{\hat\alpha\hat\beta}
       &=\int_Y \alpha_{\hat\alpha}\wedge *_6 \alpha^{\hat\beta} \ , \\
   C^{\hat\alpha\hat\beta}
       &=-\int_Y \beta^{\hat\alpha}\wedge *_6 \beta^{\hat\beta} \ , 
   \qquad & \qquad \tbi{D}{\hat\beta}{\hat\alpha}
       &=\int_Y \alpha_{\hat\alpha}\wedge *_6 \beta^{\hat\beta} \ ,
\end{aligned}
\end{equation}
or equivalently for the three-forms $*_6\alpha_{\hat\alpha}$ and $*_6\beta^{\hat\alpha}$ we find modulo exact forms the expansion
\begin{align}
   *_6\alpha_{\hat\alpha}=\bti{A}{\hat\alpha}{\hat\beta}\alpha_{\hat\beta}
        +B_{\hat\alpha\hat\beta}\beta^{\hat\beta} \ , 
   &&*_6\beta^{\hat\alpha}=C^{\hat\alpha\hat\beta}\alpha_{\hat\beta}
        +\tbi{D}{\hat\alpha}{\hat\beta}\beta^{\hat\beta} \ .
\end{align}
It is straight forward to verify that the matrices \eqref{eq:mat3} fulfill
\begin{align}
   \trans{A}=-D \ , && \trans{B}=B \ , && \trans{C}=C \ .
\end{align}

In the four dimensional democratic action \eqref{eq:4Dbulk} we could now impose the (dimensional reduced) duality condition \eqref{eq:dual} by adding Lagrangian multiplier terms and integrate out the redundant degrees of freedom \cite{Dall'Agata:2001zh,Louis:2002ny,Gukov:2002iq} to obtain the effective orientifold action of refs.~\cite{Grimm:2004uq}. However, as there arise additional couplings of the brane fields to the RR~forms, we postpone this dualization procedure until we have the combined action of both bulk and brane fields.

%%%%%%%%%%%%%%%%%%%%%%%%%%%%%%%%%%%%%%%%%%%%%%%%%%%%%%%%%%%%%%%%%%

\section{D7-brane: Spectrum and effective action} \label{sec:brane}

%%%%%%%%%%%%%%%%%%%%%%%%%%%%%%%%%%%%%%%%%%%%%%%%%%%%%%%%%%%%%%%%%%

In this section we add to the bulk theory of the previous section a space-time filling D7-brane while preserving $\mathcal{N}=1$ supersymmetry at the same time. First we introduce the new degrees of freedom arising from this extended object in section~\ref{sec:spec}. The possibility of turning on flux on the D-brane world-volume is discussed in \ref{sec:DBIflux}. In order to preserve the $\mathcal{N}=1$ supersymmetry of the bulk theory in the presence of D7-branes a geometrical consistency condition is imposed. This calibration condition is studied in section~\ref{sec:susycal}. After this interlude on supersymmetry the Dirac-Born-Infeld action and the Chern-Simons action of the D7-brane are reduced to their effective four dimensional Lagrangians in \ref{sec:D7}. The effective action of the D7-brane couples to all fields of the RR~forms, namely to both the four dimensional scalars and their dual two-forms. 

%%%%%%%%%%%%%%%%%%%%%%%%%%%%%%%%%%%%%%%%%%%%%%%%%%%%%%%%%%%%%%%%%

\subsection{D7-brane spectrum} \label{sec:spec}

%%%%%%%%%%%%%%%%%%%%%%%%%%%%%%%%%%%%%%%%%%%%%%%%%%%%%%%%%%%%%%%%%%

Before we enter the discussion of the spectrum of the D7-brane, let us clarify the geometric picture of the setup which we have in mind. We concentrate on a single space-time filling D7-brane with gauge group $U(1)$. The internal part of its world-volume is wrapped on a four cycle $S^{(1)}$ of the Calabi-Yau manifold~$Y$. Since we are working with an orientifold theory we must in addition to the brane wrapped on $S^{(1)}$ also include its image under the orientifold involution $\sigma$, i.e.\ we have an image D7-brane on the Calabi-Yau manifold~$Y$ wrapped on the four cycle $S^{(2)}=\sigma(S^{(1)})$. Note that this is the geometry in the covering space of the orientifold theory. In the orientifold space~$Y/\mathcal{O}$ the D7-brane and its image D7-brane coincide and represent a single object. Hence it is convenient to introduce the four cycle $S^\Lambda$ which is the union of the cycles $S^{(1)}$ and $S^{(2)}$ in the Calabi-Yau manifold~$Y$. $S^\Lambda$ obeys
\begin{equation}
  \sigma(S^\Lambda)=S^\Lambda \ .
\end{equation}
The Poincar\'e dual two-form $\omega_\Lambda$ of $S^{\Lambda}$ is an element of $H^2_+(Y)$. By referring to the D7-brane we mean in the following the object which wraps the internal cycle $S^{\Lambda}$ and thus describes both the D7-brane and its image in the covering space of the Calabi-Yau orientifold. For later convenience we further define $S^P$ as the union of the cycle $S^{(1)}$ and its orientation reversed image $-S^{(2)}$. This cycle obeys 
\begin{equation}
   \sigma(S^P)=-S^P \ ,
\end{equation}
and has a Poincar\'e dual two-form $\omega_P$ in $H^2_-(Y)$. In Table~\ref{tab:cycles} all these different D7-brane four cycles are listed with their associated Poincar\'e dual two-forms.
\begin{table}
\begin{center}
\begin{tabular}{|c|c|c|c|}
   \hline
      \bf description  &  \bf cycle  &  \bf relation  &  \bf Poincar\'e dual
      \rule[-1.5ex]{0pt}{4.5ex} \\
   \hline
   \hline
      D7-brane (covering space)  &  $S^{(1)}\in H_4(Y,\mathbb{Z})$  
      &  $\tfrac{1}{2}(S^\Lambda + S^P)$ 
      &  $\omega_{(1)}\in H_{\bar\partial}^{(1,1)}(Y)$ 
      \rule[-1.5ex]{0pt}{4.5ex} \\
   \hline
      D7-image-brane (covering space)  &  $S^{(2)}\in H_4(Y,\mathbb{Z})$  
      &  $\tfrac{1}{2}(S^\Lambda - S^P)$ 
      &  $\omega_{(2)}\in H_{\bar\partial}^{(1,1)}(Y)$
      \rule[-1.5ex]{0pt}{4.5ex} \\
   \hline
      D7-brane (orientifold $Y$)  &  $S^\Lambda\in H_4(Y,\mathbb{Z})$  
      &  $S^{(1)}+S^{(2)}$ 
      &  $\omega_\Lambda\in H_{\bar\partial,+}^{(1,1)}(Y)$ 
      \rule[-1.5ex]{0pt}{4.5ex} \\
   \hline
      D7-pair (opposite orientation)  &  $S^P\in H_4(Y,\mathbb{Z})$  
      &  $S^{(1)}-S^{(2)}$ 
      &  $\omega_P\in H_{\bar\partial,-}^{(1,1)}(Y)$
      \rule[-1.5ex]{0pt}{4.5ex} \\      
   \hline
\end{tabular} 
\caption{D7-brane cycles} \label{tab:cycles} 
\end{center}
\end{table}

In the above analysis we have implicitly assumed that the D7-brane does not coincide with any orientifold O7-plane, because this would imply that $S^{(1)}$ and $S^{(2)}$ represent the same cycle and that the gauge group of the world-volume theory is $SO(N)$ or $USp(N)$ \cite{Polchinski:1998}. Additionally we require that the involution $\sigma$ does not have any fixed points in $S^{(1)}$ since this would give rise to extra massless states in the twisted open string sector \cite{Berkooz:1996km}.

The spectrum of the D7-brane is comprised of two parts. The first part corresponds to fluctuations of the embedding of the internal four cycle $S^\Lambda$ in the Calabi-Yau orientifold~$Y$, and the second part describes Wilson lines of the $U(1)$ gauge field on the four cycle $S^\Lambda$. Both types of degrees of freedom give rise to bosonic components of chiral multiplets in the effective four dimensional low energy theory. The former complex bosons are members of the `matter' multiplets and we denote them by $\db(x)$. The latter Wilson line moduli are denoted by $a(x)$.

The analysis of the spectrum of the D7-brane is carried out in two steps. First we describe the massless spectrum of the D7-brane on $S^\Lambda$ and neglect the orientifold action~$\mathcal{O}$. Then this spectrum is truncated by keeping only states invariant under $\mathcal{O}$.

Mathematically, the massless open string degrees of freedom are given in terms of sections of appropriate sheaf cohomology groups \cite{Witten:1992fb,Katz:2002gh,Mayr:2001}. For the cycle $\iota:S^\Lambda\hookrightarrow Y$, where the map $\iota$ embeds $S^\Lambda$ in $Y$, we have the exact sequence
\begin{equation} \label{eq:TNexact}
   0\rightarrow\tbundle{S^\Lambda}\xrightarrow{\iota_*}\tbundle{Y}|_{S^\Lambda}
    \rightarrow\nbundle{S^\Lambda}\rightarrow 0 \ ,
\end{equation} 
which defines $\nbundle{S^\Lambda}$ the normal bundle of $S^\Lambda$ in terms of $\tbundle{S^\Lambda}$ the tangent bundle of $S^\Lambda$ and  $\tbundle{Y}|_{S^\Lambda}$ the tangent bundle of $Y$ restricted to the cycle $S^\Lambda$. Then the massless fields $\db$ and $a$ are sections of the sheaf 
cohomology groups\footnote{In general, D-branes are described by sheaves that are supported on the world-volume of the brane. Then the spectrum of marginal open string modes of strings stretching from a D-brane specified by the sheaf $\sheaf{E}$ to a D-brane specified by the sheaf $\sheaf{F}$ is given by the $\Ext$-group $\Ext^1(\sheaf{E},\sheaf{F})$ \cite{Katz:2002gh}. However, if \eqref{eq:TNexact} splits holomorphically, which we will always assume in the following, the $\Ext$-group reduces to the sheaf cohomology description, and in the case of open strings with both ends on a single brane, we have the above spectrum \eqref{eq:sp1}.} \cite{Mayr:2001}
\begin{align} \label{eq:sp1}
   \db\in H^0(S^\Lambda,\nbundle{S^\Lambda}) \ , && a\in H^1(S^\Lambda,\sheaf{O}) \ .
\end{align}

In order to generalize to the orientifold case, as a second step we must take into account the truncation of the spectrum due to the orientifold projection $\mathcal{O}=\Omega_p \sigma$. The action of $\Omega_p\sigma$ on a general open string state $\ket{\Psi,ij}$ with Chan-Paton indices $i$ and $j$ is given by \cite{PradisiGimon,Douglas:1996sw}
\begin{equation}
   \Omega_p\sigma:\ \ket{\Psi,ij}\rightarrow \left(\gamma_{\Omega_p\sigma}\right)_{ik}
       \ket{\Omega_p\sigma\cdot\Psi,lk} \left(\gamma_{\Omega_p\sigma}^{-1}\right)_{lj} \ ,
\end{equation}
where the indices $i$ and $j$ label the set of branes, and thus we have in our case $i,j=1,2$ for the brane $S^{(1)}$ and its image brane $S^{(2)}$. Note that the parity operator $\Omega_p$ has exchanged the indices $l\leftrightarrow k$. The matrix $\gamma_{\Omega_p\sigma}$ with the adjoint action on the Chan-Paton indices is unitary, and since $(\Omega_p\sigma)^2=\id$ one deduces that $\gamma_{\Omega_p\sigma}=\pm \gamma_{\Omega_p\sigma}^\mathbf{T}$ \cite{PradisiGimon}. In our basis of the Chan-Paton indices the state $\ket{\Psi,11}$ labels an open string state with both endpoints on the brane $S^{(1)}$ and $\ket{\Psi,22}$ labels an open string state with both endpoints on the image brane $S^{(2)}$. As the orientifold action~$\mathcal{O}$ maps an open string state on $S^{(1)}$ to an open string state on $S^{(2)}$ it implies that the unitary matrix $\gamma_{\Omega_p\sigma}$ must have the form
\begin{equation}
   \gamma_{\Omega_p\sigma}=
      \begin{pmatrix} 0 & e^{i\theta} \\ \pm e^{i\theta} & 0 \end{pmatrix} \ ,
\end{equation}
and therefore 
\begin{equation} \label{eq:oaction}
   \Omega_p\sigma:\
     \begin{aligned}
        \ket{\Psi,11}&\rightarrow \ket{\Omega_p\sigma\cdot\Psi,22} \\
        \ket{\Psi,22}&\rightarrow \ket{\Omega_p\sigma\cdot\Psi,11} \ .
     \end{aligned}
\end{equation}
The action of the world-sheet parity operator $\Omega_p$ acts with a minus sign on the massless vertex operators tangent and with a plus sign on the vertex operators normal to the world-volume of the brane \cite{PradisiGimon}. As the complex fields $\db$ are sections in the normal bundle and the complex fields $a$ are sections in the tangent bundle, the spectrum of the invariant massless open string states becomes in the cohomology description
\begin{align} \label{eq:sp2}
   \db\in H^0_+(S^\Lambda,\nbundle{S^\Lambda})\ , 
   && a\in H^1_-(S^\Lambda,\sheaf{O}) \ .
\end{align} 
(The fact that $\db$ lies in the positive eigenspace of $H^0(S^\Lambda,\nbundle{S^\Lambda})$ agrees with the geometric picture, because a fluctuation of the brane should give rise to the same fluctuation of the image brane.)

More generally for a stack of $N$ D7-branes wrapped on $S^\Lambda$ the gauge theory on the world-volume of these branes is enhanced to $U(N)$ \cite{Witten:1995im}. As a consequence the massless fields $\db$ and $a$ transform in the adjoint representation of $U(N)$, i.e. $\db$ is a $U(N)$ Lie algebra valued section of the normal bundle and $a$ is a $U(N)$ Lie algebra valued one-form \cite{Douglas:1997}. In the following we consider just a single D7-brane wrapped on $S^\Lambda$ with the spectrum spectrum given by \eqref{eq:sp2}.

For the derivation of the effective action let us introduce a basis $\{s_A\}$ for $H^0_+(S^\Lambda,\nbundle{S^\Lambda})$ and the complex conjugate basis $\{\bar s_{\bar B}\}$. In terms of these basis elements we can expand any fluctuation $\db(x,y)$ of the world-volume of the D7-brane into
\begin{equation} \label{eq:fluc}
   \db(x,y)=\db^A(x)\:s_A(y)+\bar\db^{\bar A}(x)\:\bar s_{\bar A}(y) \ ,
   \quad A=1,\ldots,\dim H^0_+(S^\Lambda,\nbundle{S^\Lambda}) \ .
\end{equation}   
As discussed in ref.~\cite{Mayr:2001}, we can map sections $\db$ of $H^0_+(S^\Lambda,\nbundle{S^\Lambda})$ isomorphically to $H_{\bar \partial,-}^{(2,0)}(S^\Lambda)$ via the Poincar\'e residue map. In practice this map is simply given by contracting $\db$ with the unique holomorphic three form $\Omega$ of the Calabi-Yau manifold~$Y$ \cite{Mayr:2001}
\begin{equation} \label{eq:PR}
   \Omega : \ H^0_+(S^\Lambda,\nbundle{S^\Lambda}) \rightarrow
      H_{\bar\partial,-}^{(2,0)}(S^\Lambda),\ \db \mapsto \ins{\db} \Omega \ .
\end{equation}
The Poincar\'e residue map also allows us to rewrite the basis elements $\{s_A\}$ and $\{\bar s_{\bar A}\}$ into basis elements $\{\tilde s_A\}$ and $\{\tilde s_{\bar B}\}$ of $H_{\bar\partial,-}^{(2,0)}(S^\Lambda)$ and $H_{\bar\partial,-}^{(0,2)}(S^\Lambda)$, and hence via \eqref{eq:PR} the expansion \eqref{eq:fluc} corresponds to an expansion in two-forms of $S^\Lambda$. Similarly, we use $\{A^I\}$ and $\{\bar A^{\bar J}\}$ as a basis of $H_{\bar\partial,-}^{(0,1)}(S^\Lambda)$ and $H_{\bar\partial,-}^{(1,0)}(S^\Lambda)$, and we obtain for the $U(1)$ gauge boson $A(x,y)$ localized on the world-volume of the brane the expansion in harmonics of $S^\Lambda$
\begin{equation} \label{eq:A}
   A(x,y)=A_\mu(x)\dd x^\mu\:P_-(y)+a_I(x)\:A^I(y)+\bar a_{\bar J}(x)\:\bar A^{\bar J}(y) \ , 
\end{equation}
where $P_-$ is a harmonic zero form of $S^\Lambda$ given by
\begin{equation}\label{eq:P}
   P_-(y)=\begin{cases} 1 & y\in S^{(1)} \\ -1 & y\in S^{(2)} \end{cases} \ .
\end{equation}
Note that $\sigma^*P_-=-P_-$, and hence $P_-$ is an element of $H_-^0(S^\Lambda)$. 

In Table~\ref{tab:spec} the massless open string spectrum resulting from the D7-brane is summarized. The table shows the cohomology groups, which describe the spectrum in geometric terms, and lists the basis elements thereof. Note that for the matter fields $\db$ there are alternative descriptions related by \eqref{eq:PR}.
\begin{table}
\begin{center}
\begin{tabular}{|c|c|c|c|}
   \hline
      \bf chiral multiplet  &  \bf bosonic fields  
         &  \bf geometric space &  \bf basis \rule[-1.5ex]{0pt}{4.5ex} \\
   \hline
   \hline
  matter  &  $\db^A$, $A=1,\ldots, \dim H^0_+(S^\Lambda,\nbundle{S^\Lambda})$ 
         &  $H^0_+(S^\Lambda,\nbundle{S^\Lambda})$  &  $\{s_A\}$ 
         \rule[-1.5ex]{0pt}{4.5ex} \\
   &  &  $H_{\bar\partial,-}^{(2,0)}(S^\Lambda)$  &  $\{\tilde s_A\}$ 
         \rule[-1.5ex]{0pt}{4.5ex} \\
   \hline
  Wilson lines &  $a_I$, $I=1,\ldots, \dim H_{\bar\partial,-}^{(0,1)}(S^\Lambda)$
         &  $H_{\bar\partial,-}^{(0,1)}(S^\Lambda)$   &  $\{A^I\}$ 
         \rule[-1.5ex]{0pt}{4.5ex} \\
   \hline
\end{tabular} 
\caption{Massless D7-brane spectrum} \label{tab:spec} 
\end{center}
\end{table}

%%%%%%%%%%%%%%%%%%%%%%%%%%%%%%%%%%%%%%%%%%%%%%%%%%%%%%%%%%%%%%%%%%

\subsection{Dirac-Born-Infeld action and brane fluxes} \label{sec:DBIflux}

%%%%%%%%%%%%%%%%%%%%%%%%%%%%%%%%%%%%%%%%%%%%%%%%%%%%%%%%%%%%%%%%%%

In a low energy effective description the kinetic terms of the brane degrees of freedom of a single D-brane and their couplings to the bulk NS-NS~fields are captured by the Abelian Dirac-Born-Infeld action, which reads for a D$p$-brane in the string frame 
\begin{align} \label{eq:DBIab}
   \mathcal{S}^{\text{sf}}_{\text{DBI}}
      =-\mu_p\int_\mathcal{W} \dd^{p+1}\xi
       \:e^{-\phi}\sqrt{-\det \left(\varphi^*(g_{10}+B)_{ab}-\ell F_{ab}\right)} \ , 
   && \ell=2\pi\alpha' \ ,
\end{align}
where $\alpha'$ is the string coupling constant. The action couples to the ten dimensional bulk metric $g_{10}$ and to the NS-NS two-form field $B$ via the pullback of the map $\varphi$. This map describes the embedding of the $p+1$ dimensional world-volume $\mathcal{W}$ in the ten dimensional space-time manifold. The dynamics of the D$p$-brane, namely the dynamics of the degrees of freedom which describe the fluctuations of the world-volume $\mathcal{W}$ in the space-time manifold, is encoded in the pullback $\varphi^*$. $F$ is the field strength of the $U(1)$ gauge boson $A$ localized on the world-volume $\mathcal{W}$ of the D$p$-brane. Finally the coupling constant $\mu_p$ is the tension of the D$p$-brane, which is the absolute value of its RR~charge for BPS branes.  

A D$p$-brane is not completely characterized by its embedding of the world-volume~$\mathcal{W}$ into the space-time manifold, because branes may carry lower dimensional RR~brane charges \cite{Douglas:1995bn}. As these charges are not localized but rather distributed over the world-volume of the brane, they appear in the form of background fluxes $\dbf$ of the $U(1)$ gauge theory of the brane. Thus in the presence of these fluxes, the field strength $F$ of \eqref{eq:DBIab} has the form
\begin{equation}
   F=\dbf+\dd A \ ,
\end{equation}
where $A$ is the $U(1)$ gauge boson and $\dbf$ is a harmonic two-form of the world-volume of the D$p$-brane.

In order to describe the low energy physics of the space-time filling D7-brane, we take the action \eqref{eq:DBIab}, where now the integral extends over the eight dimensional world-volume $\mathcal{W}=\mathbb{R}^{3,1}\times S^\Lambda$. On the D7-brane we also include background brane fluxes~$\dbf$ which preserve Poincar\'e invariance of the four dimensional space-time $\mathbb{R}^{3,1}$. This restricts possible background fluxes only to be non-trivial in the internal four cycle $S^\Lambda$. Due to the negative parity of the $U(1)$ gauge boson \eqref{eq:A} of the brane in orientifolds the corresponding background flux~$f$ must be an element of $H_-^2(S^\Lambda)$. For simplicity we limit ourselves in the following to fluxes~$f$, which can be solely expanded into negative parity harmonic two-forms inherited from the ambient Calabi-Yau manifold~$Y$,\footnote{It is also possible to turn on fluxes along two-cycles of $S^\Lambda$ which are trivial in the Calabi-Yau. A thorough discussion of such fluxes is beyond the scope of this paper and will be discussed elsewhere.} i.e.
\begin{equation} \label{eq:ftyp}
   \dbf=\dbf^a\:\iota^*\omega_a \ .
\end{equation}
Recall that $\omega_a$ is a basis of $H_-^2(Y)$ (c.f.\ Table~\ref{tab:coh}) and $\iota$ is the map which embeds the cycle $S^\Lambda$ into $Y$ (c.f.\ \eqref{eq:TNexact}).

For later convenience we define the quantity $\dbbf$, which is the combination of the bulk NS-NS $B$-field and the background flux $\dbf$ and which is given by
\begin{equation}
   \dbbf=\iota^*B-\ell f \ .
\end{equation}
For fluxes of the form \eqref{eq:ftyp} $\dbbf$ enjoys the expansion
\begin{equation} \label{eq:bf}
   \dbbf^a(x)\:\iota^*\omega_a = \left(b^a(x)-\ell f^a\right)\:\iota^*\omega_a \ ,
\end{equation}
where $f^a$ are the brane flux quanta and $b^a$ are the four dimensional scalar fields \eqref{eq:NS}.
 
%%%%%%%%%%%%%%%%%%%%%%%%%%%%%%%%%%%%%%%%%%%%%%%%%%%%%%%%%%%%%%%%%%

\subsection{Supersymmetry and calibration condition} \label{sec:susycal}

%%%%%%%%%%%%%%%%%%%%%%%%%%%%%%%%%%%%%%%%%%%%%%%%%%%%%%%%%%%%%%%%%%

Before entering the discussion of calibration conditions for BPS D7-branes, we first recall some general aspects of supersymmetry in orientifold models with branes. A compactification of type~IIB string theory on a Calabi-Yau manifold leaves eight supercharges unbroken corresponding to $\mathcal{N}=2$ in four dimensions. In this case one finds two linear independent supersymmetry parameters $\xi$ and $\eta$ for which $\delta_\xi(\text{fermions})=0$ and $\delta_\eta(\text{fermions})=0$. If a (Super-)D$p$-brane is included into the theory, the new fermionic degrees of freedom $\Psi$ of the brane also vary with the supersymmetry transformations but in general obey neither $\delta_\xi\Psi=0$ nor $\delta_\eta\Psi=0$. However, the (Super-)D$p$-brane has an extra fermionic local symmetry called $\kappa$-symmetry \cite{Hughes:1986fa}, and supersymmetry is unbroken if it is possible to compensate the supersymmetry variation by a $\kappa$-symmetry transformation \cite{Becker:1995kb}
\begin{equation} \label{eq:BPSbrane}
   \delta\Psi=\delta_\epsilon\Psi+\delta_\kappa\Psi=0 \ ,
\end{equation} 
where $\epsilon$ is some linear combination of $\xi$ and $\eta$. If this condition can be fulfilled for some parameter $\epsilon$ the brane breaks only half of the supercharges and saturates a BPS bound. The conditions for BPS D-branes in Calabi-Yau manifolds translate into calibration conditions in geometry \cite{Becker:1995kb,Marino:1999af,Cascales:2004qp}. Thus if we add BPS D-branes to a Calabi-Yau compactification four of eight supercharges remain unbroken. 

A Calabi-Yau orientifold compactification can be seen as gauging a discrete $\mathbb{Z}_2$ symmetry of a Calabi-Yau manifold compactification \cite{Acharya:2002ag}. In our case this symmetry is given by \eqref{eq:proj} and arises from a holomorphic involution $\sigma$ of the Calabi-Yau manifold. This $\mathbb{Z}_2$ symmetry also acts on the supercharges \cite{Brunner:2003zm}, and hence only one particular linear combination of $\xi$ and $\eta$ generate a supersymmetry transformation compatible with this discrete symmetry. Thus Calabi-Yau orientifolds are four dimensional $\mathcal{N}=1$ theories. Furthermore these orientifolds with additional branes preserves also $\mathcal{N}=1$ supersymmetry, if the condition \eqref{eq:BPSbrane} is fulfilled for the same supersymmetry parameter $\epsilon$ which also corresponds to the linear combination invariant under the orientifold $\mathbb{Z}_2$ symmetry. 

In order to determine whether the space-time filling D7-brane in a Calabi-Yau compactification is a BPS state, the condition \eqref{eq:BPSbrane} must be evaluated and as shown in ref.~\cite{Marino:1999af} implies that the internal four cycle $S^\Lambda$ of the D7-brane world-volume must satisfy the calibration equation
\begin{equation} \label{eq:cali2}
   \dd^4\xi \sqrt{\det \left(\hat g+\dbbf^a\:\iota^*\omega_a\right)}
   \ =\ \frac{1}{2}e^{-i\theta}
     \left(J+i\dbbf^a\:\iota^*\omega_a\right)
           \wedge\left(J+i\dbbf^b\:\iota^*\omega_b\right) \ .
\end{equation}
Here the real constant $\theta$ parametrizes the unbroken supersymmetry variation as a linear combination of $\xi$ and $\eta$ of the previous paragraph. $\dbbf^a$ is defined in \eqref{eq:bf}, $J$ is the K\"ahler form of the Calabi-Yau threefold~$Y$ pulled back to $S^\Lambda$, and $\hat g$ is the metric on $S^\Lambda$ induced from the Calabi-Yau manifold.\footnote{For ease of notation we use the letter $J$ for both the K\"ahler form $J$ of the Calabi-Yau manifold~$Y$ and for the K\"ahler form $\iota^*J$ of $S^\Lambda$. Similarly the metric $\hat g$ stands for $\hat g$ as well as $\iota^*\hat g$.} 

Note that the left hand side of \eqref{eq:cali2} is real, and hence for vanishing $\dbbf$ we find $\theta=0$, and therefore we recover the calibration condition 
\begin{equation} \label{eq:cali}
   \dd^4\xi \sqrt{\det \hat g}\ =\ \frac{1}{2} J\wedge J \ ,
\end{equation}
which has already been derived in refs.~\cite{Becker:1995kb}.\footnote{Recently in ref.~\cite{Cascales:2004qp} the calibration condition has also been rederived for the case of compactifications with non-trivial background fluxes for the bulk fields.} However, in the case of non-trivial $\dbbf$, reality of the calibration condition tells us that the imaginary part of the right hand side of \eqref{eq:cali2} should vanish, namely
\begin{equation} \label{eq:susy1}
   \cos\:\theta\ \left(\dbbf^a\:\iota^*\omega_a\wedge J\right)\ =\
     \sin\:\theta\ \left(\tfrac{1}{2}J\wedge J
     -\tfrac{1}{2}\:\dbbf^a\:\iota^*\omega_a
     \wedge\dbbf^b\:\iota^*\omega_b\right) \ .
\end{equation}
The latter condition has to be fulfilled for all points in $S^\Lambda$, i.e. it has to hold on the brane and its mirror brane. Since $\theta$ is a constant on $S^\Lambda$ we obtain from \eqref{eq:susy1} the condition 
\begin{equation} \label{eq:susy2}
   \cos\:\theta\ \left(\dbbf^a\:\iota^*\omega_a\wedge J\right)\ =\
     \pm\sin\:\theta\ \left(\tfrac{1}{2}J\wedge J
     -\tfrac{1}{2}\dbbf^a\:\iota^*\omega_a
     \wedge\dbbf^b\:\iota^*\omega_b\right) \ ,
\end{equation}
by applying the holomorphic involution $\sigma$. Therefore we arrive at the two supersymmetry constraints 
\begin{equation} \label{eq:susy3}
\begin{split}
   &\sin\:\theta \ \left(J\wedge J
       -\dbbf^a\:\iota^*\omega_a\wedge\dbbf^b\:\iota^*\omega_b\right)\ =\ 0 \ , \\
   &\cos\:\theta \ \left(\dbbf^a\:\iota^*\omega_a\wedge J\right)\ =\ 0 \ .
\end{split}
\end{equation}
From the first condition we deduce that $\theta$ must still vanish in the supersymmetric case.\footnote{$\dbbf$ is taken to be small and then $J\wedge J-\dbbf\wedge\dbbf\neq 0$.} 
Since $\theta$ parameterizes the linear combination of supersymmetry 
parameters preserved by the D7-brane one indeed expects
that supersymmetry fixes $\theta$ in orientifolds. 

The second condition of \eqref{eq:susy3} implies $\dbbf\wedge J = 0$, which integrated over $S^P$ and using \eqref{eq:K} reads 
\begin{equation} \label{eq:susy4}
   \mathcal{K}_{Pa}\dbbf^a\ =\ 0 \ .
\end{equation}
This is directly related to the $\omega$-stability condition $J\wedge\dbbf=\text{const.}\ J\wedge J$ of refs.~\cite{Brunner:1999jq,Harvey:1996gc} which is imposed by supersymmetry. In orientifolds $\dbbf$ is odd and $J$ is even which implies $\dbbf\wedge J=0$ as the $\omega$-stability condition. Moreover it is argued in ref.~\cite{Brunner:1999jq}, that $\omega$-stability gives rise to a D-term constraint in the low energy effective action, i.e. if supersymmetry is broken $\omega$-stability is not fulfilled and the non-vanishing D-term breaks supersymmetry spontaneously. Conversely, a $\omega$-stable configuration corresponds to a vanishing D-term in field theory. Thus in the low energy effective theory which we derive in the next section, we expect a D-term potential $V_\text{D}$ proportional to \eqref{eq:susy4}
\begin{equation} \label{eq:D1}
   V_\text{D} \sim \left(\mathcal{K}_{Pa}\dbbf^a\right)^2 \ .
\end{equation}
In section~\ref{sec:chiral} we will see that such a D-term is indeed required by supersymmetry and moreover we will determine the (field dependent) proportionality constant of \eqref{eq:D1}.

Now we are prepared to perform the Kaluza-Klein reduction of the D-brane action. For its derivation we use the calibration condition rescaled with \eqref{eq:Weyl} to Einstein frame
\begin{equation} \label{eq:cali3}
   \dd^4\xi\sqrt{\det\left(e^{\phi/2}g+\dbbf^a\:\iota^*\omega_a\right)}
   \ =\ \tfrac{1}{2}e^\phi J\wedge J
      -\tfrac{1}{2}\dbbf^a\:\iota^*\omega_a\wedge \dbbf^b\:\iota^*\omega_b \ .
\end{equation}
As we have just argued this calibration only holds for BPS D7-branes
which satisfy $\dbbf\wedge J=0$. However, in order to also derive the D-term
we allow for the possibility of small perturbations in $\dbbf$ 
which do not obey $\dbbf\wedge J =0$. The deviation from \eqref{eq:cali3} 
is then taken into account by adding \eqref{eq:D1} to the scalar potential of the effective action.

%%%%%%%%%%%%%%%%%%%%%%%%%%%%%%%%%%%%%%%%%%%%%%%%%%%%%%%%%%%%%%%%%%

\subsection{Reduction of the D7-brane action} \label{sec:D7}

%%%%%%%%%%%%%%%%%%%%%%%%%%%%%%%%%%%%%%%%%%%%%%%%%%%%%%%%%%%%%%%%%%

The first task in this section is the reduction of the bosonic part of the Abelian Dirac-Born-Infeld action \eqref{eq:DBI} for the space-time filling D7-brane with world-volume $\mathcal{W}=\mathbb{R}^{3,1}\times S^\Lambda$. To obtain the effective four dimensional fields describing the fluctuations of the internal cycle $S^\Lambda$ in the compactified six dimensions, we perform a normal coordinate expansion of the pullback metric $\varphi^*g_{10}$ and the pullback two-form $\varphi^*B$ as described in ref.~\cite{Grana:2003ek}. For convenience we recall the relevant formulae of this procedure in appendix~\ref{sec:normal}.
Applying \eqref{eq:PB} and \eqref{eq:PB2} to the metric given in
\eqref{eq:met} and the two-from $B$ we obtain after Weyl rescaling 
the metric according to \eqref{eq:Weyl}  
\begin{equation}\label{eq:nc}
\begin{split}
   \varphi^*g_{10}&=\frac{6}{\mathcal{K}}e^{\phi/2}\:\eta_{\mu\nu}\:\dd x^\mu\dd x^\nu
      +2\:e^{\phi/2}\:g_{i\bar\jmath}\:\dd y^i\dd\bar y^{\bar\jmath}
      +2\:e^{\phi/2}\:g_{i\bar\jmath}\:\partial_\mu\db^i\partial_\nu\bar\db^{\bar\jmath}
      \:\dd x^\mu\dd x^\nu \ , \\
   \varphi^*B&=b^a\:\iota^*\omega_a
      +b_{i\bar\jmath}\:\partial_\mu\db^i\partial_\nu\bar\db^{\bar\jmath}\:\dd x^\mu\dd x^\nu \ , 
\end{split}
\end{equation}
where the normal vector fields $\db$ and $\bar\db$ are part of the D7-brane spectrum given by \eqref{eq:sp2} and its conjugate. Now we insert 
\eqref{eq:nc} into \eqref{eq:DBIab} and expand the determinant
with the help of the Taylor series
\begin{equation}
   \sqrt{\det\left(\mathfrak{A}+t\mathfrak{B}\right)}
     =\sqrt{\det\mathfrak{A}}\cdot\left[1+\frac{t}{2}\tr \mathfrak{A}^{-1}\mathfrak{B}
      +\frac{t^2}{8}\left[\left(\tr\mathfrak{A}^{-1}\mathfrak{B}\right)^2
      -2\:\tr \left(\mathfrak{A}^{-1}\mathfrak{B}\right)^2\right]+\cdots \right] \ ,
\end{equation}  
evaluated for $t=1$. This yields the effective four dimensional action of the Dirac-Born-Infeld Lagrangian for the massless Kaluza-Klein modes \eqref{eq:fluc}, \eqref{eq:A} and \eqref{eq:NS}. The next task is to insert the BPS calibration condition \eqref{eq:cali3} and up to second order in derivatives we obtain in the Einstein frame 
\begin{align} \label{eq:DBI}
   \mathcal{S}^{\text{E}}_{\text{DBI}}
      =&\mu_7 \ell^2\int\left[\frac{1}{4}\left(\mathcal{K}_\Lambda
        -e^{-\phi}\mathcal{K}_{\Lambda ab}\dbbf^a\dbbf^b\right) F\wedge *_4 F 
        +\frac{12}{\mathcal{K}}i\mathcal{C}^{I\bar J}_\alpha v^\alpha
        \dd a_I\wedge *_4\dd\bar a_{\bar J}\right] \\
      +&\mu_7 \int\left[i\mathcal{L}_{A\bar B}\left(e^\phi-G_{ab}\dbbf^a\dbbf^b\right)
        \dd\db^A\wedge *_4\dd\bar\db^{\bar B}+\frac{18}{\mathcal{K}^2} 
        \left(e^\phi\mathcal{K}_\Lambda-\mathcal{K}_{\Lambda ab}\dbbf^a\dbbf^b \right) *_4 1
        \right] \ , \nonumber
\end{align}
where
\begin{equation} 
   \mathcal{L}_{A\bar B}=\frac{\int_{S^\Lambda}\tilde s_A\wedge\tilde s_{\bar B}}
                               {\int_Y\Omega\wedge\bar\Omega}  \label{eq:L} \ , 
\end{equation}
and 
\begin{equation}
   \mathcal{C}_\alpha^{I\bar J}=\int_{S^\Lambda}\iota^*\omega_\alpha\wedge A^I\wedge\bar A^{\bar J}
      \label{eq:Cint} \ .
\end{equation}
The details of the derivation of \eqref{eq:DBI} are presented in appendix~\ref{sec:ints}. 
The first term in \eqref{eq:DBI} 
is the kinetic term of the field strength~$F$ of the $U(1)$ gauge boson arising from the gauge theory of the space-time filling part of the world-volume~$\mathcal{W}$ of the D7-brane. 
The next two terms are the kinetic terms for the Wilson line moduli of the D7-brane  and the matter fields (see \eqref{eq:sp2}).
Finally the last term is a potential term.
Note that it is proportional 
to the inverse square of the gauge coupling and thus can be identified as a 
D-term potential. We further discuss this term in section~\ref{sec:tadpole}.

The next step is to analyze the couplings of the brane fields to the remaining fields of the bulk. As D$p$-branes are extended objects carrying RR~charges \cite{Polchinski:1995mt}, they must couple to the bulk RR~fields. These couplings are captured in the Chern-Simons action of the D$p$-brane which is given in the Abelian case by  
\begin{equation} \label{eq:CS}
   \mathcal{S}_{\text{CS}}=\mu_p \int_{\mathcal{W}}
       \sum_q \varphi^*\left(C^{(q)}\right)e^{\ell F-\varphi^*B} \ .
\end{equation}
The exponential in the integrand of \eqref{eq:CS} is meant as a formal power series in the two-form $\ell F-\varphi^*B$ wedged with the RR~forms of the bulk theory, which are pulled back to the world-volume of the brane. As in the case of the Dirac-Born-Infeld action \eqref{eq:DBIab} the integral is taken over the $p+1$ dimensional world-volume~$\mathcal{W}$ of the brane, and therefore the only non-vanishing contributions of the integral arise from $p+1$-forms in the power series of the integrand.

As before the Chern-Simons action of the space-time filling D7-brane is integrated over the world-volume $\mathcal{W}=\mathbb{R}^{(3,1)}\times S^\Lambda$. In order to arrive at the effective action of \eqref{eq:CS} in four dimensions the first task is to perform a normal coordinate expansion of the pullback tensors according \eqref{eq:PB2}. Next we insert the massless Kaluza-Klein modes of \eqref{eq:NS}, \eqref{eq:C}, \eqref{eq:A} and \eqref{eq:fluc}, and finally we obtain up to second order in derivatives the effective four dimensional Chern-Simons action 
\begin{align} \label{eq:CS1}
   \mathcal{S}_\text{CS}&=
      \mu_7\int\left(\tfrac{1}{4}\dd\tilde l^{(2)}
        -\dd\left(\tilde c^{(2)}_a\:\dbbf^a\right)+\tfrac{1}{2}\mathcal{K}_{\alpha bc}
        \dd\left(D^\alpha_{(2)}\:\dbbf^b\dbbf^c\right)\right)
        \wedge\mathcal{L}_{A\bar B}
        \left(\dd\db^A\bar\db^{\bar B}-\dd\bar\db^{\bar B}\db^A\right) \nonumber \\ 
      &-\mu_7\ell^2\int\left[\tfrac{1}{2}\mathcal{C}_\alpha^{I\bar J}\dd D_{(2)}^\alpha\wedge
        \left(\dd a_I\bar a_{\bar J}-\dd\bar a_{\bar J}a_I\right) 
        +\dd\left(\tilde c^{(2)}_P-\mathcal{K}_{\alpha bP}
        D^\alpha_{(2)}\dbbf^b\right)\wedge A\right]  \nonumber \\
      &+\mu_7\ell^2\int\frac{1}{2}
        \left(\rho_\Lambda-\mathcal{K}_{\Lambda ab}c^a\dbbf^b
        +\tfrac{1}{2}\mathcal{K}_{\Lambda ab}\dbbf^a\dbbf^bl\right)F\wedge F  \nonumber \\
      &-\mu_7\ell^2\int\left((a_{\hat\alpha}+\bar a_{\hat\alpha}) \,
        \dd V^{\hat\alpha}\wedge F
        +(a^{\hat\alpha}+\bar a^{\hat\alpha})\, \dd U_{\hat\alpha}\wedge F \right) \ ,
\end{align}
where
\begin{equation} \label{eq:aints}
\begin{aligned}
   a^{\hat\alpha}&
     =a_I\int_{S^P}\iota^*\beta^{\hat\alpha}\wedge A^I \ , \qquad & \qquad 
   a_{\hat\alpha}&
     =a_I \int_{S^P}\iota^*\alpha_{\hat\alpha}\wedge A^I \ , \\
   \bar a^{\hat\alpha}&
     =\bar a_{\bar J}\int_{S^P}\iota^*\beta^{\hat\alpha}\wedge\bar A^{\bar J} \ ,
   \qquad & \qquad 
   \bar a_{\hat\alpha}&
     =\bar a_{\bar J}\int_{S^P}\iota^*\alpha_{\hat\alpha}\wedge\bar A^{\bar J} \ . 
\end{aligned}
\end{equation}
Furthermore in the derivation of \eqref{eq:CS1} 
we have used  \eqref{eq:L}, \eqref{eq:Cint}, \eqref{eq:N2L} and the fact 
that for any four-form $\theta$ of $S^\Lambda$ 
\begin{equation} \label{eq:Prel}
   \int_{S^\Lambda} \theta\cdot P_- =\int_{S^P} \theta \ ,
\end{equation}
holds. This can be easily seen by using \eqref{eq:P}
and the explicit definition of the cycles $S^\Lambda$ and $S^P$ 
given in Table~\ref{tab:cycles}.

In order to proceed we need to have additional information about the pullbacked three-forms $\iota^*\alpha_{\hat\alpha}$ and $\iota^*\beta^{\hat\alpha}$. If their pullback to $S^\Lambda$ is trivial all integrals \eqref{eq:aints} vanish, and in the action \eqref{eq:CS1} the mixed terms of the field strength $F$ with field strengths of the bulk vectors $V^{\hat\alpha}$ and $U_{\hat\alpha}$ disappear. In this case the complex structure deformations of the Calabi-Yau orientifold are unobstructed and all deformations remain moduli of the effective theory.\footnote{This situation is discussed in section~\ref{sec:geom} form a mathematical point of view.} The case where the pullbacks are non-trivial and the integrals \eqref{eq:aints} do not vanish is more involved and further elaborated on in appendix~\ref{sec:3form}. However, in this case we are not able to consistently include the complex structure deformations into the effective action and leave a thorough analysis to a future publication. In this section we therefore continue with the simplified assumption that all integrals in \eqref{eq:aints} vanish or in other words with the effective Chern-Simons action
\begin{align} \label{eq:4DCS}
   \mathcal{S}_\text{CS}&=
      \mu_7\int\left(\tfrac{1}{4}\dd\tilde l^{(2)}
        -\dd\left(\tilde c^{(2)}_a\:\dbbf^a\right)+\tfrac{1}{2}\mathcal{K}_{\alpha bc}
        \dd\left(D^\alpha_{(2)}\:\dbbf^b\dbbf^c\right)\right)
        \wedge\mathcal{L}_{A\bar B}
        \left(\dd\db^A\bar\db^{\bar B}-\dd\bar\db^{\bar B}\db^A\right) \nonumber \\ 
      &-\mu_7\ell^2\int\left[\tfrac{1}{2}\mathcal{C}_\alpha^{I\bar J}\dd D_{(2)}^\alpha\wedge
        \left(\dd a_I\bar a_{\bar J}-\dd\bar a_{\bar J}a_I\right) 
        +\dd\left(\tilde c^{(2)}_P-\mathcal{K}_{\alpha bP}
        D^\alpha_{(2)}\dbbf^b\right)\wedge A\right]  \nonumber \\
      &+\mu_7\ell^2\int\frac{1}{2}
        \left(\rho_\Lambda-\mathcal{K}_{\Lambda ab}c^a\dbbf^b
        +\tfrac{1}{2}\mathcal{K}_{\Lambda ab}\dbbf^a\dbbf^bl\right)F\wedge F  \ . 
\end{align}

The action contains the topological Yang-Mills term $F\wedge F$, with a field dependent $\Theta$-angle, which due to supersymmetry must eventually be given as the imaginary part of the holomorphic gauge coupling function. All the other terms in \eqref{eq:4DCS} involve the space-time two-forms $\tilde l^{(2)}$, $\tilde c^{(2)}$ or $D_{(2)}$, resulting from the expansion of the ten dimensional RR~fields \eqref{eq:C}. The third term in \eqref{eq:4DCS} is known as a Green-Schwarz term in that, after integrating by parts, the $U(1)$ field strength $F$ couples linearly to the space-time two-forms $\tilde c^{(2)}$ and $D_{(2)}$. In the next section we eliminate the two-forms in favor of their dual scalars by imposing \eqref{eq:dual}. As we will see the Green-Schwarz terms give rise to charged dual scalars which transform non-linearly under the gauge transformation.

%%%%%%%%%%%%%%%%%%%%%%%%%%%%%%%%%%%%%%%%%%%%%%%%%%%%%%%%%%%%%%%%%%

\section{Bulk and D7-brane: Effective theory} \label{sec:total}

%%%%%%%%%%%%%%%%%%%%%%%%%%%%%%%%%%%%%%%%%%%%%%%%%%%%%%%%%%%%%%%%%%

After having discussed both the effective action of the bulk theory in section~\ref{sec:bulk} and the effective action of the D7-brane in section~\ref{sec:brane}, we now consider the combined action. In order for this theory to be stable and consistent, tadpoles resulting from the orientifold planes and the D7-branes must cancel among another, and then the whole action yields a $\mathcal{N}=1$ effective supergravity action in four space-time dimensions. For this supergravity theory we determine all its defining data that is to say the K\"ahler potential, the scalar potential and the gauge kinetic coupling functions.

%%%%%%%%%%%%%%%%%%%%%%%%%%%%%%%%%%%%%%%%%%%%%%%%%%%%%%%%%%%%%%%%%%

\subsection{Branes and orientifolds: Tadpole cancellation} \label{sec:tadpole}

%%%%%%%%%%%%%%%%%%%%%%%%%%%%%%%%%%%%%%%%%%%%%%%%%%%%%%%%%%%%%%%%%%

In type~IIB orientifold string theories there are potentially two kinds of tadpoles, namely RR~tadpoles and NS-NS~tadpoles. While the appearance of the former tadpoles render the theory inconsistent, the divergencies of NS-NS~tadpoles give rise to potentials for NS-NS~fields \cite{Dudas:2000ff,Blumenhagen:2001te} and can be absorbed in the background fields via the Fischler-Susskind mechanism \cite{Fischler:1986}. Sources for both type of tadpoles are branes, orientifold planes and background fluxes because all these objects contribute RR~charges and couple to the NS-NS~graviton and NS-NS~dilaton due to their energy density. 

The contribution of D7-branes (with internal fluxes) to the RR~tadpoles can easily be read off from the Chern-Simons action of the branes. Similarly from the analog of the Chern-Simons action for orientifold planes \cite{StefanskiScrucca}, we obtain their share of RR~tadpoles also as topological expressions. For D7-branes wrapped on $S_i^{(7)}$ with internal two-form fluxes $f_i^{(7)}$, O7-planes wrapped on $O_j^{(7)}$, D3-branes located at $s_k^{(3)}$ and O3-planes located at $o_l^{(3)}$ we find altogether two tadpole cancellation conditions\footnote{The indices $i$, $j$ , $k$ and $l$ account for several D7-branes, O7-planes, D3-branes and O3-planes respectively. $S_i^{(7)}$ and $O_j^{(7)}$ are four-cycles whereas $s_k^{(3)}$ and $o_l^{(3)}$ are points in the Calabi-Yau manifold.} \cite{Blumenhagen:2002wn}
\begin{equation} \label{eq:RRtad}
\begin{split}
   0&=\sum_i \mu_7 \int_{\mathbb{R}^{3,1}\times S_{i}^{(7)}} C^{(8)}+
   \sum_j \nu^{j}_7 \int_{\mathbb{R}^{3,1}\times O_j^{(7)}} C^{(8)} \ , \\
   0&=\sum_i \mu_7\ell^2 \int_{\mathbb{R}^{3,1}\times S_{i}^{(7)}} C^{(4)}\wedge
   f_i^{(7)}\wedge f_i^{(7)}+
   \sum_k \mu^{k}_3 \int_{\mathbb{R}^{3,1}\times \{s_k^{(3)}\}} C^{(4)}+
   \sum_l \nu^{l}_3 \int_{\mathbb{R}^{3,1}\times \{o_l^{(3)}\}} C^{(4)} \ .
\end{split}
\end{equation}
Here $\mu^{i}_7$, $\nu^{j}_7$, $\mu^{k}_3$ and $\nu^{l}_3$ are the RR~charges of the D-branes and the orientifold planes. Note that there are no six-form tadpoles because, due to the negative parity of \eqref{eq:A}, the integrals $\int C^{(6)}\wedge f_i^{(7)}$ vanish \cite{Blumenhagen:2001te}.\footnote{Recall that in our conventions each cycle $S_i^{(7)}$ includes both the D7-brane and its image.} In the presence of bulk background fluxes (which we do not turn on in this paper) there appear additional terms in \eqref{eq:RRtad} from the Chern-Simons terms of the bulk action. As we are not considering a specific orientifold compactification we cannot explicitly check the conditions \eqref{eq:RRtad}. Instead we assume that we have appropriately chosen a Calabi-Yau manifold~$Y$ with involution $\sigma$ and D7-brane cycle $S^\Lambda$ so as to meet the RR~tadpole conditions \eqref{eq:RRtad}.

In ref.~\cite{Blumenhagen:2001te} it is argued that all NS-NS~tadpoles arise as derivatives of a D-term scalar potential with respect to the corresponding NS-NS~fields. In the supersymmetric case the NS-NS~tadpoles vanish as they are related to the RR~tadpole conditions via supersymmetry. This corresponds to the vanishing of the D-term \cite{Blumenhagen:2002wn,Lust:2004fi} and thus the potential terms in the Dirac-Born-Infeld action \eqref{eq:DBI} has to be canceled by the negative tension of the orientifold planes. If the NS-NS~tadpoles do not vanish a D-term is induced leading generically to an unstable background. In our case  this occurs for a non-vanishing $\dbbf$ in eq.\ \eqref{eq:D1}. 

%%%%%%%%%%%%%%%%%%%%%%%%%%%%%%%%%%%%%%%%%%%%%%%%%%%%%%%%%%%%%%%%%%

\subsection{Bulk and brane effective action} \label{sec:effact}

%%%%%%%%%%%%%%%%%%%%%%%%%%%%%%%%%%%%%%%%%%%%%%%%%%%%%%%%%%%%%%%%%%

Now we combine the bulk action \eqref{eq:4Dbulk} with the brane action \eqref{eq:DBI} and \eqref{eq:4DCS}. The resulting action is still formulated with all RR~fields, and therefore its equations of motion must still be supplemented by the four dimensional version of the duality constraints \eqref{eq:dual}. 

In order to obtain an action in the conventional sense, namely an action without incorporating scalars and their dual two-forms simultaneously, we must eliminate these redundant two-form fields systematically. Let us pause to demonstrate how this procedure works for a simple example 
\cite{Quevedo:1996uu,Dall'Agata:2001zh,Louis:2002ny}. We start with the four dimensional action
\begin{equation} \label{eq:SSD}
   \mathcal{S}_\text{SD}=\int\left[\frac{g}{4} \dd B^{(2)} \wedge * \dd B^{(2)}  
      +\frac{1}{4g}\dd S \wedge * \dd S \right] \ ,
\end{equation}
with the coupling constant $g$, and where $B^{(2)}$ is a two-form field and $S$ is a scalar field. Moreover we impose by hand the duality condition 
\begin{equation} \label{eq:dual2}
   g\:* \dd B^{(2)} = \dd S \ .
\end{equation}
Thus $S$ is the dual scalar of the two-form $B^{(2)}$ and the action \eqref{eq:SSD} with \eqref{eq:dual2} possesses just one degree of freedom. If we introduce the field strengths $H=\dd B^{(2)}$ and $A=\dd S$, altogether we have the equations
\begin{align} \label{eq:SSDeq}
   \dd A=0 \ , && \dd H=0 \ , && \dd *A=0 \ , && \dd *H=0 \ , && g\:* H = A \ ,
\end{align}
where the first two equations are Bianchi identities, the next two equations are the equations of motion of \eqref{eq:SSD}, and the last equation is the duality condition \eqref{eq:dual2}. Now we modify the action \eqref{eq:SSD} to
\begin{equation} \label{eq:SSD2}
   \mathcal{S}_\text{SD}=\int\left[\frac{g}{4} H \wedge *  H 
      +\frac{1}{4g}\dd S \wedge * \dd S -\frac{1}{2} H \wedge \dd S - \lambda \dd H \right] \ ,
\end{equation}
where in this action $H$ is an independent three form field and $\lambda$ is a Lagrangian multiplier. This Lagrangian also yields the equations \eqref{eq:SSDeq}, however, now only the first equation arises as a Bianchi identity. All the other equations, including the duality relation, is obtained from the equations of motion of \eqref{eq:SSD2}. In this formulation we can eliminate the three form field~$H$ and arrive at the action for $S$
\begin{equation} \label{eq:SSD3}
   \mathcal{S}_\text{SD}=\int\frac{1}{2g}\dd S \wedge * \dd S \ ,
\end{equation}
without any redundant dual fields. The next task is to generalize this procedure in the presents of source terms $J$, which we add to \eqref{eq:SSD}
\begin{equation} \label{eq:SSD4}
   \mathcal{S}_\text{SD}=\int\left[\frac{g}{4} \dd B^{(2)} \wedge *\dd B^{(2)} 
      +\frac{1}{4g} A \wedge *A - \frac{1}{2} \dd B^{(2)} \wedge J \right] \ .
\end{equation}
Note that in order to be in accord with the duality condition $g*H=A$, the field strength $A$ must be adjusted to $A=\dd S+J$ and the new equations of this system are
\begin{align} \label{eq:SSDeqS}
   \dd A=\dd J \ , && \dd H=0 \ , && \dd *A=0 \ , && \dd *H=\dd J \ , && g\:* H = A \ .
\end{align}
As before we obtain this set of equations from the Lagrangian
\begin{equation} \label{eq:SSD5}
   \mathcal{S}_\text{SD}=\int\left[\frac{g}{4} H \wedge *  H 
      +\frac{1}{4g}(\dd S+J)\wedge *(\dd S+J) -\frac{1}{2} H \wedge(\dd S+J) 
      - \lambda \dd H \right] \ ,
\end{equation}
with the independent field $H$. Finally eliminating $H$ yields
\begin{equation} 
   \mathcal{S}_\text{SD}=\int\frac{1}{2g}(\dd S+J) \wedge *(\dd S+J) \ .
\end{equation}

With the above techniques we now succinctly eliminate the space-time two-forms $\tilde l^{(2)}$, $\tilde c^{(2)}_a$ and $D_{(2)}^{\alpha}$, and also the magnetic vectors $U_{\hat\alpha}$, in the whole action functional $\mathcal{S}^\text{E}_\text{Bulk}+\mathcal{S}^\text{E}_\text{DBI}+\mathcal{S}_\text{CS}$, and as a result we obtain the combined bulk and D7-brane action in four dimensional Einstein frame 
\begin{align} \label{eq:action}
   \mathcal{S}^{\text{E}}
     =&\frac{1}{2\kappa_4^2}\int \left[-R\:*_41
          +2\mathcal{G}_{\tilde a\tilde b}\dd z^{\tilde a} \wedge *_4 \dd\bar z^{\tilde b}
          +2G_{\alpha\beta}\dd v^\alpha \wedge *_4 \dd v^\beta \right. \nonumber \\
      &+\frac{1}{2}\dd(\ln \mathcal{K})\wedge *_4 \dd(\ln \mathcal{K})
          +\frac{1}{2}\dd\phi\wedge *_4 \dd\phi 
          +2e^\phi G_{ab}\dd b^a\wedge \dd b^b \nonumber \\
      &+2i\kappa_4^2\mu_7\mathcal{L}_{A\bar B}\left(e^\phi+G_{ab}\dbbf^a\dbbf^b\right)
          \dd\db^A\wedge *_4\dd\bar\db^{\bar B} 
          +\frac{24}{\mathcal{K}}\kappa_4^2\mu_7\ell^2i\mathcal{C}^{I\bar J}_\alpha v^\alpha
          \dd a_I\wedge *_4\dd\bar a_{\bar J} \nonumber \\
      &+\frac{e^{2\phi}}{2}
          \left(\dd l+\kappa_4^2\mu_7\mathcal{L}_{A\bar B}
          \left(\dd\db^A\bar\db^{\bar B}-\dd\bar\db^{\bar B}\db^A\right)\right)\wedge  
          *_4\left(\dd l+\kappa_4^2\mu_7\mathcal{L}_{A\bar B}
          \left(\dd\db^A\bar\db^{\bar B}-\dd\bar\db^{\bar B}\db^A\right)\right) \nonumber \\
      &+2e^\phi G_{ab}
          \left(\cov c^a-l\dd b^a-\kappa_4^2\mu_7\dbbf^a \mathcal{L}_{A\bar B}
          \left(\dd\db^A\bar\db^{\bar B}-\dd\bar\db^{\bar B}\db^A\right)\right)\wedge \nonumber \\
      &\qquad *_4 \left(\cov c^b-l\dd b^b-\kappa_4^2\mu_7\dbbf^b 
          \mathcal{L}_{A\bar B} 
          \left(\dd\db^A\bar\db^{\bar B}-\dd\bar\db^{\bar B}\db^A\right)\right) \nonumber \\
      &+\frac{9}{2\mathcal{K}^2}G^{\alpha\beta}
          \left(\cov\rho_\alpha-\mathcal{K}_{\alpha bc}
          c^b\dd b^c-\tfrac{1}{2}\kappa_4^2\mu_7
          \mathcal{K}_{\alpha bc}\dbbf^b\dbbf^c\mathcal{L}_{A\bar B}
          \left(\dd\db^A\bar\db^{\bar B}-\dd\bar\db^{\bar B}\db^A\right)\right. \nonumber \\
      &\qquad\qquad\left. +2\kappa_4^2\mu_7\ell^2\mathcal{C}_\alpha^{I\bar J}
          \left(a_I\dd \bar a_{\bar J}-\bar a_{\bar J}\dd a_I\right)\right)\wedge \nonumber \\
      &\qquad *_4 \left(\cov\rho_\beta-\mathcal{K}_{\beta ab} c^a\dd b^b
          -\tfrac{1}{2}\kappa_4^2\mu_7
          \mathcal{K}_{\beta bc}\dbbf^b\dbbf^c\mathcal{L}_{A\bar B}
          \left(\dd\db^A\bar\db^{\bar B}-\dd\bar\db^{\bar B}\db^A\right)\right. \nonumber \\
      &\qquad\qquad\left. +2\kappa_4^2\mu_7\ell^2\mathcal{C}_\beta^{I\bar J}
          \left(a_I\dd\bar a_{\bar J}-\bar a_{\bar J}\dd a_I\right)\right) \nonumber \\
      &+\kappa_4^2\mu_7\ell^2 \left(\tfrac{1}{2}\mathcal{K}_\Lambda
          -\tfrac{1}{2}e^{-\phi}\mathcal{K}_{\Lambda ab} \dbbf^a\dbbf^b 
          \right) F\wedge *_4F \nonumber \\
      &+\kappa_4^2\mu_7\ell^2\left(\rho_\Lambda-\mathcal{K}_{\Lambda ab}c^a\dbbf^b
          +\tfrac{1}{2}\mathcal{K}_{\Lambda ab}\dbbf^a\dbbf^bl \right) F\wedge F \nonumber \\
      &\left.+\frac{1}{2}(\Imag \mathcal{M})_{\hat\alpha\hat\beta} 
          \dd V^{\hat\alpha}\wedge *_4 \dd V^{\hat\beta}
          +\frac{1}{2}(\Real \mathcal{M})_{\hat\alpha\hat\beta}
          \dd V^{\hat\alpha}\wedge\dd V^{\hat\beta} + \frac{1}{2}V_\text{D}\:*_41
          \right] \ , 
\end{align}
where we have included the scalar potential term $V_\text{D}$ of \eqref{eq:D1}, and where the gauge kinetic matrix $\mathcal{M}_{\hat\alpha\hat\beta}$ are related to the integrals \eqref{eq:mat3} by \cite{SuzukiCeresole}
\begin{equation} \label{eq:M}
   \mathcal{M}=A\inv{C}+i\inv{C} \ .   
\end{equation}
Note that this action (also without all the terms resulting from the D7-brane) has a set of global shift symmetries
\begin{align} \label{eq:shift}
   &c^a\rightarrow c^a+\theta^a \ , 
   &\rho_\alpha\rightarrow\rho_\alpha+\mathcal{K}_{\alpha bc}\dbbf^b\theta^c \ .
\end{align}
In the presence of a D7-brane wrapped on the cycle $S^\Lambda$ one of these symmetries is gauged, and therefore the action \eqref{eq:action} contains covariant derivatives for the charged fields $c^P$ and $\rho_\alpha$, i.e.
\begin{align} \label{eq:cd1}
   &\cov c^a=\partial_\mu c^a\:\dd x^\mu-4\kappa_4^2\mu_7\ell\delta^a_P A \ ,
   &\cov \rho_\alpha=\partial_\mu\rho_\alpha\:\dd x^\mu-4\kappa_4^2\mu_7
     \ell\mathcal{K}_{\alpha bP}\dbbf^b A \ .
\end{align} 
The gauging of the shift symmetry is a direct consequence of the Green-Schwarz term in eq.~\eqref{eq:4DCS}. In \eqref{eq:action} the Green-Schwarz term has disappeared since we eliminated the two-forms $\tilde c^{(2)}_P$ and $D^\alpha_{(2)}$ in favor of the dual scalars $c^P$ and $\rho_\alpha$. The presence of the Green-Schwarz term in the original democratic version of the action is responsible for the fact that in the scalar field basis of \eqref{eq:action} the dual scalars are charged and transform non-linearly under the $U(1)$ gauge theory of the D7-brane.

In the derivation of the action~\eqref{eq:action} we have treated the complex structure deformations~$z^{\tilde a}$ and the D7-brane matter fields~$\db^A$ independently. As a consequence the target space metric of the fields $z^{\tilde a}$ and $\db^A$ exhibits a product structure. However, as the complex structure of the D7-brane world-volume is induced from the ambient Calabi-Yau space both types of fields are interlinked \cite{Mayr:2001} and this product structure is only maintained for small complex structure deformations $z^{\tilde a}$ and small D7-brane fluctuations $\db^A$. In this limit the action~\eqref{eq:action} describes the theory adequately. In section~\ref{sec:geom} we analyze the common target space of these two kinds of fields and adjust the target space metric accordingly. 

Before we do that we first rewrite the action \eqref{eq:action} in the standard $\mathcal{N}=1$ form and determine the K\"ahler potential and the gauge kinetic functions.

%%%%%%%%%%%%%%%%%%%%%%%%%%%%%%%%%%%%%%%%%%%%%%%%%%%%%%%%%%%%%%%%%%

\subsection{$\mathcal{N}=1$ effective action in chiral coordinates} \label{sec:chiral}

%%%%%%%%%%%%%%%%%%%%%%%%%%%%%%%%%%%%%%%%%%%%%%%%%%%%%%%%%%%%%%%%%%

Any $\mathcal{N}=1$ supergravity action with chiral multiplets $M^M$ and vector multiplets $V^\Gamma$ is completely specified in terms of the K\"ahler potential~$K$, the superpotential~$W$ and the gauge kinetic coupling functions~$f_{\Gamma\Delta}$ \cite{Cremmer:1982en,Wess:1992}. All this data can already be determined unambiguously from the bosonic part of the supergravity action 
\begin{multline} \label{eq:N_1}
   \mathcal{S}^{\mathcal{N}=1}=\frac{1}{2\kappa_4^2}
      \int\left[-R\:*_41+2\:K_{M\bar N}\cov M^M\wedge*_4\cov\bar M^{\bar N}\right. \\
      \left.+\left(\Real f\right)_{\Gamma\Delta}F^\Gamma\wedge*_4F^\Delta
      +\left(\Imag f\right)_{\Gamma\Delta}F^\Gamma\wedge F^\Delta
      +\left(V_\text{F}+V_\text{D}\right)\:*_41 \right] \ ,
\end{multline}
where
\begin{align} \label{eq:V}
   V_\text{F}=e^K\left(K^{M\bar N}\mathcal{D}_MW
      \mathcal{D}_{\bar N}\bar W-3\left|W\right|^2\right) \ , &&
   V_\text{D}=\frac{1}{2}\left(\Real f\right)^{-1\:\Gamma\Delta}\text{D}_\Gamma\text{D}_\Delta \ .
\end{align}
$F^\Gamma$ is the field strength of the vector field $V^\Gamma$, $K_{M\bar N}=\partial_M\partial_{\bar N}K$ is the K\"ahler metric and the potential $V_\text{F}$ is expressed in terms of the K\"ahler covariant derivatives $\mathcal{D}_MW=\partial_MW+(\partial_MK)W$ of the superpotential. The potential $V_\text{D}$ involves the inverse matrix $\left(\Real f\right)^{-1\:\Gamma\Delta}$ of the real part of the holomorphic gauge kinetic coupling matrix $f_{\Gamma\Delta}$. 

In order to specify the K\"ahler potential in the standard form, we must first identify the correct K\"ahler variables, which are the lowest bosonic components in the $\mathcal{N}=1$ chiral multiplets. Then in terms of this variables the metric of the scalar fields in \eqref{eq:action} becomes manifest K\"ahler. Geometrically this corresponds to finding the correct complex structure of the K\"ahler manifold, which is the target space of the scalar fields. We know already that the metric $\mathcal{G}_{\tilde a\tilde b}$ of the complex structure moduli fields $z^{\tilde a}$ defined in \eqref{eq:CSt} is K\"ahler \cite{Candelas:1990pi}, and thus the complex scalar fields $z^{\tilde a}$ are good K\"ahler coordinates. We find that this also holds for the D7-brane fields $\db^A$ and $a_I$. For the remaining fields it is not so obvious how they combine to K\"ahler variables. However, guided by refs.~\cite{Haack:1999zv,BBHL,Grana:2003ek,Grimm:2004uq} the other chiral fields turn out to be $\dil$, $G^a$ and $T_\alpha$ defined by 
\begin{align}
   \dil&=\tau+\kappa_4^2\mu_7\mathcal{L}_{A\bar B}\db^A\bar\db^{\bar B} \label{eq:dil} \ , \\
   G^a&=c^a-\tau \dbbf^a \ , \\ 
   T_\alpha&=\frac{3i}{2}\left(\rho_\alpha
      -\tfrac{1}{2}\mathcal{K}_{\alpha bc}c^b\dbbf^c\right) +\frac{3}{4}\mathcal{K}_\alpha
      +\frac{3i}{4(\tau-\bar\tau)} \mathcal{K}_{\alpha bc}G^b(G^c-\bar G^c)
      +3i\kappa_4^2\mu_7\ell^2 \mathcal{C}^{I\bar J}_\alpha a_I \bar a_{\bar J} \label{eq:T} \ ,
\end{align}
where $\tau=l+ie^{-\phi}$ is the original complex type IIB dilaton field,
the intersection numbers $ \mathcal{K}_{\alpha bc}$ are defined in
\eqref{eq:K} while $\mathcal{L}_{A\bar B}, \mathcal{C}^{I\bar J}_\alpha$
are defined in \eqref{eq:L} and \eqref{eq:Cint}.

Note that $\dil$, $G^a$ and $T_\alpha$ are closely related to the K\"ahler variables of orientifold compactifications \cite{BBHL,Grimm:2004uq}. However, the dilaton field $\tau$, which is a K\"ahler variable in orientifold compactification even with D3-branes \cite{Grana:2003ek}, is not a K\"ahler variable anymore due to the coupling of the D7-brane to the two-form field~$\tilde l^{(2)}$ which is dual to the axion~$l$. This leads to a shift in the definition of the dilaton field with $\dil$ playing the role of a shifted `new' dilaton. In the K\"ahler potential and the definitions \eqref{eq:dil}--\eqref{eq:T} we should therefore think of $\tau$ as a function $\tau(\dil,\db)$ depending on both the new dilaton $\dil$ and the brane fields $\db$. Furthermore, the K\"ahler variables $T_\alpha$ of ref.~\cite{BBHL,Grimm:2004uq} are modified by the D7-brane fields $a$ and through $\tau$ also by $\db$. A similar adjustment occurs in orientifold models with D3-branes \cite{Grana:2003ek}, where the D3-brane matter fields also enter the definition of $T_\alpha$. In fact the D3-brane matter fields couple very similarly as the Wilson-line moduli~$a$.\footnote{The reason for this similarity comes about as follows: Fluctuations of the space-time filling D3-brane world-volume are captured by the D3-brane matter fields \cite{Grana:2003ek}, whereas the D7-brane Wilson-line fields~$a$ are the moduli of the internal gauge theory of the D7-brane, and this gauge theory accounts for the lower dimensional D3-brane charges of the D7-brane \cite{Hughes:1986fa}.} Note that if internal D7-brane fluxes $f^a$ are turned on, the K\"ahler variables $G^a$ are adjusted by a shift proportional to the flux as dictated by eq.~\eqref{eq:bf}.

In terms of these K\"ahler coordinates the K\"ahler potential for the supergravity action \eqref{eq:action} is found to be
\begin{multline} \label{eq:K1}
   K(\dil, G, T, z, \db, a)=K_\text{CS}(z) 
   -\log\left[-i\left(\dil-\bar\dil\right)+2i\kappa_4^2\mu_7\mathcal{L}_{A\bar B}
   \db^A\bar\db^{\bar B}\right] \\
     -2 \log\left[\tfrac{1}{6}\mathcal{K}(\dil,G,T,\db,a)\right] \ ,
\end{multline}
where $K_\text{CS}$ of the complex structure moduli $z^{\tilde a}$ is defined in \eqref{eq:CSt} and $\mathcal{K}=\mathcal{K}_{\alpha\beta\gamma}v^\alpha v^\beta v^\gamma$. This K\"ahler potential \eqref{eq:K1} reproduces all kinetic terms of \eqref{eq:action}. However, it is given as an implicit expression since $\mathcal{K}$ is explicitly known only in terms of the $v^\alpha$ which are no K\"ahler coordinates. Instead they are determined in terms of $\dil, G^a, T_\alpha, \db^A$ and $a_I$ by solving \eqref{eq:T} for $v^\alpha(\dil, G^a, T_\alpha, \db^A,a_I)$. Unfortunately, this solution cannot be given explicitly in general. 

Before we continue let us discuss a few instructive limits of the K\"ahler potential \eqref{eq:K1}. First of all, for a Calabi-Yau orientifold with $h_+^{1,1}=1$, we have a single harmonic two-form $\omega_\Lambda$ with positive parity under the involution~$\sigma$. By Poincar\'e duality we associate to $\omega_\Lambda$ the four cycle $S^\Lambda$. Let us further assume that this cycle $S^\Lambda$ is suitable to wrap a D7-brane, namely it does not intersect with any orientifold fixed points. Then with a D7-brane wrapped on $S^\Lambda$ we obtain a model with K\"ahler variables $\dil$, $G^a$, $\db^A$, $a_I$ and a single $T_\Lambda$. Furthermore $\mathcal{K}=\mathcal{K}_{\Lambda\Lambda\Lambda}(v^\Lambda)^3$ and thus \eqref{eq:T} can be solved for $v^\Lambda$ resulting in 
\begin{equation}
   2\log\mathcal{K}=3\log \left[T_\Lambda+\bar T_\Lambda
     -\frac{3i\mathcal{K}_{\Lambda ac} (G^a-\bar G^a)(G^c-\bar G^c)}
      {4(\dil-\bar\dil-2\kappa_4^2\mu_7\mathcal{L}_{A\bar B}\db^A\bar\db^{\bar B})}
     -6i\kappa_4^2\mu_7\ell^2\mathcal{C}^{I\bar J}_\Lambda a_I\bar a_{\bar J}\right] 
     +\text{const} \ .
\end{equation}
This can be further simplified by setting $G^a=0$ and $a_I=0$ leading to
\begin{equation}
   K(\dil,T_\Lambda,z,\db)=K_\text{CS}(z)-3\log\left[T_\Lambda+\bar T_\Lambda\right]
   -\log\left[-i\left(\dil-\bar\dil\right)+2i\kappa_4^2\mu_7\mathcal{L}_{A\bar B}
   \db^A\bar\db^{\bar B}\right] \ .
\end{equation}
The K\"ahler metric resulting from this K\"ahler potential is block diagonal in the modulus $T_\Lambda$ and the brane fluctuations $\db$. This particular feature of the K\"ahler potential was already anticipated for D7-brane models in ref.~\cite{Hsu:2003cy}, although we stress that it does not hold in the general case \eqref{eq:K1}.

As a second limit of \eqref{eq:K1} we consider the case $h_-^{1,1}=0$ and $h_+^{1,1}=3$ with a suitable four-cycle $S^\Lambda$ wrapped by a D7-brane. Then the K\"ahler variables of this example are $\dil$, $T_\alpha$, $z^{\tilde a}$, $\db^A$ and $a_I$ with $\alpha=\Lambda,1,2$. Moreover we suppose in analogy to the six dimensional torus that $C^{I\bar J}_\Lambda=0$ and that $\mathcal{K}_{\Lambda 12}$ is up to permutations the only non-vanishing triple intersection number, i.e. $\mathcal{K}=6\:\mathcal{K}_{\Lambda 12}v^\Lambda v^1 v^2$. Then as in the previous example we can specify $v^\alpha(\dil,T_\alpha,\db^A,a_I)$ explicitly and the K\"ahler potential \eqref{eq:K1} becomes
\begin{multline} \label{eq:Ktor}
   K(S,T,z,\db,a)=K_\text{CS}(z)-\log\left[-i\left(\dil-\bar\dil\right)
      +2i\kappa_4^2\mu_7\mathcal{L}_{A\bar B}\db^A\bar\db^{\bar B}\right]
      -\log\left[T_\Lambda+\bar T_\Lambda\right] \\
      -\log\left[T_1+\bar T_1-6i\kappa_4^2\mu_7\ell^2
       \mathcal{C}_1^{I\bar J}a^I\bar a^{\bar J}\right]
      -\log\left[T_2+\bar T_2-6i\kappa_4^2\mu_7\ell^2
       \mathcal{C}_2^{I\bar J}a^I\bar a^{\bar J}\right] \ .
\end{multline}
This form of the K\"ahler potential was first derived in ref.~\cite{Ibanez:1998rf}. We can expand this K\"ahler potential up to second order in the D7-brane fields $\db$ and $a$, and obtain
\begin{multline} \label{eq:Kexp}
   K(\dil,T_\Lambda,z,\db,a)=K_\text{CS}(z)-\log\left[-i\left(\dil-\bar\dil\right)\right]
    -\sum_{\alpha=\Lambda,1,2}\log\left(T_\alpha+\bar T_\alpha\right) \\
    +\frac{\kappa_4^2\mu_7\mathcal{L}_{A\bar B}}{\dil-\bar\dil}\ \db^A\bar\db^{\bar B}
    +\frac{3i\kappa_4^2\mu_7\ell^2\mathcal{C}^{I\bar J}_1}{T_1+\bar T_1}\
    a_I\bar a_{\bar J} 
    +\frac{3i\kappa_4^2\mu_7\ell^2\mathcal{C}^{I\bar J}_2}{T_2+\bar T_2}\
    a_I\bar a_{\bar J} \ .
\end{multline}
This expansion can be compared with the result of ref.~\cite{Lust:2004fi}, where the K\"ahler potential of a certain toroidal orientifold was derived to second order in the brane fields $\db$ and $a$. In this torus model the structure of the couplings of the brane fluctuations~$\db$ to the bulk field~$\dil$ agrees with eq.~\eqref{eq:Kexp}. Moreover the structure of the couplings of the Wilson-line moduli~$a$ to the K\"ahler moduli $T_1$ and $T_2$ also complies with \eqref{eq:Kexp}.  However, the toroidal orientifold of ref.~\cite{Lust:2004fi} also exhibits couplings of the complex structure moduli $z^{\tilde a}$ with the brane fields. These couplings can not be seen in the K\"ahler potential \eqref{eq:Ktor} because as explained at the end of section~\ref{sec:total} the action \eqref{eq:action} and thus also the K\"ahler potential \eqref{eq:K1} describe the theory only reliably in the limit of small complex structure deformations $z^{\tilde a}$ and small D7-brane fluctuations $\db^A$. Note also that in ref.~\cite{Lust:2004fi} $K$ depends on internal fluxes of the D7-brane. These fluxes, however, arise from two-forms of the D7-brane world-volume, which are not inherited from the ambient space, and therefore they are not captured by the fluxes considered in eq.~\eqref{eq:ftyp}.

From the action \eqref{eq:action} we can also read off the gauge kinetic functions by comparing it with \eqref{eq:N_1}. For the vector fields $V^\alpha$ arising from the bulk RR~four-form \eqref{eq:C} the gauge kinetic coupling matrix $f_{\hat\alpha\hat\beta}$ is given by 
\begin{equation}
   f_{\hat\alpha\hat\beta}=-\left.\frac{i}{2}\mathcal{\bar M}_{\hat\alpha\hat\beta} 
    \right|_{z^{\tilde\alpha}=\bar z^{\tilde\alpha}=0} \ ,
\end{equation}
in terms of the $\mathcal{N}=2$ gauge kinetic matrix $\mathcal{M}$ defined in \eqref{eq:M}. Due to the fact that the subset $z^{\tilde\alpha}$ of bulk complex structure deformations of Calabi-Yau manifolds are projected out by the orientifold involution $\sigma$, it is necessary to evaluate in the $\mathcal{N}=1$ orientifold context the matrix $\mathcal{M}$ at $z^{\tilde\alpha}=\bar z^{\tilde\alpha}=0$. The coupling matrix $f_{\hat\alpha\hat\beta}$ is completely independent of the brane fields and appears already in this form in Calabi-Yau orientifold compactifications without branes. As demonstrated in ref.~\cite{Grimm:2004uq} for orientifold compactifications with O3/O7 planes one has the identity
\begin{equation}
   \left.\mathcal{\bar M}_{\hat\alpha\hat\beta}
   \right|_{z^{\tilde\alpha}=\bar z^{\tilde\alpha}=0}
   =\left.\mathcal{F}_{\hat\alpha\hat\beta}
   \right|_{z^{\tilde\alpha}=\bar z^{\tilde\alpha}=0} \ ,
\end{equation}
where $\mathcal{F}_{\hat\alpha\hat\beta}$ is the second derivative of the $\mathcal{N}=2$ prepotential $\mathcal{F}$. Thus one arrives for the gauge kinetic couplings of the bulk at \cite{Grimm:2004uq}
\begin{equation} \label{eq:kinbulk}
   f_{\hat\alpha\hat\beta}(z)
      =-\left.\frac{i}{2}\mathcal{F}_{\hat\alpha\hat\beta}
        \right|_{z^{\hat\alpha}=\bar z^{\hat\alpha}=0} \ ,
\end{equation}
which manifestly shows that $f_{\hat\alpha\hat\beta}$ is holomorphic in the orientifold complex structure deformations $z^{\tilde a}$ because of the holomorphicity of the prepotential~$\mathcal{F}$.

In addition to the gauge kinetic couplings of the bulk vectors, the couplings of the D7-brane gauge degrees of freedom must also be specified. If the D7-brane has no Wilson line moduli $a$, we readily extract from the action \eqref{eq:action} using \eqref{eq:T} the coupling function 
\begin{equation} \label{eq:fbrane}
   f^\text{D7}=\frac{2\kappa_4^2\mu_7\ell^2}{3}\ T_\Lambda \ ,
\end{equation}
which is clearly holomorphic in the chiral fields. As expected the gauge coupling of the D7-brane is not the dilaton but the modulus controlling the size of the wrapped four-cycle \cite{Kakushadze:1998wp,Ibanez:1998rf}. If, however, the internal brane cycle $S^\Lambda$ has one forms, which give rise to Wilson line moduli $a$, the reduction of the D7-brane action \eqref{eq:DBIab} does not reproduce the Wilson line term in the gauge kinetic coupling function \eqref{eq:fbrane} which appears in the definition of $T_\Lambda$ \eqref{eq:T}. This mismatch has already been observed in \cite{Hsu:2003cy}. The reason for this seeming discrepancy is due to the fact, that the Dirac-Born-Infeld action is only an effective description comprising the open string tree level amplitudes \cite{Berg:2004ek}. Using a CFT approach, the open string one loop amplitudes for toroidal orientifolds with branes are computed in ref.~\cite{Berg:2004ek}, and the analysis shows that the missing quadratic Wilson line terms in the gauge kinetic coupling function of the D7-brane vector fields do indeed appear at the open string one loop level. In our case, we also expect that the coupling function \eqref{eq:fbrane} is corrected in the presence of Wilson line moduli, and that the missing terms are also generated at the one loop level of open string amplitudes.

The next task is to describe the scalar potential of the action \eqref{eq:action}. In a generic $\mathcal{N}=1$ supergravity action the scalar potential consists of F-terms and D-terms. A D-term scalar potential arises in the presence of charged chiral fields and takes the form given in \eqref{eq:V}, whereas the D-terms itself are computed from the equation \cite{Wess:1992}
\begin{equation} \label{eq:Dgen}
   \partial_N\partial_{\bar M}K\:\bar X^{\bar M}_\Gamma = i \partial_N \text{D}_\Gamma \ .
\end{equation}
Here $X_\Gamma=X^M_\Gamma\partial_M$ is the holomorphic Killing vector field of the corresponding gauged isometry of the target space K\"ahler manifold.

We have noted in \eqref{eq:cd1} that some of the bosonic fields are charged under a Peccei-Quinn symmetry. With the expressions \eqref{eq:dil} to \eqref{eq:T} we can identify the only charged chiral field $G^P$, for which the gauge covariant derivative is given by
\begin{equation} \label{eq:cd2}
   \cov_\mu G^P=\partial_\mu G^P-4\kappa^2_4\mu_7\ell A_\mu \ .
\end{equation}
The chiral fields $T_\alpha$ remain neutral because the non-linear gauge transformations of $\rho_\alpha$ and $c^P$ in the definition \eqref{eq:T} of $T_\alpha$ cancel among another. From \eqref{eq:cd2} we can read off the Killing vector to be
\begin{equation}
   X=4\kappa_4^2\mu_7\ell\:\partial_{G^P} \ ,
\end{equation}
and with \eqref{eq:Dgen} one infers
\begin{equation}
   \text{D}=-4i\kappa_4^2\mu_7\ell\:\partial_{\bar G^{\bar P}} K \ .
\end{equation}
Using \eqref{eq:K1}, \eqref{eq:T} and repeatedly the chain rule
one arrives at
\begin{equation}
   \partial_{\bar G^{\bar P}}K=-\frac{3i\mathcal{K}_{Pa}\dbbf^a}{2\mathcal{K}} \ .
\end{equation}
Finally, with \eqref{eq:fbrane} inserted in \eqref{eq:V} the scalar potential reads
\begin{equation} \label{eq:D2}
   V_\text{D}\ =\ \frac{18\kappa_4^2\mu_7}{\mathcal{K}^2 \Real{T_\Lambda}}\
     \left(\mathcal{K}_{Pa}\dbbf^a\right)^2 \ .
\end{equation}

Note, that we have not obtained \eqref{eq:D2} in the reduction of the D7-brane action. This is due to the fact, that we have used the calibration condition \eqref{eq:cali2}, which holds in the supersymmetric case. However, from general considerations we argued in section~\ref{sec:susycal} that the deviation from the supersymmetry condition \eqref{eq:susy4} should appear in the form of a D-term potential \eqref{eq:D1}, which measures spontaneous supersymmetry breaking in the effective action. The above supergravity analysis shows that this is indeed the case, and furthermore it has determined the previously unknown coefficient of \eqref{eq:D1}. Of course the minimum of $V_\text{D}$ is obtained for $\dbbf^a=0$ which in our setup is always a possible solution.

This completes the description of the scalar potential of \eqref{eq:action}, which does not contain any F-terms, and hence the perturbative superpotential $W$ vanishes. Now we have specified all the defining quantities for the $\mathcal{N}=1$ supergravity action \eqref{eq:action}, which are valid in the regime of small complex structure deformations $z^{\tilde a}$ and small brane fluctuations $\db$. In the next section we analyze the underlying geometric structure of the moduli space of these fields beyond this limit and find an adequate adjustment of the K\"ahler potential.

Before we conclude this section, let us briefly mention how to include D3-branes in these D7-brane orientifold models. In ref.~\cite{Grana:2003ek} the K\"ahler potential of type IIB orientifold models with a stack of $N$ D3-branes was derived. The inclusion of D3-branes adds chiral matter fields $\Phi$ to the other chiral fields of the theory. The chiral fields $\Phi$ parametrize the position of the stack of D3-branes in the internal Calabi-Yau space~$Y$ and transform in the adjoint representation of the gauge group $U(N)$. We take the stack of D3-branes to be distant from the D7-brane in order to avoid additional fields in the low energy effective theory. Besides adding $\Phi$ to the theory the D3-branes change the definition \eqref{eq:T} of the K\"ahler variables $T_\alpha$ further to \cite{Grana:2003ek} 
\begin{multline}
   T_\alpha=\frac{3i}{2}\left(\rho_\alpha
      -\tfrac{1}{2}\mathcal{K}_{\alpha bc}c^b\dbbf^c\right) +\frac{3}{4}\mathcal{K}_\alpha
      +\frac{3i}{4(\tau-\bar\tau)} \mathcal{K}_{\alpha bc}G^b(G^c-\bar G^c) \\
      +3i\kappa_4^2\mu_7\ell^2 \mathcal{C}^{I\bar J}_\alpha a_I \bar a_{\bar J}
      +\frac{3}{2}i\mu_3\ell^2\left(\omega_\alpha\right)_{i\bar\jmath}
        \tr \Phi^i\left(\bar\Phi^{\bar\jmath}-\tfrac{i}{2}\bar z^{\tilde a}
        \left(\bar\chi_{\tilde a}\right)^{\bar\jmath}_l\Phi^l\right) \ ,
\end{multline}
where $\mu_3$ is the D3-brane charge. Then the K\"ahler potential becomes
\begin{multline}
   K(\dil, G, T, z, \db, a,\Phi)=K_\text{CS}(z) 
     -\log\left[-i\left(\dil-\bar\dil\right)+2i\kappa_4^2\mu_7\mathcal{L}_{A\bar B}
     \db^A\bar\db^{\bar B}\right] \\
     -2 \log\left[\tfrac{1}{6}\mathcal{K}(\dil,G,T,\db,a,\Phi,z)\right] \ .
\end{multline}
Note that through the definition of $T_\alpha$ the quantity $-2\log\mathcal{K}$ becomes now also a function of the complex structure deformations~$z^{\tilde a}$. In ref.~\cite{Grana:2003ek} it is also demonstrated that the gauge kinetic coupling function of the $U(N)$ gauge boson resulting from the D3-branes is proportional to $\tau$. In the presence of D7-branes the variable $\tau$, however, is not a K\"ahler variable anymore, but instead it is a function of $\dil$, $\db$ and $\bar\db$. Thus there is a slight puzzle how holomorphicity of the D3-brane coupling function is restored. One possibility is that similar to ref.~\cite{Berg:2004ek} there arise additional terms at the open string one loop level, which contribute to this gauge kinetic coupling function to render it holomorphic in the chiral fields. In a similar way the holomorphicity of the superpotential $W(\tau,z)$ which arises in the presence of bulk background fluxes is obscured. It is suggestive to render this superpotential holomorphic by replacing the field $\tau$ by the new dilaton $\dil$.  

%%%%%%%%%%%%%%%%%%%%%%%%%%%%%%%%%%%%%%%%%%%%%%%%%%%%%%%%%%%%%%%%%%

\section{Geometry of the $\mathcal{N}=1$ moduli space} \label{sec:geom}

%%%%%%%%%%%%%%%%%%%%%%%%%%%%%%%%%%%%%%%%%%%%%%%%%%%%%%%%%%%%%%%%%%

In section~\ref{sec:total} we have derived the $\mathcal{N}=1$ effective action for orientifold theories with D7-branes and we have discussed the corresponding K\"ahler potential. So far in this analysis the D7-brane matter fields $\db$ and the complex structure deformations $z$ have been treated independently. The matter fields $\db$ are geometrically governed by $(2,0)$-forms of the four-cycle $S^\Lambda$. As the complex structure of the submanifold $S^\Lambda$ is induced from the complex structure of the ambient Calabi-Yau space $Y$, we expect that the moduli space of the matter fields $\db$ and the moduli space of the bulk complex structure deformations $z$ is not of product type \cite{Mayr:2001}. We should rather have in mind a common moduli space $\mathcal{M}_{\mathcal{N}=1}$, which is parametrized by both the complex structure deformations $z$ and the matter fields $\db$. In section~\ref{sec:relform} we summarize the concept of relative cohomology, which is used in section~\ref{sec:var} to describe the moduli space $\mathcal{M}_{\mathcal{N}=1}$. Finally in section~\ref{sec:kcs} we extend the K\"ahler potential of the previous section adequately to incorporate the moduli space $\mathcal{M}_{\mathcal{N}=1}$ in the supergravity description of section~\ref{sec:chiral}.

%%%%%%%%%%%%%%%%%%%%%%%%%%%%%%%%%%%%%%%%%%%%%%%%%%%%%%%%%%%%%%%%%%

\subsection{Relative cohomology and relative homology} \label{sec:relform}

%%%%%%%%%%%%%%%%%%%%%%%%%%%%%%%%%%%%%%%%%%%%%%%%%%%%%%%%%%%%%%%%%%

In order to describe the bulk complex structure deformations~$z$ and the D7-brane matter fields~$\db$ in their common moduli space $\mathcal{M}_{\mathcal{N}=1}$ one needs to find a mathematical formulation which captures both types of fields simultaneously. As these fields are respectively expanded into three-forms of the bulk and into two-forms of the internal D7-brane cycle~$S^\Lambda$ the relative cohomology group $H^3(Y,S^\Lambda)$ proves to be the adequate framework \cite{Mayr:2001}. 

First of all we introduce relative forms $\rel{\Theta}\in\Omega^n(Y,S^\Lambda)$.\footnote{For an introduction on relative forms e.g. ref.~\cite{Karoubi:1987}.} These forms are $n$-forms of the Calabi-Yau manifold~$Y$ in the kernel of $\iota^*$. Recall that the map $\iota$ embeds $S^\Lambda$ into $Y$, i.e. $\iota:S^\Lambda\hookrightarrow Y$. Hence the set of relative forms $\Omega^n(Y,S^\Lambda)$ fits into the exact sequence
\begin{equation} \label{eq:RFexact}
   0\rightarrow\Omega^n(Y,S^{\Lambda})\hookrightarrow\Omega^n(Y)
    \xrightarrow{\iota^*}\Omega^n(S^\Lambda)\rightarrow 0 \ .
\end{equation}
Then the cohomology of these relative forms with respect to the exterior differential $\dd$ defines the relative cohomology groups $H^n(Y,S^\Lambda)$, namely
\begin{equation}
   H^n(Y,S^\Lambda)=\frac{\{\rel{\Theta}\in\Omega^n(Y,S^\Lambda)|\dd\rel{\Theta}=0\}}
     {\dd\left(\Omega^{n-1}(Y,S^\Lambda)\right)} \ .
\end{equation}
As the duality of the cohomology group $H^n(Y)$ to the homology group $H_n(Y)$, each relative cohomology group $H^n(Y,S^\Lambda)$ has a dual description in terms of a relative homology group $H_n(Y,S^\Lambda)$. The elements of $H_n(Y,S^\Lambda)$ are $n$-cycles $\rel{\Gamma}$, which are not necessarily closed anymore, but may have boundaries $\partial\rel{\Gamma}$ in $\iota(S^\Lambda)$. Furthermore the pairing of a relative $n$-cycle with a relative $n$-form is given by the integral
\begin{equation} \label{eq:rpair}
   \langle\rel{\Gamma},\rel{\Theta}\rangle\:=\:\int_{\rel{\Gamma}} \rel{\Theta} \ .
\end{equation}
Note that this bilinear product is independent of the choice of representative of the relative cohomology element $\rel{\Theta}$ and the relative homology element $\rel{\Gamma}$.

In the following we concentrate on the relative cohomology group $H^3(Y,S^\Lambda)$, which is relevant for the moduli space $\mathcal{M}_{\mathcal{N}=1}$. In order to get a better handle on this space of relative three-forms one constructs from the short exact sequence \eqref{eq:RFexact} the long exact sequence
\begin{equation}
   \ldots\rightarrow H^2(Y)\xrightarrow{\iota^*} H^2(S^\Lambda)\xrightarrow{\delta}
   H^3(Y,S^\Lambda)\rightarrow H^3(Y)\xrightarrow{\iota^*}H^3(S^\Lambda)\rightarrow\ldots \ .
\end{equation} 
From this sequence one extracts
\begin{equation} \label{eq:RC}
   H^3(Y,S^\Lambda)\cong\ker\left(H^3(Y)\xrightarrow{i^*}H^3(S^\Lambda)\right) 
     \oplus\coker\left(H^2(Y)\xrightarrow{i^*}H^2(S^\Lambda)\right) \ .
\end{equation}
Thus we can think of a representative $\rel{\Theta}$ of $H^3(Y,S^\Lambda)$ as a pair of a three-form $\Theta_Y$ of $Y$ and a two-form $\theta_{S^\Lambda}$ of $S^\Lambda$, where $\Theta_Y$ is in the kernel of $\iota^*$ and $\theta_{S^\Lambda}$ is in the cokernel of $\iota^*$.

Now the relative cohomology elements of $H^3(Y,S^\Lambda)$ can be used to address the complex structure deformations~$z^{\tilde a}$ and the D7-brane fluctuations~$\db^A$ on an equal footing \cite{Mayr:2001}. One identifies the two-forms $H^{(2,0)}_{\bar\partial,-}(S^\Lambda)$ of the cycle $S^\Lambda$, which form a basis for the matter fields $\db$ (see Table~\ref{tab:spec}), with the two-form elements of \eqref{eq:RC}. Note that these two-forms are automatically in the cokernel of $\iota^*$ as there are no harmonic $(2,0)$-forms in the ambient manifold~$Y$. Similarly the three-forms $H^{(2,1)}_{\bar\partial,-}(Y)$, which form a basis for the complex structure deformations~$z^{\tilde a}$, are identified with the three-form elements of \eqref{eq:RC}. However, in general there can be three-forms of $H^{(2,1)}_{\bar\partial,-}(Y)$ which are not in the kernel of $\iota^*$. Then the corresponding complex structure deformations are obstructed. In the following we limit the complex structure deformations to those fields~$z^{\tilde a}$ where the associated $(2,1)$-forms are in the kernel of $\iota^*$.\footnote{If the internal D-brane cycle $S^\Lambda$ has no Wilson-line degrees of freedom then $H^1(S^\Lambda)$ and by Poincar\'e duality $H^3(S^\Lambda)$ are trivial. Hence in this case any three-form of $H^3(Y)$ is in the kernel of $\iota^*$ and thus all complex structure deformations $z^{\tilde a}$ are unobstructed.} For ease of notation we denote the set of unobstructed complex structure deformations also by $z^{\tilde a}$. Note that the unique holomorphic $(3,0)$-form of the Calabi-Yau manifold~$Y$ is always in the kernel of $i^*$ as there are no $(3,0)$-forms on the four-cycle~$S^\Lambda$.

Recall that in the context of Calabi-Yau orientifolds with a holomorphic involution~$\sigma$ the complex structure deformations~$z^{\tilde a}$ and the D7-brane matter fields~$\db^A$ are expanded into odd forms with respect to the involution~$\sigma$. Therefore the appropriate relative cohomology space is $H^3_-(Y,S^\Lambda)$ that is to say the elements are also odd relative forms with respect to the involution~$\sigma$. In this case the relation \eqref{eq:RC} becomes
\begin{equation} \label{eq:RC2}
   H^3_-(Y,S^\Lambda)\cong \widetilde H^3_-(Y) \oplus \widetilde H^2_-(S^\Lambda) \ ,
\end{equation}
where $\widetilde H^3_-(Y)=\ker\left(H^3_-(Y)\xrightarrow{i^*}H^3_-(S^\Lambda)\right)$ and $\widetilde H^2_-(S^\Lambda)=\coker\left(H^2_-(Y)\xrightarrow{i^*}H^2_-(S^\Lambda)\right)$.

%%%%%%%%%%%%%%%%%%%%%%%%%%%%%%%%%%%%%%%%%%%%%%%%%%%%%%%%%%%%%%%%%%

\subsection{Variation of Hodge structure} \label{sec:var}

%%%%%%%%%%%%%%%%%%%%%%%%%%%%%%%%%%%%%%%%%%%%%%%%%%%%%%%%%%%%%%%%%%

The complex structure deformations of the bulk theory is mathematical captured in the language of variation of Hodge structure, which describes how the definition of $(p,3-p)$-forms in $H^3(Y)$ varies over the complex structure moduli space $\mathcal{M}_\text{CS}$ \cite{Candelas:1990pi,Morrison:1992,Aspinwall:1993nu,Greene:1993vm}. In ref.~\cite{Brunner:2003zm} it is further argued that the deformation theory of orientifolds with holomorphic involution~$\sigma$ is unobstructed and hence the framework of variation of Hodge structure also applies for $H^3_-(Y)$. For orientifold theories with D7-branes the concept of Hodge structures is further extended to $H^3_-(Y,S^\Lambda)$ \cite{Mayr:2001}, where now certain relative forms vary with both the complex structure deformations~$z^{\tilde a}$ and the D7-brane matter fields~$\db^A$. That is to say we consider (locally) the variation of relative forms over the moduli space $\mathcal{M}_{\mathcal{N}=1}$ which has the complex coordinates $(z^{\tilde a},\db^A)$.

Part of the definition of the Hodge structure of $H^3(Y,S^\Lambda)$ is the Hodge filtration $\{F^p\}$ \cite{Morrison:1992,Cox:1999,Mayr:2001}, which is a decomposition of $H^3_-(Y,S^\Lambda)$ into\footnote{In this section, we think of relative forms as a pair of a three-form on $Y$ and a two-form on $S^\Lambda$ according to \eqref{eq:RC2}.} 
\begin{equation}
   H^3(Y,S^\Lambda)=F^0\supset \ldots \supset F^3 \ ,
\end{equation}
where
\begin{equation} \label{eq:filt}
\begin{split}
   F^3&=\widetilde H^{(3,0)}_-(Y) \ , \\
   F^2&=F^3\oplus\widetilde H^{(2,1)}_{\bar\partial,-}(Y)
           \oplus\widetilde H^{(2,0)}_{\bar\partial,-}(S^\Lambda) \ , \\
   F^1&=F^2\oplus\widetilde H^{(1,2)}_{\bar\partial,-}(Y)
           \oplus\widetilde H^{(1,1)}_{\bar\partial,-}(S^\Lambda) \ , \\
   F^0&=F^1\oplus\widetilde H^{(0,3)}_{\bar\partial,-}(Y)
           \oplus\widetilde H^{(0,2)}_{\bar\partial,-}(S^\Lambda) \ .
\end{split}
\end{equation}
Note that this filtration looks almost like the Hodge filtration of $H^3_-(Y)$ for orientifold models except for the additional two-forms of $S^\Lambda$, i.e. if one considers the case of a vanishing four-cycle $S^\Lambda$ then all relative forms reduce to ordinary three forms of $Y$ and the Hodge filtration simplifies to the orientifold case without D7-branes. 

The spaces $\widetilde H^{(3-p,p)}_{\bar\partial,-}(Y)$ for $p>3$ and $\widetilde H^{(2-q,q)}_{\bar\partial,-}(S^\Lambda)$ do not vary holomorphically with respect to $(z^{\tilde a},\db^A)$, instead $F^p$ are the fibers of holomorphic fiber bundles $\mathcal{F}^p$ over the moduli space $\mathcal{M}_{\mathcal{N}=1}$ \cite{Morrison:1992,Greene:1993vm,Cox:1999,Mayr:2001}.
\begin{table}
\begin{center}
\begin{tabular}{|c|c|c|}
   \hline
      \bf filtration  
      \rule[-1.5ex]{0pt}{4.5ex} &  \multicolumn{2}{|c|}{\bf basis} \\ \cline{2-3}
      \bf &  \bf holomorphic section  &  \bf fiber at $z^{\tilde a}=\db^A=0$
      \rule[-1.5ex]{0pt}{4.5ex} \\
   \hline
   \hline
      $\mathcal{F}^3$  &  $\rel{\Omega}$  &  $\Omega$
      \rule[-1.5ex]{0pt}{4.5ex} \\
   \hline
      $\mathcal{F}^2$  
        &  $\rel{\Omega}$, $\rel{\chi_{\tilde a}}$, $\rel{\tilde s_{A}}$
        &  $\Omega$, $\chi_{\tilde a}$, $\tilde s_{A}$
      \rule[-1.5ex]{0pt}{4.5ex} \\
   \hline
      $\mathcal{F}^1$
        &  $\rel{\Omega}$, $\rel{\chi_{\tilde a}}$, $\rel{\tilde s_{A}}$, 
           $\rel{\bar\chi_{\tilde a}}$, $\rel{\tilde\eta_{\tilde A}}$
        &  $\Omega$, $\chi_{\tilde a}$, $\tilde s_{A}$,
           $\bar\chi_{\tilde a}$, $\tilde\eta_{\tilde A}$
      \rule[-1.5ex]{0pt}{4.5ex} \\
   \hline
      $\mathcal{F}^0$
        &  $\rel{\Omega}$, $\rel{\chi_{\tilde a}}$, $\rel{\tilde s_{A}}$, 
           $\rel{\bar\chi_{\tilde a}}$, $\rel{\tilde\eta_{\tilde A}}$,
           $\rel{\bar\Omega}$, $\rel{\tilde s_{\bar A}}$
        &  $\Omega$, $\chi_{\tilde a}$, $\tilde s_{A}$,
           $\bar\chi_{\tilde a}$, $\tilde\eta_{\tilde A}$,
           $\bar\Omega$, $\tilde s_{\bar A}$
      \rule[-1.5ex]{0pt}{4.5ex} \\
   \hline
\end{tabular} 
\caption{D7-brane cycles} \label{tab:basis} 
\end{center}
\end{table}
For each holomorphic fiber bundle~$\mathcal{F}^p$ we choose a (local) basis of sections summarized in Table~\ref{tab:basis}. The fibers at $z^{\tilde a}=\db^A=0$ of these local sections coincide with the form basis of Table~\ref{tab:coh}, Table~\ref{tab:spec} and the basis $\{\eta_{\tilde A}\}$ of $\widetilde H^{(1,1)}_{\bar\partial,-}(S^\Lambda)$. Note that at a generic point in the moduli space $\mathcal{M}_{\mathcal{N}=1}$ the fibers of these sections are a mixture of various three- and two-forms due to the non-holomorphicity of the bundles $\widetilde H^{(3-p,p)}_{\bar\partial,-}(Y)$ and $\widetilde H^{(2-q,q)}_{\bar\partial,-}(S^\Lambda)$ over $\mathcal{M}_{\mathcal{N}=1}$. 

As the space $H^3_-(Y,S^\Lambda)$ is purely topological the bundle $\mathcal{F}^0=H^3_-(Y,S^\Lambda)$ is locally constant over the moduli space $\mathcal{M}_{\mathcal{N}=1}$. Thus this bundle has a canonically flat connection $\nabla$ called Gauss-Manin connection and it fulfills Griffith's transversality \cite{Morrison:1992,Greene:1993vm,Cox:1999,Forbes:2003ki} 
\begin{equation} \label{eq:connect}
   \nabla \mathcal{F}^p \subseteq \mathcal{F}^{p-1}\otimes 
      \Omega^1\mathcal{M}_{\mathcal{N}=1} \ .
\end{equation}     
Note that the covariant derivatives $\nabla_{z^{\tilde a}}$ and $\nabla_{\db^A}$ acting on sections of $\mathcal{F}^p$ differ form the ordinary derivatives $\partial_{z^{\tilde a}}$ and $\partial_{\db^A}$ only by sections in $\mathcal{F}^p\otimes\Omega^1\mathcal{M}_{\mathcal{N}=1}$ \cite{Mayr:2001}. As a consequence and with eq.~\eqref{eq:connect} one reaches local sections of all $\mathcal{F}^p$ by taking derivatives of the unique section $\rel{\Omega}(z^{\tilde a},\db^A)$ of $\mathcal{F}^3$.
\begin{figure}
\begin{center}
\begin{displaymath}
   \xymatrix{
     &  \text{\underline{3-forms of $Y\vphantom{S^\Lambda}$}}  
        &  \text{\underline{2-forms of $S^\Lambda$}} \\
     \mathcal{F}^3 \ar@{}[d]|{\bigcap}  
        &  \rel{\Omega}\ar[d]_{\partial_z}\ar[dr]^{\partial_\db} \\
     \mathcal{F}^2 \ar@{}[d]|{\bigcap}  
        &  \rel{\chi_{\tilde a}}\ar[d]_{\partial_z}\ar[dr]^{\partial_\db}
        &  \rel{\tilde s_A}\ar[d]^{\partial_z,\partial_\db} \\
     \mathcal{F}^1 \ar@{}[d]|{\bigcap}  
        &  \rel{\bar\chi_{\tilde a}}\ar[d]_{\partial_z}\ar[dr]^{\partial_\db}
        &  \rel{\tilde\eta}\ar[d]^{\partial_z,\partial_\db} \\ 
     \mathcal{F}^0  &  \rel{\bar\Omega}\ar[dr]^{\partial_z,\partial_\db} 
        &  \rel{\tilde s_{\bar A}}\ar[d]^{\partial_z,\partial_\db} \\
     &&  0 } 
\end{displaymath} 
\end{center}
\caption{Variation of Hodge structure of $H^3_-(Y,S^\Lambda)$.} \label{fig:var}
\end{figure}
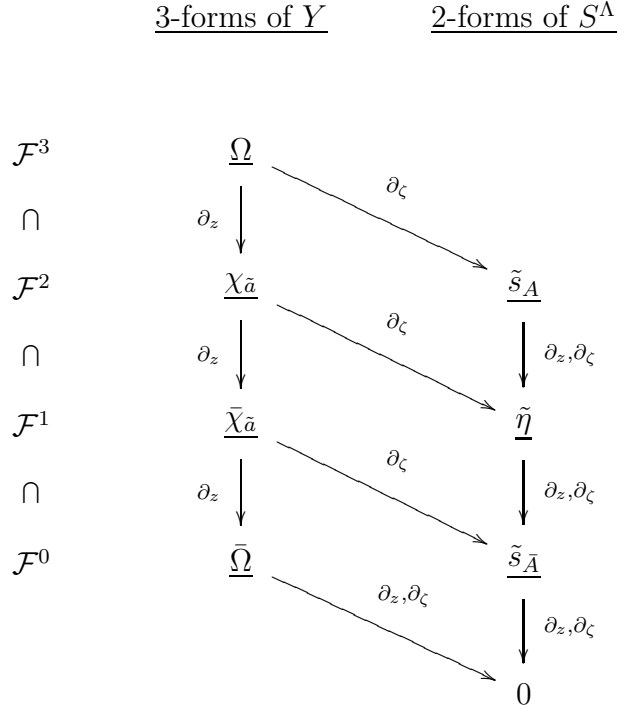
This procedure is schematically depicted in Figure~\ref{fig:var} and one obtains the extended
Kodaira formulae \cite{Candelas:1990pi}, i.e.
\begin{equation} \label{eq:Kod}
\begin{aligned}
   \partial_{z^{\tilde a}}\rel{\Omega}=k_{\tilde a}\rel{\Omega}+i \rel{\chi_{\tilde a}} \ , 
   \qquad && \qquad\partial_{\db^A}\rel{\Omega}=k_A\rel{\Omega}+\rel{\tilde s_A} \ .
\end{aligned}
\end{equation}

As discussed in section~\ref{sec:relform} the variation of relative forms captures only a subset of complex structure deformations. That is to say those complex structure deformations for which the corresponding $(2,1)$-forms of an infinitesimal deformation must be trivial on $S^\Lambda$. Now we are able to understand this condition from a different point of view: Let us assume that we perform a bulk complex structure deformations where the associated $(2,1)$-form is not trivial in $S^\Lambda$ and hence this form is not an element of $\widetilde H^3_-(Y)$. For a finite deformation in the direction of this $(2,1)$-form we obtain a new unique $(3,0)$-form of the deformed Calabi-Yau manifold~$Y$, which according to the Kodaira formula \eqref{eq:Kod} contains part of the $(2,1)$-form in terms of the old complex structure. This, however, implies that the new $(3,0)$-form does not pull back trivially to the cycle~$S^\Lambda$, and hence the four-cycle $S^\Lambda$ cannot be K\"ahler with respect to the K\"ahler-form of the bulk. Therefore the four-cycle $S^\Lambda$ is not calibrated with respect to the K\"ahler form of the bulk anymore and the theory becomes non-supersymmetric. Thus if one includes these obstructed bulk complex structure deformations in the four dimensional effective theory there should arise a potential for the associated complex structure moduli.

As alluded in Figure~\ref{fig:var} one generates sections of $\mathcal{F}^p$ for all $p$ by acting with the connection $\nabla$ on the unique relative form $\rel{\Omega}$. Since $H^3(Y,S^\Lambda)$ is a finite dimensional space one obtains linear relations among $\rel{\Omega}$ and its covariant derivatives \cite{Greene:1993vm,Hosono:1994av}, i.e. 
\begin{equation} \label{eq:GKZ}
   \mathcal{L}(z,\db,\partial_z,\partial_\db)\rel{\Omega}(z,\db) \sim 0 \ ,
\end{equation}
where $\mathcal{L}(z,\db,\partial_z,\partial_\db)$ are fourth order differential operators, and where $\sim$ means modulo exact relative forms.

Similar to the derivation of the GKZ equations for the bulk complex structure deformations, the system of differential equations of forms \eqref{eq:GKZ} can be transformed into a set of differential equations over relative periods \cite{Mayr:2001}. These relative periods arise as integrals of the relative three-form $\rel{\Omega}$ over a fixed homology basis of relative three-cycles. For this basis we choose $\{\rel{A^{\hat a}}, \rel{B_{\hat a}},\rel{\Gamma^{\hat A}}\}$ which is dual to the relative forms associated to the three-forms $\alpha_{\hat a}$, $\beta^{\hat b}$ and the two-forms $\gamma_{\hat A}$ where the latter forms are a basis of $\widetilde H^2_-(S^\Lambda)$. With this choice eq.~\eqref{eq:GKZ} gives rise to the system of differential equations for the relative periods
\begin{equation} \label{eq:GKZ2}
\begin{aligned}
   \mathcal{L}(z,\db,\partial_z,\partial_\db) \Pi^{\hat a}(z,\db)=0 \ , 
   &&\mathcal{L}(z,\db,\partial_z,\partial_\db) \Pi_{\hat a}(z,\db)=0 \ , 
   &&\mathcal{L}(z,\db,\partial_z,\partial_\db)\Pi^{\hat A}(z,\db)=0 \ , 
\end{aligned}
\end{equation}
where 
\begin{align}
   \Pi^{\hat a}(z,\db)=\langle \rel{A^{\hat a}},\rel{\Omega}\rangle \ , &&
   \Pi_{\hat a}(z,\db)=\langle \rel{B_{\hat a}},\rel{\Omega}\rangle \ , &&
   \Pi^{\hat A}(z,\db)=\langle \rel{\Gamma^{\hat A}},\rel{\Omega}\rangle \ . 
\end{align}
The solution to the system of partial differential equations \eqref{eq:GKZ2} takes the form
\begin{equation} \label{eq:sol1}
   \rel{\Omega}(z,\db)=X^{\hat a}(z,\db)\rel{\alpha_{\hat a}}
      +\mathcal{F}_{\hat a}(z,\db)\rel{\beta^{\hat a}}
      +\mathcal{G}^{\hat A}(z,\db)\rel{\gamma_{\hat A}} \ ,
\end{equation}
with holomorphic functions $X^{\hat a}(z,\db)$, $\mathcal{F}_{\hat a}(z,\db)$ and $\mathcal{G}^{\hat A}(z,\db)$. Note that eq.~\eqref{eq:sol1} reduces for $\db=0$ to the known bulk part where the solution is given by the prepotential $\mathcal{F}$ of $\mathcal{N}=2$ special geometry. In general we do not expect that the solution of the system of differential equations \eqref{eq:GKZ2} can be expressed in terms of a single holomorphic function $\mathcal{F}$. This reflects the fact that the structure of $\mathcal{N}=1$ is less restrictive than $\mathcal{N}=2$ supersymmetry.  

%%%%%%%%%%%%%%%%%%%%%%%%%%%%%%%%%%%%%%%%%%%%%%%%%%%%%%%%%%%%%%%%%%

\subsection{K\"ahler potential of the $\mathcal{N}=1$ moduli space} \label{sec:kcs}

%%%%%%%%%%%%%%%%%%%%%%%%%%%%%%%%%%%%%%%%%%%%%%%%%%%%%%%%%%%%%%%%%%

In this section we recall the definitions of the metrics for the bulk complex structure deformations and the D7-brane fluctuations independently. That is to say in the limit of small fields $z^{\tilde a}$ and $\db^A$ where the metric remains block diagonal. Then we apply the mathematical tools of the previous section in order to obtain a K\"ahler metric for the moduli space $\mathcal{M}_{\mathcal{N}=1}$ which is not block diagonal anymore but holds for higher orders in $z^{\tilde a}$ and $\db^A$ as well. This extension turns out to be also suitable to generalize the supergravity K\"ahler potential of section~\ref{sec:chiral}. 

In refs.~\cite{Candelas:1990pi} it is demonstrated that the metric of the bulk complex structure deformations $\delta z$ reads 
\begin{equation} \label{eq:Gz}
   \mathcal{G}_{\tilde a\tilde b}\delta z^{\tilde a}\delta\bar z^{\tilde b} 
    =\frac{3}{2\mathcal{K}}\int_Y\dd^6y \sqrt{\det g} g^{i\bar\jmath}g^{l\bar k}
      \ \delta g_{il} \delta g_{\bar\jmath\bar k} 
    =-\frac{\int_Y \chi_{\tilde a}\wedge\bar\chi_{\tilde b}}
       {\int_Y \Omega\wedge\bar\Omega}\ \delta z^{\tilde a}\delta\bar z^{\tilde b} \ .
\end{equation}
Analogously we can ask for the metric of the fluctuations $\delta\db$ which describe how the four-cycle $S^\Lambda$ is deformed in the normal direction of the ambient space $Y$. This metric is obtained by examining the variation of the volume element of $S^\Lambda$ with respect to $\delta\db$, namely one performs a normal coordinate expansion of the volume element according to \eqref{eq:PB} and with \eqref{eq:fluc} arrives at
\begin{equation}
   \mathcal{G}_{A\bar B}\:\delta\db^A\delta\db^{\bar B}
   =\frac{6}{\mathcal{K}}\int_{S^\Lambda}\dd^4\xi\sqrt{\det g}\:
   s^{i}_A s^{\bar\jmath}_{\bar B} g_{i\bar\jmath}\ \delta\db^A\delta\db^{\bar B} \ .
\end{equation}
Furthermore with \eqref{eq:G2L} one finds
\begin{equation} \label{eq:Gphi}
   \mathcal{G}_{A\bar B}\:\delta\db^A\delta\db^{\bar B}
      =i\mathcal{L}_{A\bar B}\:\delta\db^A\delta\db^{\bar B} \ .
\end{equation}

Without any D7-brane the metric of the complex structure $\mathcal{G}_{\tilde a\tilde b}$ is K\"ahler with the K\"ahler potential \eqref{eq:CSt} \cite{Candelas:1990pi}. This expression must now be modified to take account for the generalized concept of the variation of Hodge structure of relative forms. In the limit of small complex structure deformations~$\delta z^{\tilde a}$ and small D7-brane fluctuations~$\delta\db^A$ the modified K\"ahler potential needs to reproduce the metrics \eqref{eq:Gz} and \eqref{eq:Gphi}. Moreover, in the limit where the D7-brane cycle~$S^\Lambda$ disappears the extended K\"ahler potential must simplify to eq.~\eqref{eq:CSt}. Guided by these observations the K\"ahler potential for the common moduli space of $z^{\tilde a}$ and $\db^A$ becomes 
\begin{equation} \label{eq:kcsx}
   K_{CS}(z,\bar z,\db,\bar\db)=-\log\left[-i\int_{(Y,S^\Lambda)} \rel{\Omega}(z,\db)
      \bullet_g\rel{\bar\Omega}(\bar z,\bar\db)\right]  \ ,
\end{equation}
where the integral over relative three-forms $\rel{A}$ and $\rel{B}$ is defined as 
\begin{equation} \label{eq:sp}
   \int_{(Y,S^\Lambda)} \rel{A}\bullet_g\rel{B}=
      g \int_Y P^{(3)}\rel{A}\wedge P^{(3)}\rel{B}
      -i \int_{S^\Lambda} P^{(2)}\rel{A}\wedge P^{(2)}\rel{B} \ .
\end{equation}
Here $g$ is a coupling constant which is needed for dimensional reasons. $P^{(3)}$ and $P^{(2)}$ are projection operators that extract the three-form and the two-form part of the relative form according to eq.~\eqref{eq:RC2}. 

The next task is to check that in the limit of small bulk fields $z^{\tilde a}$ and small D7-brane matter fields $\db^A$ the K\"ahler potential \eqref{eq:kcsx} is contained in the supergravity K\"ahler potential \eqref{eq:K1}. In order to perform the comparison eq.~\eqref{eq:kcsx} is rewritten to
\begin{multline}
   K_{CS}(z,\bar z,\db,\bar\db)=-\log\left[-i\int_Y P^{(3)}\rel{\Omega}(z,\db)
     \wedge P^{(3)}\rel{\bar\Omega}(\bar z,\bar\db)\right] \\
     -\log\left[g-i
         \frac{\int_{S^\Lambda}P^{(2)}\rel{\Omega}(z,\db)\wedge P^{(2)}\rel{\bar\Omega}(\bar z,\bar\db)}
         {\int_Y P^{(3)}\rel{\Omega}(z,\db)\wedge P^{(3)}\rel{\bar\Omega}(\bar z,\bar\db)}\right] \ .
\end{multline}
Taking now the limit and using eqs.~\eqref{eq:Kod} and \eqref{eq:L} one finds agreement with the supergravity K\"ahler potential \eqref{eq:K1} for the coupling constant
\begin{equation} \label{eq:coupl}
   g(\dil)=\frac{i(\dil-\bar\dil)}{2\kappa_4^2\mu_7} \ .
\end{equation}

Thus on the common moduli space $\mathcal{M}_{\mathcal{N}=1}$ of the complex structure deformations~$z^{\tilde a}$ and of the D7-brane matter fields~$\db^A$ the supergravity K\"ahler potential \eqref{eq:K1} is modified to
\begin{equation} \label{eq:K2}
   K(\dil,G,T,\db,z,a)=
      -\log\left[-i\int_{(Y,S^\Lambda)}\rel{\Omega}(z,\db)
      \:\bullet_{g(\dil)}\:\rel{\bar\Omega}(\bar z,\bar\db)\right]
      -2\log\left[\tfrac{1}{6}\mathcal{K}(\dil,G,T,\db,z,a)\right] \ ,
\end{equation} 
with the coupling constant \eqref{eq:coupl}. In general in this K\"ahler potential $\mathcal{K}=\mathcal{K}_{\alpha\beta\gamma}v^\alpha v^\beta v^\gamma$ also depends on the bulk complex structure deformations $z^{\tilde a}$ which enter in the process of solving for $v^\alpha(\dil,G^a,T_\alpha,z^{\tilde a},\db^A,a_I)$ because $\mathcal{L}_{A\bar B}$ has become a function of $z^{\tilde a}$. The K\"ahler potential \eqref{eq:K2} still constitutes all the scalar kinetic terms of \eqref{eq:action} but in addition it generates new terms, which are of higher order in the fields $z^{\tilde a}$ and $\db^A$, and which are not captured by the Kaluza-Klein reduction of section~\ref{sec:brane}.

%%%%%%%%%%%%%%%%%%%%%%%%%%%%%%%%%%%%%%%%%%%%%%%%%%%%%%%%%%%%%%%%%%

\section{Conclusions} \label{sec:conc}

%%%%%%%%%%%%%%%%%%%%%%%%%%%%%%%%%%%%%%%%%%%%%%%%%%%%%%%%%%%%%%%%%%

In this paper we computed the effective action of an orientifold theory with a D7-brane. Instead of specifying a particular orientifold we performed the analysis for a generic Calabi-Yau orientifold with O3/O7 planes and a D7-brane wrapped on a generic cycle thereof. The effective action was obtained by a Kaluza-Klein compactification of the bulk theory and by a reduction of the Abelian Dirac-Born-Infeld and Abelian Chern-Simons action of the D7-brane. For the resulting $\mathcal{N}=1$ supergravity action we determined the K\"ahler potential, the gauge kinetic coupling functions and investigated the structure of the D-term. In specifying this data we showed that the effective action is indeed in accord with $\mathcal{N}=1$ supergravity.

The calculated K\"ahler potential is similar to the K\"ahler potential of orientifold compactifications without branes \cite{Giddings:2001yu,BBHL,Grimm:2004uq}. In addition to the new K\"ahler variables resulting from the D7-brane matter fields and the Wilson-line moduli, the definitions of the K\"ahler variables of the pure bulk theory are also modified by the D7-brane fields. Or in other words the complex structure of the target space of the bulk scalar fields is altered in the presence of D7-branes. For instance the dilaton field of the IIB~bulk theory is not a K\"ahler coordinate anymore but is replaced by a new dilaton which differs from the old dilaton by a term depending on the D7-brane matter fields. As this K\"ahler potential for a generic Calabi-Yau orientifold with D7-branes is rather complicated we explicitly examined the K\"ahler potential for some instructive examples and compared our results with refs.~\cite{Hsu:2003cy,Lust:2004fi}.

The gauge kinetic coupling of the D7-brane gauge boson is not the dilaton as for the gauge theory of space-time filling D3-branes in Calabi-Yau orientifolds but the K\"ahler modulus which controls the volume of the D7-brane cycle. In this paper we have only considered orientifold theories with a single D7-brane which gave rise to a $U(1)$ gauge theory on the world-volume of the brane. However, more generally for the compactification of Calabi-Yau orientifolds with a stack of $N$ D7-branes the effective four dimensional $U(1)$ gauge theory of a single D7-brane is enhanced to $U(N)$. This non-Abelian gauge group implies via gaugino condensation a non-perturbative superpotential which couples to the K\"ahler modulus of the gauge kinetic function. As bulk background fluxes can only stabilize the complex dilaton and complex structure moduli \cite{Giddings:2001yu,Kachru:2002he,Blumenhagen:2002mf,Blumenhagen:2003vr}, a stack of D7-branes may be used to lift the flat directions of the K\"ahler moduli fields by gaugino condensation \cite{KKLMT:2003,Gorlich:2004qm}. This mechanism but also non-perturbative superpotentials arising from Euclidean D3-brane instantons have also been used in refs.~\cite{KKLMT:2003,Burgess:2003ic} as an ingredient for the construction of (metastable) deSitter vacua. The calculated supergravity data of this work serves as a good starting point to further examine these features for generic orientifold theories with D7-branes. It would also be interesting to include perturbative $\alpha'$ corrections into the analysis along the lines of refs.~\cite{BBHL,Balasubramanian:2004uy}.

In the computed effective action we have also identified a D-term resulting from charged fields which transform non-linearly under the $U(1)$ gauge theory of the D7-brane. This D-term appears already without any fluxes, however it is further modified if we turn on internal fluxes on the world-volume of the D7-brane. The analysis of these D-terms and the effect of more general internal D7-brane fluxes is interesting in its own right and will be presented elsewhere.

Motivated by the structure of $\mathcal{N}=1$ special geometry in the context of D5-branes \cite{Mayr:2001}, we analogously analyzed the common moduli space of the bulk complex structure deformations and the D7-brane fluctuations. We found a K\"ahler potential for this moduli space and adjusted the supergravity K\"ahler potential accordingly. The K\"ahler metric for this extended K\"ahler potential exhibits not the block-diagonal structure of the complex structure moduli and the D7-brane matter moduli anymore but includes also off-diagonal entries. We showed that these entries vanish in the limit of small complex structure and small D7-brane matter field excitations.  

%%%%%%%%%%%%%%%%%%%%%%%%%%%%%%%%%%%%%%%%%%%%%%%%%%%%%%%%%%%%%%%%%%
%%%%%%%%%%%%%%%%%%%%%%%%%%%%%%%%%%%%%%%%%%%%%%%%%%%%%%%%%%%%%%%%%%
%                                                                %
%    Appendices                                                  %
%                                                                %
%%%%%%%%%%%%%%%%%%%%%%%%%%%%%%%%%%%%%%%%%%%%%%%%%%%%%%%%%%%%%%%%%%
%%%%%%%%%%%%%%%%%%%%%%%%%%%%%%%%%%%%%%%%%%%%%%%%%%%%%%%%%%%%%%%%%%

\vskip 3cm
\bigskip
\noindent {\Large {\bf Appendix}}
\appendix
\bigskip

%%%%%%%%%%%%%%%%%%%%%%%%%%%%%%%%%%%%%%%%%%%%%%%%%%%%%%%%%%%%%%%%%%

\section{Normal coordinate expansion} \label{sec:normal}

%%%%%%%%%%%%%%%%%%%%%%%%%%%%%%%%%%%%%%%%%%%%%%%%%%%%%%%%%%%%%%%%%%

In the Dirac-Born-Infeld action and the Chern-Simons action of the D$p$-brane various tensors fields of the bulk theory are pulled back from the space-time manifold $M$ to the world-volume $\mathcal{W}$ of the brane via $\varphi:\mathcal{W}\hookrightarrow M$. As the embedding map $\varphi$ is not rigid but fluctuates due to the dynamics of the brane, a normal coordinate expansion has to be performed so as to extract the couplings of these brane fluctuations to the bulk fields. The details of this procedure are described in appendix~C of ref.~\cite{Grana:2003ek}, however, for convenience the relevant formulae are recalled here.

The fluctuation of the world-volume $\mathcal{W}$ embedded in the space-time manifold $M$ can be described by considering a displacement vector field $\db$ in the normal bundle of the world-volume. The world-volume shifted by $\db$ is embedded via the map $\varphi_\db$. Note that for $\db=0$ the two maps $\varphi$ and $\varphi_\db$ coincide. For small fluctuations $\db$ any bulk tensor field~$T$ pulled back with the map $\varphi_\db$ can be expanded in terms of tensor fields pulled back with the map $\varphi$, i.e.
\begin{equation} \label{eq:normexp}
   \varphi_\db^*T=\varphi^*\left(e^{\nabla_{\db}}T\right) 
     =\varphi^*\left(T\right)+\varphi^*\left(\nabla_\db T\right)+
       \frac{1}{2}\varphi^*\left(\nabla_\db\nabla_\db T \right)+\ldots  \  ,
\end{equation}
where $\nabla$ is the Levi-Cevita connection of the manifold~$M$.

For local coordinates $x^\mu$ on the world-volume $\mathcal{W}$ where $\mu=1,\ldots,\dim\mathcal{W}$, we have the associated vector fields $\partial_\mu$, and since the Levi-Cevita connection has no torsion one can show that
\begin{align} \label{eq:normprop}
   \nabla_\db \partial_\mu = \nabla_{\partial_\mu} \db  \ , &&
   R(\db,\partial_\mu)\db=\nabla_\db\nabla_{\partial_\mu}\db
      =\nabla_\db\nabla_\db \partial_\mu \ ,
\end{align}
where $R(\cdot,\cdot)\cdot$ is the Riemann tensor. 

Applying \eqref{eq:normexp} to the metric tensor $g(\cdot,\cdot)$ of the manifold~$M$, we obtain the induced metric on the world-volume of the brane subject to the fluctuations $\db$. With the identity \eqref{eq:normprop} the expansion up to second order in derivatives yields
\begin{multline}
   \varphi^*_\db\left(g(\partial_\mu,\partial_\nu)\right)
     =\varphi^*\left(g(\partial_\mu,\partial_\nu)\right)
      +\varphi^*\left(g(\nabla_{\partial_\mu}\db,\partial_\nu)\right)
      +\varphi^*\left(g(\partial_\mu,\nabla_{\partial_\nu}\db)\right) \\
      +\varphi^*\left(g(\nabla_{\partial_\mu}\db,\nabla_{\partial_\nu}\db) \right)
      +\varphi^*\left(g(R(\db,\partial_\mu )\db,\partial_\nu )\right) +\ldots \ .
\end{multline}
This index free notation translates in a slightly abusive way of notation into the component expression
\begin{equation}\label{eq:PB}
   \varphi^*_\db(g)_{\mu\nu}=g_{\mu\nu}+g_{\mu n}
     \nabla_\nu\db^n+g_{\nu n}\nabla_\mu\db^n \\
   +g_{nm}\nabla_\mu\db^n\nabla_\nu\db^m+g_{\mu\tau}\Riem{n}{\tau}{\nu}{m}\db^n\db^m +\ldots \ ,
\end{equation}
where now $\nabla$ is the connection of the normal bundle of the world-volume~$\mathcal{W}$, which is induced form the Levi-Cevita connection of the ambient space~$M$. The Greek indices $\mu,\nu,\ldots$ denote directions tangent to the world-volume~$\mathcal{W}$ whereas the Roman indices $n,m,\ldots$ stand for directions normal to the world-volume~$\mathcal{W}$.

Analogously, one computes with \eqref{eq:normexp} and \eqref{eq:normprop} the pullback of a $q$-form of the manifold~$M$ to the world-volume~$\mathcal{W}$ and obtains up to second order in derivatives
\begin{align}\label{eq:PB2}
   \varphi^*_\db C^{(q)}
   =&\left(\tfrac{1}{q!}C^{(q)}_{\nu_1\ldots\nu_q} 
    +\tfrac{1}{q!}\db^n\partial_n(C^{(q)}_{\nu_1\ldots\nu_q})
    -\tfrac{1}{(q-1)!}\nabla_{\nu_1}\db^n C^{(q)}_{n\nu_2\ldots\nu_q} \right. \nonumber \\
   &+\tfrac{1}{2q!}\db^n\partial_n(\db^m\partial_m(C^{(q)}_{\nu_1\ldots\nu_q}))
    -\tfrac{1}{(q-1)!}\nabla_{\nu_1}\db^n\cdot\db^m\partial_m(C^{(q)}_{n\nu_2\ldots\nu_q}) \\
   &+\tfrac{1}{2(q-2)!}\nabla_{\nu_1}\db^n\nabla_{\nu_2}\db^m C^{(q)}_{nm\nu_3\ldots\nu_q}
   \left.+\tfrac{q-2}{2q!}\Riem{n}{\tau}{\nu_1}{m}\db^n\db^m
    C^{(q)}_{\tau\nu_2\ldots\nu_q}\right)\dd x^{\nu_1}\wedge\ldots\wedge\dd x^{\nu_q}  \ . \nonumber
\end{align}

%%%%%%%%%%%%%%%%%%%%%%%%%%%%%%%%%%%%%%%%%%%%%%%%%%%%%%%%%%%%%%%%%%

\section{Reduction of Dirac-Born-Infeld action} \label{sec:ints}

%%%%%%%%%%%%%%%%%%%%%%%%%%%%%%%%%%%%%%%%%%%%%%%%%%%%%%%%%%%%%%%%%%

In section~\ref{sec:D7} we have expanded the Abelian Dirac-Born-Infeld action \eqref{eq:DBIab} for the space-time filling D7-brane. Before obtaining the expanded action in its final form \eqref{eq:DBI} as an intermediate step one arrives at 
\begin{align} \label{eq:DBI1}
   \mathcal{S}^{\text{E}}_{\text{DBI}}
      &=\mu_7\ell^2\int\left[\frac{1}{4}\left(\mathcal{K}_\Lambda
        -e^{-\phi}\mathcal{K}_{\Lambda ab}\dbbf^a\dbbf^b\right) F\wedge *_4 F 
        +\frac{6}{\mathcal{K}}\mathcal{H}^{I\bar J}\dd a_I\wedge *_4\dd\bar a_{\bar J} \right] \\
      &-\mu_7 \int\left[\left(e^\phi\mathcal{G}_{A\bar B}-\tfrac{1}{4}
        \mathcal{N}_{A\bar B ab}\dbbf^a\dbbf^b\right)
        \dd\db^A\wedge *_4\dd\bar\db^{\bar B}+\frac{18}{\mathcal{K}^2} 
        \left(e^\phi\mathcal{K}_\Lambda-\mathcal{K}_{\Lambda ab}\dbbf^a\dbbf^b \right) *_4 1
        \right] \ , \nonumber
\end{align}
where $\dbbf^a$ is the background flux~$f$ introduced in \eqref{eq:bf}, and where 
\begin{equation} \label{eq:ints}
\begin{aligned}
   \mathcal{G}_{A\bar B}&=\frac{3}{\mathcal{K}}\int_{S^\Lambda} J\wedge J
       \:g_{i\bar\jmath} s_A^i s_{\bar B}^{\bar\jmath}  \ ,\qquad  
   && \mathcal{N}_{A\bar Bab}=\frac{3}{\mathcal{K}}\int_{S^\Lambda} \omega_a\wedge\omega_b 
       \:g_{i\bar\jmath} s_A^i s_{\bar B}^{\bar\jmath}  \ , \\ 
   \mathcal{H}^{I\bar J}&=\int_{S^\Lambda} A^I\wedge *_4 \bar A^{\bar J} \ . 
\end{aligned}
\end{equation}
The integral for $\mathcal{G}_{A\bar B}$ can be rewritten according to
\begin{equation} 
\begin{split}
   \mathcal{G}_{A\bar B}
      &=\frac{3}{\mathcal{K}}\int_{S^\Lambda} J\wedge J\:g_{i\bar\jmath}s_A^i\bar s_{\bar B}^{\bar\jmath} 
       =-\frac{3i}{\mathcal{K}}\int_{S^\Lambda} J\wedge J \cdot \ins{s_A}\ins{\bar s_{\bar B}}J \\
      &=-\frac{i}{\mathcal{K}}\int_{S^\Lambda}
        \ins{s_A}\ins{\bar s_{\bar B}}\left(J\wedge J\wedge J\right) 
       =i\frac{\int_{S^\Lambda} 
         \ins{s_A}\Omega\wedge\ins{\bar s_{\bar B}}\bar\Omega}{\int_Y \Omega\wedge\bar\Omega} \ ,
\end{split}
\end{equation}
where we used that the sections $s_A$ and $\bar s_{\bar B}$ are in the normal bundle of $S^\Lambda$ and that $\Omega\wedge\bar\Omega$ as well as $J\wedge J\wedge J$ are proportional to the volume form of the Calabi-Yau manifold~$Y$. Finally with \eqref{eq:PR} and \eqref{eq:L} we find
\begin{equation} \label{eq:G2L}
   \mathcal{G}_{A\bar B}=i\mathcal{L}_{A\bar B} \ .
\end{equation} 
Similarly one shows with \eqref{eq:metK} that
\begin{equation} \label{eq:N2L}
   \mathcal{N}_{A\bar Bab}=-4i\:G_{ab}\mathcal{L}_{A\bar B} \ .
\end{equation}

With the definition of the Hodge star operator $*$ on a four dimensional K\"ahler manifold with K\"ahler form $J$ we find for a one form $\bar A$ of type $(0,1)$ the identity $*\bar A=2i \bar A\wedge J$. And thus $\mathcal{H}^{I\bar J}$ of \eqref{eq:ints} fulfills with \eqref{eq:Cint}
\begin{equation} \label{eq:C2H}
   \mathcal{H}^{I\bar J}=2iv^\alpha \mathcal{C}_\alpha^{I\bar J} \ .
\end{equation}

If we now insert the expressions \eqref{eq:G2L}, \eqref{eq:N2L} and \eqref{eq:C2H} into \eqref{eq:DBI1}, we arrive at the action \eqref{eq:DBI} of section~\ref{sec:D7}.

%%%%%%%%%%%%%%%%%%%%%%%%%%%%%%%%%%%%%%%%%%%%%%%%%%%%%%%%%%%%%%%%%%

\section{Mixed gauge kinetic coupling functions} \label{sec:3form}

%%%%%%%%%%%%%%%%%%%%%%%%%%%%%%%%%%%%%%%%%%%%%%%%%%%%%%%%%%%%%%%%%%

In section~\ref{sec:D7} we have assumed that all the integrals \eqref{eq:aints} vanish, and that therefore the topological Yang-Mills terms $\dd V^{\hat\alpha}\wedge F$ and $\dd U_{\hat\alpha}\wedge F$ in the Chern-Simons action \eqref{eq:CS1} disappear. Here we drop this assumption and examine how these additional terms in the action modify the gauge kinetic coupling functions \eqref{eq:kinbulk} and \eqref{eq:fbrane}. However, the presence of non-zero integrals \eqref{eq:aints} also implies that there are three forms of the Calabi-Yau manifold~$Y$, which pull back non-trivially to $S^\Lambda$. As a consequence there are bulk complex structure deformations, which are not captured by the variation of mixed Hodge structure of relative forms (cf. section~\ref{sec:geom} and \eqref{eq:RC}). In order to avoid these subtleties of the complex structure deformations these moduli are kept fixed in this appendix. 

As the appearance of non-zero integrals \eqref{eq:aints} only involves the topological Yang-Mills terms of the bulk and brane vector fields, it can only affect the gauge kinetic coupling functions of the defining data of the $\mathcal{N}=1$ supergravity action, that is to say the K\"ahler potential and the superpotential remain unchanged.\footnote{In general the D-terms of the scalar potential are also modified through their dependence on the gauge kinetic coupling functions \eqref{eq:V}.} Therefore our first task is to collect all the terms relevant for the gauge kinetic coupling functions
\begin{equation}
\begin{split}
   \mathcal{S}^\text{E}_\text{YM}=&\frac{1}{2\kappa_4^2}\int\left[ 
      \frac{1}{4} B_{\hat\alpha\hat\beta} \dd V^{\hat\alpha}\wedge *_4\dd V^{\hat\beta}-
        \frac{1}{4} C^{\hat\alpha\hat\beta} \dd U_{\hat\alpha}\wedge *_4\dd U_{\hat\beta} 
        -\frac{1}{2}\bti{A}{\hat\beta}{\hat\alpha} 
        \dd U_{\hat\alpha}\wedge *_4\dd V^{\hat\beta} \right. \\
      &+\kappa_4^2\mu_7\ell^2\left(\tfrac{1}{2}\mathcal{K}_\Lambda
        -\tfrac{1}{2}e^{-\phi}\mathcal{K}_{\Lambda ab}\dbbf^a\dbbf^b\right) F\wedge *_4 F \\
      &+\kappa_4^2\mu_7\ell^2\left(\rho_\Lambda-\mathcal{K}_{\Lambda ab}c^a\dbbf^b
        +\tfrac{1}{2}\mathcal{K}_{\Lambda ab}\dbbf^a\dbbf^bl\right)F\wedge F  \\
      &-\left.2\kappa_4^2\mu_7\ell^2\left((a_{\hat\alpha}+\bar a_{\hat\alpha}) 
        \dd V^{\hat\alpha}\wedge F
        +(a^{\hat\alpha}+\bar a^{\hat\alpha})\dd U_{\hat\alpha}\wedge F \right)\vphantom{\frac{1}{2}}\right] \ .
\end{split}
\end{equation}
Note that in this action the vector fields $V^{\hat\alpha}$ and $U_{\hat\alpha}$ are not independent but via \eqref{eq:dual} dual among another. Therefore we eliminate the fields $U_{\hat\alpha}$ by the procedure described at the beginning of section~\ref{sec:effact} and obtain 
\begin{equation} \label{eq:YM} 
\begin{split}
   \mathcal{S}^\text{E}_\text{YM}=&\frac{1}{2\kappa_4^2}\int\left[ 
      \frac{1}{2}(\Imag \mathcal{M})_{\hat\alpha\hat\beta} 
          \dd V^{\hat\alpha}\wedge *_4 \dd V^{\hat\beta}
          +\frac{1}{2}(\Real \mathcal{M})_{\hat\alpha\hat\beta}
          \dd V^{\hat\alpha}\wedge\dd V^{\hat\beta} \right. \\
      &+\kappa_4^2\mu_7\ell^2 \left(\tfrac{1}{2}\mathcal{K}_\Lambda
          -\tfrac{1}{2}e^{-\phi}\mathcal{K}_{\Lambda ab}\dbbf^a\dbbf^b
          +8\kappa_4^2\mu_7\ell^2 C_{\hat\alpha\hat\beta}
          \left(2\bar a^{\hat\alpha}a^{\hat\beta}+a^{\hat\alpha}a^{\hat\beta}
          +\bar a^{\hat\alpha}\bar a^{\hat\beta}\right) \right) F\wedge *_4 F \\
      &+\kappa_4^2\mu_7\ell^2\left(\rho_\Lambda-\mathcal{K}_{\Lambda ab}c^a\dbbf^b
          +\tfrac{1}{2}\mathcal{K}_{\Lambda ab}\dbbf^a\dbbf^bl-8i\kappa_4^2\mu_7\ell^2
          C_{\hat\alpha\hat\beta}
          \left(a^{\hat\alpha}a^{\hat\beta}-\bar a^{\hat\alpha}\bar a^{\hat\beta}\right)
          \right) F\wedge F  \\
      &-4\kappa_4^2\mu_7\ell^2 C_{\hat\alpha\hat\beta}\left((a^{\hat\beta}+\bar a^{\hat\beta})
          \:\dd V^{\hat\alpha}\wedge *_4 F -i(a^{\hat\beta}-\bar a^{\hat\beta})
          \:\dd V^{\hat\alpha}\wedge F\right) \left. \vphantom{\frac{1}{2}}\right] \ , 
\end{split}
\end{equation}
where $C_{\hat\alpha\hat\beta}$ is the inverse matrix of \eqref{eq:mat3}, which is a constant for the case of fixed complex structure moduli.

The dualization of the vectors $U_{\hat\alpha}$ has also generated kinetic terms, which are mixtures of the bulk vectors $V^{\hat\alpha}$ and the D7-brane vector. From the action \eqref{eq:YM} we can now read off the gauge kinetic coupling matrix, which for the field strength vector $F^\Gamma=\left(\dd V^{\hat\alpha}, F\right)$ reads in terms of chiral fields\footnote{As in section~\ref{sec:chiral} extracting the real part of the chiral field $T_\Lambda$ we find a mismatch by one term involving the Wilson line fields $a^I$. Here we also expect this disagreement to be corrected at the one loop level of open string amplitudes \cite{Berg:2004ek}.}
\begin{equation} \label{eq:fmix}
   f_{\Gamma\Delta}=\begin{pmatrix}
        -\frac{i}{2}\mathcal{\bar M}_{\hat\alpha\hat\beta} 
            & -4\kappa_4^2\mu_7\ell^2 C_{\hat\alpha\hat\gamma}a^{\hat\gamma} \\
        -4\kappa_4^2\mu_7\ell^2 C_{\hat\gamma\hat\beta}a^{\hat\gamma} 
            & \frac{2}{3}\kappa_4^2\mu_7\ell^2 T_\Lambda
              +16\kappa_4^4\mu_7^2\ell^4 C_{\hat\gamma\hat\delta} a^{\hat\gamma}a^{\hat\delta} 
     \end{pmatrix} \ .
\end{equation}
First one observes that for vanishing integrals \eqref{eq:aints} the gauge kinetic coupling functions \eqref{eq:kinbulk} and \eqref{eq:fbrane} are recovered. However, for this more general case the gauge kinetic coupling functions are not anymore diagonal in the bulk and D7-brane vector fields, as there arise off-diagonal entries depending on the Wilson line fields. Furthermore the coupling matrix \eqref{eq:fmix} is still holomorphic in the chiral fields at least as long as the complex structure moduli are kept fixed. 

%%%%%%%%%%%%%%%%%%%%%%%%%%%%%%%%%%%%%%%%%%%%%%%%%%%%%%%%%%%%%%%%%%

\vskip 1cm
\subsection*{Acknowledgments}

%%%%%%%%%%%%%%%%%%%%%%%%%%%%%%%%%%%%%%%%%%%%%%%%%%%%%%%%%%%%%%%%%%

We would like to thank Thomas Grimm, Michael Haack, Olaf Hohm, Renata Kallosh, Dieter L\"ust, Peter Mayr, Andrei Micu, Sakura Sch\"afer-Nameki, Uwe Semmelmann, Angel Uranga, Nick Warner and Marco Zagermann for helpful conversations and correspondences.  This work is supported by DFG -- The German Science Foundation, the European RTN Program HPRN-CT-2000-00148, the DAAD -- the German Academic Exchange Service. 

%%%%%%%%%%%%%%%%%%%%%%%%%%%%%%%%%%%%%%%%%%%%%%%%%%%%%%%%%%%%%%%%%%

\newpage

%%%%%%%%%%%%%%%%%%%%%%%%%%%%%%%%%%%%%%%%%%%%%%%%%%%%%%%%%%%%%%%%%%

\providecommand{\href}[2]{#2}

\begingroup
\begin{thebibliography}{99}

%%%%%%%%%%%%%%%%%%%%%%%%%%%%%%%%
% Introduction                 %
%%%%%%%%%%%%%%%%%%%%%%%%%%%%%%%%

%\cite{Polchinski:1995mt}
\bibitem{Polchinski:1995mt}
J.~Polchinski,
``Dirichlet-Branes and Ramond-Ramond Charges,''
Phys.\ Rev.\ Lett.\  {\bf 75} (1995) 4724
[arXiv:hep-th/9510017].
%%CITATION = HEP-TH 9510017;%%

%\cite{reviewPP}
\bibitem{reviewPP}
For a review see, for example,
E.~Kiritsis,
``D-branes in standard model building, gravity and cosmology,''
Fortsch.\ Phys.\  {\bf 52} (2004) 200
[arXiv:hep-th/0310001];\\
A.~M.~Uranga,
``Chiral four-dimensional string compactifications with intersecting
D-branes,''
Class.\ Quant.\ Grav.\  {\bf 20}, S373 (2003)
[arXiv:hep-th/0301032];\\
D.~L\"ust,
``Intersecting brane worlds: A path to the standard model?,''
Class.\ Quant.\ Grav.\  {\bf 21} (2004) S1399
[arXiv:hep-th/0401156];\\
L.~E.~Ib\'a\~nez,
``The fluxed MSSM,''
arXiv:hep-ph/0408064, and references therein.

%\cite{reviewcosmo}
\bibitem{reviewcosmo}
For a review see, for example,
A.~Linde,
``Prospects of inflation,''
arXiv:hep-th/0402051;\\
V.~Balasubramanian,
``Accelerating universes and string theory,''
Class.\ Quant.\ Grav.\  {\bf 21} (2004) S1337
[arXiv:hep-th/0404075];\\
C.~P.~Burgess,
``Inflationary String Theory?,''
arXiv:hep-th/0408037,
and references therein.

%\cite{JP}
\bibitem{JP}
A.~Sagnotti,
``Open Strings And Their Symmetry Groups,''
arXiv:hep-th/0208020;\\
J.~Dai, R.~G.~Leigh and J.~Polchinski,
``New Connections Between String Theories,''
Mod.\ Phys.\ Lett.\ A {\bf 4} (1989) 2073;\\
R.~G.~Leigh,
``Dirac-Born-Infeld Action From Dirichlet Sigma Model,''
Mod.\ Phys.\ Lett.\ A {\bf 4} (1989) 2767;\\
M.~Bianchi and A.~Sagnotti,
``On The Systematics Of Open String Theories,''
Phys.\ Lett.\ B {\bf 247} (1990) 517;
``Twist Symmetry And Open String Wilson Lines,''
Nucl.\ Phys.\ B {\bf 361} (1991) 519;\\
P.~Horava,
``Strings On World Sheet Orbifolds,''
Nucl.\ Phys.\ B {\bf 327} (1989) 461.

%\cite{PradisiGimon}
\bibitem{PradisiGimon}
G.~Pradisi and A.~Sagnotti,
``Open String Orbifolds,''
Phys.\ Lett.\ B {\bf 216}, 59 (1989);\\
E.~G.~Gimon and J.~Polchinski,
``Consistency Conditions for Orientifolds and D-Manifolds,''
Phys.\ Rev.\ D {\bf 54} (1996) 1667
[arXiv:hep-th/9601038].

%\cite{Frey:2002hf}
\bibitem{Frey:2002hf}
A.~R.~Frey and J.~Polchinski,
``N = 3 warped compactifications,''
Phys.\ Rev.\ D {\bf 65}, 126009 (2002)
[arXiv:hep-th/0201029].
%%CITATION = HEP-TH 0201029;%%

%\cite{Ferrara}
\bibitem{Ferrara}
R.~D'Auria, S.~Ferrara, F.~Gargiulo, M.~Trigiante and S.~Vaula,
``N = 4 supergravity Lagrangian for type IIB on T**6/Z(2) in presence of  fluxels and D3-branes,''
JHEP {\bf 0306} (2003) 045
[arXiv:hep-th/0303049];\\
C.~Angelantonj, S.~Ferrara and M.~Trigiante,
``New D = 4 gauged supergravities from N = 4 orientifolds with fluxes,''
JHEP {\bf 0310} (2003) 015
[arXiv:hep-th/0306185];\\
for a review see,
L.~Andrianopoli, S.~Ferrara and M.~Trigiante,
``Fluxes, supersymmetry breaking and gauged supergravity,''
arXiv:hep-th/0307139.

%\cite{BHK}
\bibitem{BHK}
M.~Berg, M.~Haack and B.~K{\"o}rs,
``An orientifold with fluxes and branes via T-duality,''
Nucl.\ Phys.\ B {\bf 669} (2003) 3
[arXiv:hep-th/0305183];\\
M.~Berg, M.~Haack and B.~K\"ors,
``Brane/Flux Interactions in Orientifolds,''
arXiv:hep-th/0312172.

%\cite{Bachas:1995ik}
\bibitem{Bachas:1995ik}
C.~Bachas,
``A Way to break supersymmetry,''
arXiv:hep-th/9503030.
%%CITATION = HEP-TH 9503030;%%

%\cite{Polchinski:1995sm}
\bibitem{Polchinski:1995sm}
J.~Polchinski and A.~Strominger,
``New Vacua for Type II String Theory,''
Phys.\ Lett.\ B {\bf 388}, 736 (1996)
[arXiv:hep-th/9510227].
%%CITATION = HEP-TH 9510227;%%

%\cite{Michelson:1996pn}
\bibitem{Michelson:1996pn}
J.~Michelson,
``Compactifications of type IIB strings to four dimensions with  non-trivial
classical potential,''
Nucl.\ Phys.\ B {\bf 495}, 127 (1997)
[arXiv:hep-th/9610151].
%%CITATION = HEP-TH 9610151;%%

%\cite{Gukov:1999ya}
\bibitem{Gukov:1999ya}
S.~Gukov, C.~Vafa and E.~Witten,
``CFT's from Calabi-Yau four-folds,''
Nucl.\ Phys.\ B {\bf 584}, 69 (2000)
[Erratum-ibid.\ B {\bf 608}, 477 (2001)]
[arXiv:hep-th/9906070].
%%CITATION = HEP-TH 9906070;%%

%\cite{Dasgupta:1999ss}
\bibitem{Dasgupta:1999ss}
K.~Dasgupta, G.~Rajesh and S.~Sethi,
``M theory, orientifolds and G-flux,''
JHEP {\bf 9908}, 023 (1999)
[arXiv:hep-th/9908088].
%%CITATION = HEP-TH 9908088;%%

%\cite{Taylor:1999ii}
\bibitem{Taylor:1999ii}
T.~R.~Taylor and C.~Vafa,
``RR flux on Calabi-Yau and partial supersymmetry breaking,''
Phys.\ Lett.\ B {\bf 474}, 130 (2000)
[arXiv:hep-th/9912152].
%%CITATION = HEP-TH 9912152;%%

%\cite{Mayr:2000hh}
\bibitem{Mayr:2000hh}
P.~Mayr,
``On supersymmetry breaking in string theory and its realization in brane
worlds,''
Nucl.\ Phys.\ B {\bf 593}, 99 (2001)
[arXiv:hep-th/0003198].
%%CITATION = HEP-TH 0003198;%%

%\cite{Curio:2000sc}
\bibitem{Curio:2000sc}
G.~Curio, A.~Klemm, D.~L\"ust and S.~Theisen,
``On the vacuum structure of type II string compactifications on  Calabi-Yau
spaces with H-fluxes,''
Nucl.\ Phys.\ B {\bf 609}, 3 (2001)
[arXiv:hep-th/0012213].
%%CITATION = HEP-TH 0012213;%%

%\cite{Giddings:2001yu}
\bibitem{Giddings:2001yu}
S.~B.~Giddings, S.~Kachru and J.~Polchinski,
``Hierarchies from fluxes in string compactifications,''
Phys.\ Rev.\ D {\bf 66}, 106006 (2002)
[arXiv:hep-th/0105097].
%%CITATION = HEP-TH 0105097;%%

\bibitem{BB}
K.~Becker and M.~Becker,
``Supersymmetry breaking, M-theory and fluxes,''
JHEP {\bf 0107} (2001) 038
[arXiv:hep-th/0107044].

%\cite{Kachru:2002he}
\bibitem{Kachru:2002he}
S.~Kachru, M.~B.~Schulz and S.~Trivedi,
``Moduli stabilization from fluxes in a simple IIB orientifold,''
JHEP {\bf 0310}, 007 (2003)
[arXiv:hep-th/0201028].
%%CITATION = HEP-TH 0201028;%%

\bibitem{BBHL}
K.~Becker, M.~Becker, M.~Haack and J.~Louis,
``Supersymmetry breaking and alpha'-corrections to flux induced  potentials,''
JHEP {\bf 0206} (2002) 060
[arXiv:hep-th/0204254].

\bibitem{DWG} O.~DeWolfe and S.~B.~Giddings,
``Scales and hierarchies in warped compactifications and brane worlds,''
Phys.\ Rev.\ D {\bf 67}, 066008 (2003)
[arXiv:hep-th/0208123].

%\cite{Giryavets:2003vd}
\bibitem{Giryavets:2003vd}
A.~Giryavets, S.~Kachru, P.~K.~Tripathy and S.~P.~Trivedi,
``Flux compactifications on Calabi-Yau threefolds,''
JHEP {\bf 0404}, 003 (2004)
[arXiv:hep-th/0312104].
%%CITATION = HEP-TH 0312104;%%

%\cite{Grana:2002}
\bibitem{Grana:2002}
M.~Grana,
``D3-brane action in a supergravity background: The fermionic story,''
Phys.\ Rev.\ D {\bf 66} (2002) 045014
[arXiv:hep-th/0202118]; \\
M.~Gra\~na,
``MSSM parameters from supergravity backgrounds,''
Phys.\ Rev.\ D {\bf 67} (2003) 066006
[arXiv:hep-th/0209200].

%\cite{Kors:2003wf}
\bibitem{Kors:2003wf}
B.~K\"ors and P.~Nath,
``Effective action and soft supersymmetry breaking for intersecting D-brane
models,''
Nucl.\ Phys.\ B {\bf 681}, 77 (2004)
[arXiv:hep-th/0309167].
%%CITATION = HEP-TH 0309167;%%

%\cite{Camara:2003ku}
\bibitem{Camara:2003ku}
P.~G.~C\'amara, L.~E.~Ib\'a\~nez and A.~M.~Uranga,
``Flux-induced SUSY-breaking soft terms,''
Nucl.\ Phys.\ B {\bf 689} (2004) 195
[arXiv:hep-th/0311241].
%%CITATION = HEP-TH 0311241;%%

%\cite{Grana:2003ek}
\bibitem{Grana:2003ek}
M.~Gra\~na, T.~W.~Grimm, H.~Jockers and J.~Louis,
``Soft supersymmetry breaking in Calabi-Yau orientifolds with D-branes and
fluxes,''
Nucl.\ Phys.\ B {\bf 690}, 21 (2004)
[arXiv:hep-th/0312232].
%%CITATION = HEP-TH 0312232;%%

%\cite{Lawrence:2004zk}
\bibitem{Lawrence:2004zk}
A.~Lawrence and J.~McGreevy,
``Local string models of soft supersymmetry breaking,''
JHEP {\bf 0406} (2004) 007
[arXiv:hep-th/0401034].
%%CITATION = HEP-TH 0401034;%%

%\cite{Lust:2004fi}
\bibitem{Lust:2004fi}
D.~L\"ust, S.~Reffert and S.~Stieberger,
``Flux-induced soft supersymmetry breaking in chiral type IIb orientifolds with
D3/D7-branes,''
arXiv:hep-th/0406092.
%%CITATION = HEP-TH 0406092;%%

%\cite{Camara:2004jj}
\bibitem{Camara:2004jj}
P.~G.~C\'amara, L.~E.~Ib\'a\~nez and A.~M.~Uranga,
``Flux-induced SUSY-breaking soft terms on D7-D3 brane systems,''
arXiv:hep-th/0408036.
%%CITATION = HEP-TH 0408036;%%

%\cite{Blumenhagen:2002mf}
\bibitem{Blumenhagen:2002mf}
R.~Blumenhagen, B.~K\"ors and D.~L\"ust,
``Moduli stabilization for intersecting brane worlds in type 0' string
theory,''
Phys.\ Lett.\ B {\bf 532}, 141 (2002)
[arXiv:hep-th/0202024].
%%CITATION = HEP-TH 0202024;%%

%\cite{Blumenhagen:2003vr}
\bibitem{Blumenhagen:2003vr}
R.~Blumenhagen, D.~L\"ust and T.~R.~Taylor,
``Moduli stabilization in chiral type IIB orientifold models with fluxes,''
Nucl.\ Phys.\ B {\bf 663}, 319 (2003)
[arXiv:hep-th/0303016].
%%CITATION = HEP-TH 0303016;%%

%\cite{Cascales:2003pt}
\bibitem{Cascales:2003pt}
J.~F.~G.~Cascales and A.~M.~Uranga,
``Chiral 4d string vacua with D-branes and moduli stabilization,''
arXiv:hep-th/0311250.
%%CITATION = HEP-TH 0311250;%%

%\cite{Cascales:2003wn}
\bibitem{Cascales:2003wn}
J.~F.~G.~Cascales, M.~P.~Garc\'ia del Moral, F.~Quevedo and A.~M.~Uranga,
``Realistic D-brane models on warped throats: Fluxes, hierarchies and moduli
stabilization,''
JHEP {\bf 0402} (2004) 031
[arXiv:hep-th/0312051].
%%CITATION = HEP-TH 0312051;%%

%\cite{D'Auria:2004qv}
\bibitem{D'Auria:2004qv}
R.~D'Auria, S.~Ferrara and M.~Trigiante,
``Orientifolds, brane coordinates and special geometry,''
arXiv:hep-th/0407138.
%%CITATION = HEP-TH 0407138;%%

%\cite{KKLMT:2003}
\bibitem{KKLMT:2003}
S.~Kachru, R.~Kallosh, A.~Linde and S.~P.~Trivedi,
``De Sitter vacua in string theory,''
Phys.\ Rev.\ D {\bf 68} (2003) 046005
[arXiv:hep-th/0301240];\\
S.~Kachru, R.~Kallosh, A.~Linde, J.~Maldacena, L.~McAllister and S.~P.~Trivedi,
``Towards inflation in string theory,''
JCAP {\bf 0310} (2003) 013
[arXiv:hep-th/0308055].

%\cite{Denef:2004dm}
\bibitem{Denef:2004dm}
F.~Denef, M.~R.~Douglas and B.~Florea,
``Building a better racetrack,''
JHEP {\bf 0406} (2004) 034
[arXiv:hep-th/0404257].
%%CITATION = HEP-TH 0404257;%%

%\cite{Gorlich:2004qm}
\bibitem{Gorlich:2004qm}
L.~G\"orlich, S.~Kachru, P.~K.~Tripathy and S.~P.~Trivedi,
``Gaugino condensation and nonperturbative superpotentials in flux
compactifications,''
arXiv:hep-th/0407130.
%%CITATION = HEP-TH 0407130;%%

%\cite{Burgess:2003ic}
\bibitem{Burgess:2003ic}
C.~P.~Burgess, R.~Kallosh and F.~Quevedo,
``de Sitter string vacua from supersymmetric D-terms,''
JHEP {\bf 0310}, 056 (2003)
[arXiv:hep-th/0309187].
%%CITATION = HEP-TH 0309187;%%

%\cite{Balasubramanian:2004uy}
\bibitem{Balasubramanian:2004uy}
V.~Balasubramanian and P.~Berglund,
``Stringy corrections to Kaehler potentials, SUSY breaking, and the
cosmological constant problem,''
arXiv:hep-th/0408054.
%%CITATION = HEP-TH 0408054;%%

%\cite{Lust:2004cx}
\bibitem{Lust:2004cx}
D.~L\"ust, P.~Mayr, R.~Richter and S.~Stieberger,
``Scattering of gauge, matter, and moduli fields from intersecting branes,''
arXiv:hep-th/0404134.
%%CITATION = HEP-TH 0404134;%%

%\cite{Grimm:2004uq}
\bibitem{Grimm:2004uq}
T.~W.~Grimm and J.~Louis,
``The effective action of N = 1 Calabi-Yau orientifolds,''
arXiv:hep-th/0403067.
%%CITATION = HEP-TH 0403067;%%

%\cite{Bergshoeff:2001pv}
\bibitem{Bergshoeff:2001pv}
E.~Bergshoeff, R.~Kallosh, T.~Ortin, D.~Roest and A.~Van Proeyen,
``New formulations of D = 10 supersymmetry and D8 - O8 domain walls,''
Class.\ Quant.\ Grav.\  {\bf 18} (2001) 3359
[arXiv:hep-th/0103233].
%%CITATION = HEP-TH 0103233;%%

%\cite{Mayr:2001}
\bibitem{Mayr:2001}
P.~Mayr,
``N = 1 mirror symmetry and open/closed string duality,''
Adv.\ Theor.\ Math.\ Phys.\  {\bf 5} (2002) 213
[arXiv:hep-th/0108229];\\
W.~Lerche and P.~Mayr,
``On N = 1 mirror symmetry for open type II strings,''
arXiv:hep-th/0111113;\\
W.~Lerche, P.~Mayr and N.~Warner,
``N = 1 special geometry, mixed Hodge variations and toric geometry,''
arXiv:hep-th/0208039;\\
W.~Lerche, P.~Mayr and N.~Warner,
``Holomorphic N = 1 special geometry of open-closed type II strings,''
arXiv:hep-th/0207259;\\
for a review see, W.~Lerche,
``Special geometry and mirror symmetry for open string backgrounds with N = 1
supersymmetry,''
arXiv:hep-th/0312326.

%%%%%%%%%%%%%%%%%%%%%%%%%%%%%%%%
% Section 2                    %
%%%%%%%%%%%%%%%%%%%%%%%%%%%%%%%%

%\cite{Acharya:2002ag}
\bibitem{Acharya:2002ag}
B.~Acharya, M.~Aganagic, K.~Hori and C.~Vafa,
``Orientifolds, mirror symmetry and superpotentials,''
arXiv:hep-th/0202208.
%%CITATION = HEP-TH 0202208;%%

%\cite{Brunner:2003zm}
\bibitem{Brunner:2003zm}
I.~Brunner and K.~Hori,
``Orientifolds and mirror symmetry,''
arXiv:hep-th/0303135.
%%CITATION = HEP-TH 0303135;%%

%\cite{Marcus:1982yu}
\bibitem{Marcus:1982yu}
N.~Marcus and J.~H.~Schwarz,
``Field Theories That Have No Manifestly Lorentz Invariant Formulation,''
Phys.\ Lett.\ B {\bf 115} (1982) 111;\\
J.~H.~Schwarz,
``Covariant Field Equations Of Chiral N=2 D = 10 Supergravity,''
Nucl.\ Phys.\ B {\bf 226} (1983) 269;\\
P.~S.~Howe and P.~C.~West,
``The Complete N=2, D = 10 Supergravity,''
Nucl.\ Phys.\ B {\bf 238}, 181 (1984).

%\cite{Dall'Agata:1998va}
\bibitem{Dall'Agata:1998va}
G.~Dall'Agata, K.~Lechner and M.~Tonin,
``D = 10, N = IIB supergravity: Lorentz-invariant actions and duality,''
JHEP {\bf 9807}, 017 (1998)
[arXiv:hep-th/9806140].
%%CITATION = HEP-TH 9806140;%%

%\cite{Gukov:2002iq}
\bibitem{Gukov:2002iq}
S.~Gukov and M.~Haack,
``IIA string theory on Calabi-Yau fourfolds with background fluxes,''
Nucl.\ Phys.\ B {\bf 639} (2002) 95
[arXiv:hep-th/0203267].
%%CITATION = HEP-TH 0203267;%%

%\cite{deAlwisBuchel}
\bibitem{deAlwisBuchel}
S.~P.~de Alwis,
``On potentials from fluxes,''
Phys.\ Rev.\ D {\bf 68} (2003) 126001
[arXiv:hep-th/0307084];\\
S.~P.~de Alwis,
``Brane worlds in 5D and warped compactifications in IIB,''
arXiv:hep-th/0407126;\\
A.~Buchel,
``On effective action of string theory flux compactifications,''
Phys.\ Rev.\ D {\bf 69}, 106004 (2004)
[arXiv:hep-th/0312076].

%\cite{Andrianopoli:2001}
\bibitem{Andrianopoli:2001}
L.~Andrianopoli, R.~D'Auria and S.~Ferrara,
``Supersymmetry reduction of N-extended supergravities in four  dimensions,''
JHEP {\bf 0203} (2002) 025
[arXiv:hep-th/0110277];\\
L.~Andrianopoli, R.~D'Auria and S.~Ferrara,
``Consistent reduction of N = 2 $\to$ N = 1 four dimensional supergravity
coupled to matter,''
Nucl.\ Phys.\ B {\bf 628} (2002) 387
[arXiv:hep-th/0112192];\\
R.~D'Auria, S.~Ferrara and M.~Trigiante,
``c-map,very special quaternionic geometry and dual Kaehler spaces,''
Phys.\ Lett.\ B {\bf 587}, 138 (2004)
[arXiv:hep-th/0401161].

%\cite{Strominger:1985ks}
\bibitem{Strominger:1985ks}
A.~Strominger,
``Yukawa Couplings In Superstring Compactification,''
Phys.\ Rev.\ Lett.\  {\bf 55} (1985) 2547.
%%CITATION = PRLTA,55,2547;%%

%\cite{Candelas:1990pi}
\bibitem{Candelas:1990pi}
P.~Candelas and X.~de la Ossa,
``Moduli Space Of Calabi-Yau Manifolds,''
Nucl.\ Phys.\ B {\bf 355} (1991) 455.
%%CITATION = NUPHA,B355,455;%%

%\cite{SuzukiCeresole}
\bibitem{SuzukiCeresole}
H.~Suzuki,
``Calabi-Yau compactification of type IIB string and a mass formula of the
extreme black holes,''
Mod.\ Phys.\ Lett.\ A {\bf 11} (1996) 623
[arXiv:hep-th/9508001];\\
A.~Ceresole, R.~D'Auria and S.~Ferrara,
``The Symplectic Structure of N=2 Supergravity and its Central Extension,''
Nucl.\ Phys.\ Proc.\ Suppl.\  {\bf 46}, 67 (1996)
[arXiv:hep-th/9509160].

%\cite{Dall'Agata:2001zh}
\bibitem{Dall'Agata:2001zh}
G.~Dall'Agata,
``Type IIB supergravity compactified on a Calabi-Yau manifold with  H-fluxes,''
JHEP {\bf 0111} (2001) 005
[arXiv:hep-th/0107264].
%%CITATION = HEP-TH 0107264;%%

%\cite{Louis:2002ny}
\bibitem{Louis:2002ny}
J.~Louis and A.~Micu,
``Type II theories compactified on Calabi-Yau threefolds in the presence  of
background fluxes,''
Nucl.\ Phys.\ B {\bf 635}, 395 (2002)
[arXiv:hep-th/0202168].
%%CITATION = HEP-TH 0202168;%%

%%%%%%%%%%%%%%%%%%%%%%%%%%%%%%%%
% Section 3                    %
%%%%%%%%%%%%%%%%%%%%%%%%%%%%%%%%

%\cite{Polchinski:1998}
\bibitem{Polchinski:1998}
J.~Polchinski,
``String theory'',
Cambridge University Press (1998).

%\cite{Berkooz:1996km}
\bibitem{Berkooz:1996km}
M.~Berkooz, M.~R.~Douglas and R.~G.~Leigh,
``Branes intersecting at angles,''
Nucl.\ Phys.\ B {\bf 480} (1996) 265
[arXiv:hep-th/9606139].
%%CITATION = HEP-TH 9606139;%%

%\cite{Witten:1992fb}
\bibitem{Witten:1992fb}
E.~Witten,
``Chern-Simons gauge theory as a string theory,''
Prog.\ Math.\  {\bf 133} (1995) 637
[arXiv:hep-th/9207094].
%%CITATION = HEP-TH 9207094;%%

%\cite{Katz:2002gh}
\bibitem{Katz:2002gh}
S.~Katz and E.~Sharpe,
``D-branes, open string vertex operators, and Ext groups,''
Adv.\ Theor.\ Math.\ Phys.\  {\bf 6} (2003) 979
[arXiv:hep-th/0208104].
%%CITATION = HEP-TH 0208104;%%

%\cite{Douglas:1996sw}
\bibitem{Douglas:1996sw}
M.~R.~Douglas and G.~W.~Moore,
``D-branes, Quivers, and ALE Instantons,''
arXiv:hep-th/9603167.
%%CITATION = HEP-TH 9603167;%%

%\cite{Witten:1995im}
\bibitem{Witten:1995im}
E.~Witten,
``Bound states of strings and p-branes,''
Nucl.\ Phys.\ B {\bf 460} (1996) 335
[arXiv:hep-th/9510135].
%%CITATION = HEP-TH 9510135;%%

%\cite{Douglas:1997}
\bibitem{Douglas:1997}
M.~R.~Douglas,
``D-branes and matrix theory in curved space,''
Nucl.\ Phys.\ Proc.\ Suppl.\  {\bf 68}, 381 (1998)
[arXiv:hep-th/9707228]; \\
M.~R.~Douglas,
``D-branes in curved space,''
Adv.\ Theor.\ Math.\ Phys.\  {\bf 1}, 198 (1998)
[arXiv:hep-th/9703056]; \\
M.~R.~Douglas, A.~Kato and H.~Ooguri,
``D-brane actions on Kaehler manifolds,''
Adv.\ Theor.\ Math.\ Phys.\  {\bf 1} (1998) 237
[arXiv:hep-th/9708012].

%\cite{Douglas:1995bn}
\bibitem{Douglas:1995bn}
M.~R.~Douglas,
``Branes within branes,''
arXiv:hep-th/9512077.
%%CITATION = HEP-TH 9512077;%%

%\cite{Hughes:1986fa}
\bibitem{Hughes:1986fa}
J.~Hughes, J.~Liu and J.~Polchinski,
``Supermembranes,''
Phys.\ Lett.\ B {\bf 180} (1986) 370;\\
M.~Aganagic, C.~Popescu and J.~H.~Schwarz,
``D-brane actions with local kappa symmetry,''
Phys.\ Lett.\ B {\bf 393}, 311 (1997)
[arXiv:hep-th/9610249];\\
M.~Cederwall, A.~von Gussich, B.~E.~Nilsson, P.~Sundell and A.~Westerberg,
``The Dirichlet super-p-branes in ten-dimensional type IIA and IIB  supergravity,''
Nucl.\ Phys.\ B {\bf 490}, 179 (1997)
[arXiv:hep-th/9611159];\\
E.~Bergshoeff and P.~K.~Townsend,
``Super D-branes,''
Nucl.\ Phys.\ B {\bf 490} (1997) 145
[arXiv:hep-th/9611173].

%\cite{Becker:1995kb}
\bibitem{Becker:1995kb}
K.~Becker, M.~Becker and A.~Strominger,
``Five-branes, membranes and nonperturbative string theory,''
Nucl.\ Phys.\ B {\bf 456}, 130 (1995)
[arXiv:hep-th/9507158];\\
E.~Bergshoeff, R.~Kallosh, T.~Ortin and G.~Papadopoulos,
``$\kappa$-Symmetry, supersymmetry and intersecting branes,''
Nucl.\ Phys.\ B {\bf 502}, 149 (1997)
[arXiv:hep-th/9705040].
%%CITATION = HEP-TH 9705040;%%

%\cite{Marino:1999af}
\bibitem{Marino:1999af}
M.~Marino, R.~Minasian, G.~W.~Moore and A.~Strominger,
``Nonlinear instantons from supersymmetric p-branes,''
JHEP {\bf 0001} (2000) 005
[arXiv:hep-th/9911206].
%%CITATION = HEP-TH 9911206;%%

%\cite{Cascales:2004qp}
\bibitem{Cascales:2004qp}
J.~F.~G.~Cascales and A.~M.~Uranga,
``Branes on generalized calibrated submanifolds,''
arXiv:hep-th/0407132.
%%CITATION = HEP-TH 0407132;%%

%\cite{Brunner:1999jq}
\bibitem{Brunner:1999jq}
I.~Brunner, M.~R.~Douglas, A.~E.~Lawrence and C.~Romelsberger,
``D-branes on the quintic,''
JHEP {\bf 0008} (2000) 015
[arXiv:hep-th/9906200].
%%CITATION = HEP-TH 9906200;%%

%\cite{Harvey:1996gc}
\bibitem{Harvey:1996gc}
J.~A.~Harvey and G.~W.~Moore,
``On the algebras of BPS states,''
Commun.\ Math.\ Phys.\  {\bf 197}, 489 (1998)
[arXiv:hep-th/9609017].
%%CITATION = HEP-TH 9609017;%%

%%%%%%%%%%%%%%%%%%%%%%%%%%%%%%%%
% Section 4                    %
%%%%%%%%%%%%%%%%%%%%%%%%%%%%%%%%

%\cite{Dudas:2000ff}
\bibitem{Dudas:2000ff}
E.~Dudas and J.~Mourad,
``Brane solutions in strings with broken supersymmetry and dilaton  tadpoles,''
Phys.\ Lett.\ B {\bf 486} (2000) 172
[arXiv:hep-th/0004165].
%%CITATION = HEP-TH 0004165;%%

%\cite{Blumenhagen:2001te}
\bibitem{Blumenhagen:2001te}
R.~Blumenhagen, B.~K\"ors, D.~L\"ust and T.~Ott,
``The standard model from stable intersecting brane world orbifolds,''
Nucl.\ Phys.\ B {\bf 616}, 3 (2001)
[arXiv:hep-th/0107138].
%%CITATION = HEP-TH 0107138;%%

%\cite{Fischler:1986}
\bibitem{Fischler:1986}
W.~Fischler and L.~Susskind,
``Dilaton Tadpoles, String Condensates And Scale Invariance,''
Phys.\ Lett.\ B {\bf 171} (1986) 383;\\
W.~Fischler and L.~Susskind,
``Dilaton Tadpoles, String Condensates And Scale Invariance. 2,''
Phys.\ Lett.\ B {\bf 173} (1986) 262.

%\cite{StefanskiScrucca}
\bibitem{StefanskiScrucca}
B.~J.~Stefanski,
``Gravitational couplings of D-branes and O-planes,''
Nucl.\ Phys.\ B {\bf 548} (1999) 275
[arXiv:hep-th/9812088];\\
C.~A.~Scrucca and M.~Serone,
``Anomalies and inflow on D-branes and O-planes,''
Nucl.\ Phys.\ B {\bf 556} (1999) 197
[arXiv:hep-th/9903145].

%\cite{Blumenhagen:2002wn}
\bibitem{Blumenhagen:2002wn}
R.~Blumenhagen, V.~Braun, B.~K\"ors and D.~L\"ust,
``Orientifolds of K3 and Calabi-Yau manifolds with intersecting D-branes,''
JHEP {\bf 0207} (2002) 026
[arXiv:hep-th/0206038].
%%CITATION = HEP-TH 0206038;%%

\bibitem{Quevedo:1996uu}
F.~Quevedo and C.~A.~Trugenberger,
``Phases of antisymmetric tensor field theories,''
Nucl.\ Phys.\ B {\bf 501} (1997) 143
[arXiv:hep-th/9604196].

%\cite{Cremmer:1982en}
\bibitem{Cremmer:1982en}
E.~Cremmer, S.~Ferrara, L.~Girardello and A.~Van Proeyen,
``Yang-Mills Theories With Local Supersymmetry: Lagrangian, Transformation Laws
And Superhiggs Effect,''
Nucl.\ Phys.\ B {\bf 212} (1983) 413.
%%CITATION = NUPHA,B212,413;%%

%\cite{Wess:1992}
\bibitem{Wess:1992}
J.~Wess and J.~Bagger,
``Supersymmetry And Supergravity,''
Princeton University Press, Princeton, 1992.

%\cite{Haack:1999zv}
\bibitem{Haack:1999zv}
M.~Haack and J.~Louis,
``Duality in heterotic vacua with four supercharges,''
Nucl.\ Phys.\ B {\bf 575} (2000) 107
[arXiv:hep-th/9912181].
%%CITATION = HEP-TH 9912181;%%

%\cite{Ibanez:1998rf}
\bibitem{Ibanez:1998rf}
L.~E.~Ib\'a\~nez, C.~Mu\~noz and S.~Rigolin,
``Aspects of type I string phenomenology,''
Nucl.\ Phys.\ B {\bf 553}, 43 (1999)
[arXiv:hep-ph/9812397].
%%CITATION = HEP-PH 9812397;%%

%\cite{Kakushadze:1998wp}
\bibitem{Kakushadze:1998wp}
Z.~Kakushadze and S.~H.~H.~Tye,
``Brane world,''
Nucl.\ Phys.\ B {\bf 548}, 180 (1999)
[arXiv:hep-th/9809147].
%%CITATION = HEP-TH 9809147;%%

%\cite{Hsu:2003cy}
\bibitem{Hsu:2003cy}
J.~P.~Hsu, R.~Kallosh and S.~Prokushkin,
``On brane inflation with volume stabilization,''
JCAP {\bf 0312} (2003) 009
[arXiv:hep-th/0311077].
%%CITATION = HEP-TH 0311077;%%

%\cite{Berg:2004ek}
\bibitem{Berg:2004ek}
M.~Berg, M.~Haack and B.~K\"ors,
``Loop corrections to volume moduli and inflation in string theory,''
arXiv:hep-th/0404087.
%%CITATION = HEP-TH 0404087;%%

%%%%%%%%%%%%%%%%%%%%%%%%%%%%%%%%
% Section 5                    %
%%%%%%%%%%%%%%%%%%%%%%%%%%%%%%%%

%\cite{Karoubi:1987}
\bibitem{Karoubi:1987}
M.~Karoubi and C.~Leruste,
``Algebraic topology via differential geometry'',
Cambridge University Press (1987).

%\cite{Morrison:1992}
\bibitem{Morrison:1992}
D.~R.~Morrison,
``Mirror symmetry and rational curves on quintic
threefolds: A guide for mathematicians'',
Jour.\ AMS. {\bf 6} (1993) 223
[arXiv:alg-geom/9202004].

%\cite{Aspinwall:1993nu}
\bibitem{Aspinwall:1993nu}
P.~S.~Aspinwall, B.~R.~Greene and D.~R.~Morrison,
``Calabi-Yau moduli space, mirror manifolds and spacetime topology  change in
string theory,''
Nucl.\ Phys.\ B {\bf 416} (1994) 414
[arXiv:hep-th/9309097].
%%CITATION = HEP-TH 9309097;%%

%\cite{Greene:1993vm}
\bibitem{Greene:1993vm}
B.~R.~Greene, D.~R.~Morrison and M.~R.~Plesser,
``Mirror manifolds in higher dimension,''
Commun.\ Math.\ Phys.\  {\bf 173}, 559 (1995)
[arXiv:hep-th/9402119].
%%CITATION = HEP-TH 9402119;%%

%\cite{Cox:1999}
\bibitem{Cox:1999}
D.~Cox and S.~Katz,
``Mirror symmetry and algebraic geometry'',
Mathematical Surveys and Monographs 68,
American Mathematical Society (1999).

%\cite{Forbes:2003ki}
\bibitem{Forbes:2003ki}
B.~Forbes,
``Open string mirror maps from Picard-Fuchs equations on relative cohomology,''
arXiv:hep-th/0307167.
%%CITATION = HEP-TH 0307167;%%

%\cite{Hosono:1994av}
\bibitem{Hosono:1994av}
For a review see, for example, 
S.~Hosono, A.~Klemm and S.~Theisen,
``Lectures on mirror symmetry,''
arXiv:hep-th/9403096.
%%CITATION = HEP-TH 9403096;%%

\end{thebibliography}
\endgroup
\end{document}